\documentclass[12pt]{article}
\usepackage{amssymb}
\usepackage{epsfig}
\usepackage{amsmath}

\newcommand{\lsim}   {\mathrel{\mathop{\kern 0pt \rlap
  {\raise.2ex\hbox{$<$}}}
  \lower.9ex\hbox{\kern-.190em $\sim$}}}
\newcommand{\gsim}   {\mathrel{\mathop{\kern 0pt \rlap
  {\raise.2ex\hbox{$>$}}}
  \lower.9ex\hbox{\kern-.190em $\sim$}}}
\numberwithin{equation}{section}

\newcommand{\beq}{\begin{equation}}
\newcommand{\eeq}{\end{equation}}
\newcommand{\bea}{\begin{eqnarray}}
\newcommand{\eea}{\end{eqnarray}}

\def\N{{\scriptscriptstyle N}}

\def\W{{\scriptscriptstyle W}}
\def\al{\alpha}
\def\be{\beta}
\def\gam{\gamma}
\def\psib{\overline{\psi}}

\def\half{\frac{1}{2}}
\def\nubar{\overline{\nu}}
\def\qbar{\overline{q}}
\def\lbar{\overline{\ell}}
\def\ubar{\overline{u}}
\def\dbar{\overline{d}}
\def\Nbar{\overline{N}}
\def\sbar{\overline{s}}
\def\psla{\rlap{/}{p}}
\def\Psla{\rlap{/}{P}}
\def\ksla{\rlap{/}{k}}
\def\qsla{\rlap{/}{q}}
\def\la{\langle}
\def\ra{\rangle}
\def\out{_{\mathrm{out}}\la}
\def\in{\ra_{\mathrm{in}}}
\def\Ncal{\mathcal{N}}

\def\Ucal{\mathcal{U}}
\def\Ocal{\mathcal{O}}
\def\Acal{\mathcal{A}}
\def\MSbar{$\overline{\rm{MS}}$\ }
\def\ff{form factor\ }
\def\ffs{form factors\ }
\def\FVem{F_1}
\def\FMem{F_2}
\def\GEem{G_E}
\def\GMem{G_M}
\def\FVV{F_1^V}
\def\FMV{F_2^V}
\def\GMV{G_M^V}
\def\GEV{G_E^V}
\def\GENC{G_E^{NC}}
\def\GMNC{G_M^{NC}}
\def\GANC{G_A^{NC}}
\def\GACC{G_A}
\def\GPCC{G_P^{CC}}
\def\GTCC{G_T^{CC}}
\def\GMCC{G_M^{CC}}
\def\GECC{G_E^{CC}}
\def\GAV{G_A^{V}}
\def\GPV{G_P^{V}}
\def\GTV{G_T^{V}}
\def\FVs{F_1^s}
\def\FMs{F_2^s}
\def\FTs{F_3^s}
\def\GAs{G_A^s}
\def\GPs{G_P^s}
\def\GTs{G_T^s}
\def\GMs{G_M^s}
\def\GEs{G_E^s}
\def\qv{{\vec q}}

\begin{document}

\title{Strangeness in the nucleon:
neutrino--nucleon and polarized
electron--nucleon scattering}
\author{W.M. Alberico$^\mathrm{a}$, 
S.M. Bilenky$^\mathrm{a,b}$,\footnote{On leave of absence from 
Joint Institute for Nuclear Research, Dubna, Russia}
C. Maieron$^\mathrm{a}$\\
\small\it $^\mathrm{a}$ Dipartimento di Fisica Teorica, 
Universit\`a di Torino\\
\small and INFN, Sezione di Torino\\
\small via P. Giuria 1, 10125 Torino, Italy\\
\small $^\mathrm{b}$ Scuola Internazionale Superiore 
di Studi Avanzati (SISSA)\\
\small I-34014 Trieste, Italy}

\date{\empty}

\maketitle

%%%%%%%%%%%%%%%%%%%%%%%%%%%%%%%%%%%%%%%%%%%%%%%%%%%%%%%%%%%%%%%%%%%%%%%
\begin{abstract}

After the EMC and subsequent experiments at CERN, SLAC and DESY
on the deep inelastic scattering of polarized leptons on
polarized nucleons, it
is now established that the $Q^{2}=0$ value of the axial strange
form factor of the nucleon, a quantity which is connected with
the spin of the proton and is quite relevant from the
theoretical point of view, 
is relatively large.

In this review we consider different
methods and observables that allow to obtain information on the strange
axial and vector form factors of the nucleon at different
values of  $Q^{2}$. These methods
are based on the investigation of the Neutral Current
induced effects such as the P-odd asymmetry in the  scattering of
 polarized electrons on protons and nuclei, the elastic
neutrino (antineutrino) scattering on protons and the quasi--elastic
neutrino (antineutrino) scattering on nuclei.
We discuss in details the phenomenology of these
processes and the existing experimental data.

\vspace{0.8cm}

%\end{frontmatter}
%\typeout{SET RUN AUTHOR to \@runauthor}
\end{abstract}

\newpage
\tableofcontents{}
\newpage

%%%%%%%%%%%%%%%%%%%%%%%%%%%%%%%%%%%%%%%%%%%%%%%%%%%%%%%%%%
\section{ Introduction}
\label{sec.intro}

In this review we will discuss the strange form factors
 of the nucleon.  As it is well known,
the net strangeness of the nucleon is equal to zero.
However, according to quantum field theory, in the cloud of a physical
nucleon there must be pairs of strange particles. From the point of view of
QCD the nucleon consists of valence $u$ and $d$ quarks 
and of a  sea of quark--antiquark pairs $\bar{u}u$, $\bar{d}d$,
$\bar{s}s$,.... produced by virtual gluons.

In the region of large $Q^2$  information about the
$\bar{s}s$ sea can be obtained from the experiments on the 
production of charmed particles in 
charged current interactions of neutrinos and antineutrinos with
nucleons in the deep inelastic region.
The charmed particles can be produced in $d-c$ and  $s-c$
transitions. The probability of  $d-c$ transitions is proportional to
$\sin^2\theta_C$, while the  probability of  $s-c$ 
transition is proportional to $\cos^2\theta_C$ ($\theta_C$ being the 
Cabibbo angle). Due to the smallness of $\theta_C$ 
($\sin^2\theta_C\simeq 4\cdot 10^{-2}$) the $d-c$
transition is a Cabibbo--suppressed
one. This enhances the possibility of studying the strange sea in the 
nucleon  by observing  two--muon neutrino events (one muon is produced 
by neutrinos and another muon is produced in the decay of charmed 
particles)~\cite{Vilain98,Rabino93,Abramo82,Adams99,Bazarko95}. 
In the latest NuTeV experiment at Fermilab~\cite{Adams99} the following
value was found for the ratio of the total momentum fraction carried by 
the strange (and anti-strange) sea quarks in the nucleon to the total
momentum fraction carried by $\bar{u}$ and $\bar{d}$:
\begin{equation}
\kappa=\frac{S+\overline{S}}{\overline{U}+\overline{D}}
=0.42\pm 0.07\pm 0.06\,.
\end{equation}
Here $\overline{Q}=\int_0^1 dx\,x\qbar(x)$, $\qbar$ being the number 
density of antiquarks $\qbar$ which carry the fraction $x$ of the
proton momentum $p$ (in the infinite momentum frame).
A recent analysis of deep inelastic scattering data found a larger 
value of $\kappa$~\cite{Barone00}.

The investigation of the matrix elements $\la p'|\sbar\Ocal s|p\ra$
($|p\ra$ being the state of a nucleon with momentum $p$
and $\Ocal$  some spin operator) in the confinement region  
$Q^2\lesssim 1$~GeV$^2$ is a very important 
subject~\cite{Ansel95,Lampe00,Ji01}.
Some information on  this matrix element 
can be obtained from the pion--nucleon scattering data and from the 
masses of strange baryons (see Ref.~\cite{Ellis00}).

Let us consider the scalar form factor
\begin{equation}
{\hat m} \la p'|(\ubar u + \dbar d)|p\ra = 
\ubar(p')u(p)\sigma_{\pi N}(t)
\end{equation}
where
\[ {\hat m}=\half(m_u+m_d),\quad t=(p'-p)^2\,.\]
Chiral perturbation theory allows one to connect the value of the scalar 
form factor in the Cheng--Dashen point $s=u=M^2$, $t=2m_{\pi}^2$
with the isospin--even amplitude of pion--nucleon 
scattering~\cite{Brown71,Cheng71,Gasser88,Cheng84}
($s$, $u$, $t$ are the customary Mandelstam variables, $M$ is the mass of 
the nucleon, $m_{\pi}$ the mass of the pion). From the results of the
phase--shift analysis of the low energy pion--nucleon data
it is possible to obtain the value of the form factor at the point $t=0$,
$\sigma_{\pi N}(0)$, which  is called the $\sigma$--term.

The calculation of the $\sigma$--term requires an extrapolation from the
point $t=2m_{\pi}^2$ to the point $t=0$. This procedure is based on 
dispersion relations and  chiral perturbation theory.
In Ref.~\cite{Gasser91} the value\footnote{
Let us notice that in a recent lattice calculation~\cite{Wright00}, made 
within two--flavor QCD, the range $45\div 55$~MeV for the value of the 
$\sigma$--term was obtained. The authors of Ref.~\cite{Wright00} obtained
this range of values, compatible with the one given by 
Eq.~(\ref{sigmaterm1}) and derived from experimental 
data, by using an extrapolation procedure in the 
quark masses which, at variance with previous attempts, respects the 
correct chiral behavior of QCD.}
\begin{equation}
\sigma_{\pi N}\simeq 45\pm 8\,\mathrm{MeV}
\label{sigmaterm1}
\end{equation}
was found for the $\sigma$--term.

Let us define the quantity
\begin{equation}
y_N= \frac{\la p|\sbar s|p\ra}{\half\la p|(\ubar u +\dbar d)|p\ra},
\label{yN}
\end{equation}
which characterizes the strange content of the nucleon.
If one assumes that the breaking of SU(3) is
due to the quark masses, the following relation, which connects the
parameters $y_N$ and $\sigma_{\pi N}$ with the mass difference of the 
$\Lambda$ and $\Xi$ hyperons, can be derived~\cite{Cheng88}:
\begin{equation}
\frac{1}{3}\left(1-\frac{m_s}{{\hat m}}\right)\left(1-y_N\right)
\sigma_{\pi N} \simeq M_{\Lambda}-M_{\Xi}
\label{sigmaterm2}
\end{equation}
Taking into account higher--order corrections and assuming that
the mass ratio $m_s/{\hat m}$ has the standard value $\simeq 25$, 
from (\ref{sigmaterm2}) we have
\begin{equation}
(1-y_N)\sigma_{\pi N}\simeq 31.8\,\mathrm{MeV}
\label{sigmaterm3}
\end{equation}
and by combining (\ref{sigmaterm1}) and (\ref{sigmaterm2}) we 
obtain $y_N\simeq 0.3$.

Let us stress that there are many uncertainties in the determination 
of the value of the $\sigma_{\pi N}$--term and of the parameter $y_N$. 
They are mainly  connected  with the pion--nucleon
experimental data and the extrapolation procedure. Larger values for 
the parameter $y_N$ up to $y_N=0.5$
were obtained by different authors (for a recent discussion, see
Ref.~\cite{Bottino00}).

The most convincing evidence in favor of a non--zero value of the axial
strange constant $g_{A}^s$ which characterizes the matrix
element $\la p|\sbar\gam_{\al}\gam_5 s|p\ra$,  was found from the
data of experiments on deep inelastic 
scattering of polarized leptons on polarized nucleons. 
The first indication in favor of $g_{A}^s\ne 0$
was obtained in the EMC experiment at CERN~\cite{Ashm89}.
Subsequent experiments at CERN~\cite{Adeva98},
SLAC~\cite{Abe95a,Abe95b,Abe98,Anthon00}
and DESY~\cite{Hermes98,Hermes97} confirmed the EMC result. 

These experiments triggered a large number of
theoretical papers in which the problem of the strangeness of the
nucleon was investigated in detail (see the recent 
reviews~\cite{Lampe00,Ji01,Ellis00,Hai00,Hugh99}).

The cross section of the scattering of longitudinally polarized leptons
on polarized nucleons is characterized by four dimensionless
structure functions of the variables $x$ and $Q^2$:
$F_{1}(x,Q^{2})$,  $F_{2}(x,Q^{2})$,  $g_{1}(x,Q^{2})$ and
$g_{2}(x,Q^{2})$ (here $x=Q^2/2p\cdot q$, $Q^2=-q^2$, $p$ is the
momentum of the initial nucleon, $q$ the momentum of the virtual photon).
The functions  $F_{1}(x,Q^{2})$ and $F_{2}(x,Q^{2})$
determine the unpolarized cross section, while the functions
$g_{1}(x,Q^{2})$ and  $g_{2}(x,Q^{2})$ characterize the part of the cross
section which is proportional to the product of the polarizations
of leptons and nucleons.

The measurement of the asymmetry in the deep inelastic scattering of
longitudinally polarized leptons on
longitudinally polarized nucleons allows one to determine the structure
function  $g_{1}(x,Q^{2})$. 

In the framework of the naive parton model,
based on the assumption that in the infinite momentum frame
($|\vec{p}\,|\to\infty$) partons (quarks) can be considered as a free
particles, all structure functions depend only on the scaling variable $x$.
Let us consider, in the infinite momentum frame, a nucleon 
with helicity equal to one.
For the function  $g_{1}(x)$ we have
\begin{equation}
g_1(x)= \half \sum_q e_q^2\left[q^{(+1)}(x)+\qbar^{(+1)}(x)
-q^{(-1)}(x)-\qbar^{(-1)}(x)\right]
\end{equation}
where $e_q$ is the charge of the quark (in the unit of the proton charge),
$q^{(\pm1)}(x)$ ($\qbar^{(\pm1)}(x)$)
is the number--density of the quarks $q$
(antiquarks $\bar{q}$) with momentum $xp$ and helicity equal
(respectively, opposite) to the helicity of the nucleon.
Thus, the structure function  $g_{1}$ is determined
by the differences of the number of quarks and antiquarks with positive
and negative helicities. Notice that in the naive parton model
the structure functions $F_{1}(x)$ and  $F_{2}(x)$ are given by
\begin{equation}
F_2(x)=x\sum_q e_q^2 q(x) = 2xF_1(x)\, ,
\end{equation}
where
\bea
q(x)&=& q^{(+1)}(x) +q^{(-1)}(x),
\nonumber\\
\qbar(x)&=& \qbar^{(+1)}(x)+\qbar^{(-1)}(x)
\nonumber
\eea
are the total numbers of quarks and antiquarks with momentum $xp$.

From the theoretical point of view, the important quantity is the first
moment of the structure function  $g_{1}$:
\begin{equation}
\Gamma_1(Q^2)=\int_0^1 g_1(x,Q^2) dx\,.
\end{equation}
In the region  $Q^2\lesssim 10$~GeV$^2$ the main contribution to 
$\Gamma_{1}$ comes from the light $u,d,s$ quarks.

In the naive parton model, for the first moment of the proton we have
\begin{equation}
\Gamma_1^p=\half \left(\frac{4}{9}\Delta u +\frac{1}{9}\Delta d +
\frac{1}{9}\Delta s\right)
\label{Gamma1a}
\end{equation}
where
\begin{equation}
\Delta q = \int_0^1\sum_{r=\pm 1} r\left[q^{(r)}(x) +\qbar^{(r)}(x)\right]dx
\label{deltaq}
\end{equation}
is the difference of the total numbers of quarks and antiquarks in the
nucleon with helicity equal and opposite to the helicity of the nucleon.
Thus $\Delta q$ is the contribution of the 
$q$-quarks and $\bar q$-antiquarks to the spin of the proton.

The first moment $\Gamma_{1}^p$
can be determined from the
measurement of the deep inelastic scattering of
polarized leptons on longitudinally polarized protons.
In the EMC experiment at  $Q^2 = 10.7$~GeV$^2$ the value
\begin{equation}
\Gamma_1^p(10.7)=0.126\pm 0.010\pm 0.015
\label{Gamma1b}
\end{equation}
was found, while in the latest CERN SMC~\cite{Adeva98}, 
SLAC 155~\cite{Anthon00}
and DESY HERMES~\cite{Hermes98,Hermes97} experiments the following values of
$\Gamma_1^p$ were determined:
\bea
\Gamma_1^p(10)&=& 0.120\pm 0.005\pm 0.006\pm 0.014
\nonumber\\
\Gamma_1^p(5)&=& 0.118\pm 0.004\pm 0.007
\label{Gamma1c}\\
\Gamma_1^p(3)&=& 0.122\pm 0.003\pm 0.010
\nonumber
\eea

Let us stress that the experimental data can be obtained in a
limited interval of the variable $x$ which does not include the
region of very small and very large values of $x$.
In order to determine the value of $\Gamma_{1}^p$ it is necessary to 
make extrapolations of the data to the points  $x=0$ and $x=1$. 
The small $x$--behavior of the structure function $g_{1}$ 
is the most contradictory issue (see Ref.~\cite{Lampe00}).
Usually a Regge--behavior of the function $g_{1}$ is assumed at small $x$.
Recent extrapolations are based on  NLO (next to leading order) QCD fits.
Notice that in some non--perturbative approaches a
singular behavior of the function $g_{1}(x)$ at small $x$ was
obtained~\cite{Lampe00}.

Let us now discuss the possibilities of determining  the 
axial strange constant  $g_{A}^s$ from these data.

In the framework of the naive parton model the quantities 
$\Delta q$, which enter into the sum rule (\ref{Gamma1a}), are determined
by the one--nucleon matrix element of the axial quark current
$\qbar\gam_{\al}\gam_5 q$ (see Section \ref{sec.matrixNC}):
\begin{equation}
{_p\la}p|\qbar\gam_{\al}\gam_5 q|p\ra_p = 2Ms_{\al}\Delta q
\label{axialcur}
\end{equation}
Here
$s_{\alpha}$ is the polarization vector of the nucleon and $|p\ra_p$
($|p\ra_n$) is the state vector of a proton (neutron) with momentum $p$. 
The relation (\ref{axialcur}) allows one to obtain two constraints on the
quantities $\Delta u$, $\Delta d$ and $\Delta s$.
The first one comes from the isotopic SU(2) invariance of strong
interactions, which implies:
\begin{equation}
{_p\la}p|\ubar\gam_{\al}\gam_5 d|p\ra_n =
{_p\la}p|\left(\ubar\gam_{\al}\gam_5 u
-\dbar\gam_{\al}\gam_5 d\right)|p\ra_p\,.
\label{axialcur1}
\end{equation}
From (\ref{axialcur1}) and (\ref{axialcur}) it follows that
\begin{equation}
\Delta u - \Delta d =g_A
\label{deltaud}
\end{equation}
where $g_{A}$ is the weak axial constant. 
From the data on the $\beta$--decay of the neutron it follows
that~\cite{PDG00}:
\begin{equation}
g_A= 1.2670\pm 0.0035\,.
\label{gA}
\end{equation}

The second constraint follows from SU(3) symmetry. Assuming exact
SU(3) symmetry we have
\begin{equation}
\Delta u +\Delta d -2\Delta s =3F-D\equiv g_A^8
\label{gA8}
\end{equation}
where $F$ and $D$ are the constants which determine the matrix elements
of the axial weak current for the states of different hyperons belonging
to the SU(3) octet. From the fit of the experimental data it was found
that~\cite{PDG00}:
\begin{equation}
F=0.463\pm 0.008;\qquad D=0.804\pm 0.008
\label{FD}
\end{equation}
hence
\begin{equation}
g_A^8=0.585\pm 0.025
\label{gA8b}
\end{equation}

Now from Eqs.~(\ref{Gamma1a}), (\ref{gA}) and (\ref{FD}) we can express the
first moment as follows:
\begin{equation}
\label{Gamma1d}
\Gamma_1^p= 0.187 \pm 0.004 +\frac{1}{3}\Delta s\,.
\end{equation}
If we compare (\ref{Gamma1d}) with the values of $\Gamma_{1}^{p}$
which were obtained in experiments [see (\ref{Gamma1b}) and
(\ref{Gamma1c})], we come to the conclusion that the quantity
 $\Delta s$, which determines the matrix element of the strange axial
current, is different from zero and negative. Using, for example, the
EMC result (\ref{Gamma1b}), from (\ref{Gamma1d}) we find
\[ \Delta s= -0.18\pm 0.05 \,.\]
This conclusion is based on the naive parton model and was obtained
about 10 years ago (see Ref.~\cite{Ansel95}).
After the EMC result was obtained,
a lot of experimental and theoretical works were published.
The LO (leading order) and NLO
QCD corrections to the sum rule (\ref{Gamma1a}) and to the
relation (\ref{axialcur1})
were calculated and many different effects were taken into account
 (see the 
reviews~\cite{Lampe00,Ji01,Ellis00,Hai00,Hugh99,Shore98,LlSmith98,Mallot99,Keh-Liu00}).

Let us introduce the constants $A_i$
\begin{equation}
{_p\la}p|\psib\gam_{\al}\gam_5\lambda_i\psi|p\ra_p=2Ms_{\al}A_i
\quad (i=3,8)
\label{Ai}
\end{equation}
and
\begin{equation}
{_p\la}p|\psib\gam_{\al}\gam_5\psi|p\ra_p=2Ms_{\al}A_0
\label{A0}
\end{equation}
where $\psi=\left(\begin{array}{c}u\\d\\s\end{array}\right)$ 
is the flavor SU(3) triplet, $\lambda_i$ are the Gell--Mann matrices,
$\psib\gam_{\al}\gam_5\lambda_i\psi$  is the SU(3) octet of axial
currents and $\psib\gam_{\al}\gam_5\psi$ the axial singlet current.

In the naive parton model the following relations hold:
\bea
A_3 &=&\Delta u -\Delta d = g_A^3 \equiv g_A
\nonumber\\
A_8 &=&\Delta u +\Delta d -2\Delta s\equiv g_A^8
\label{Aieqs}\\
A_0 &=&\Delta u +\Delta d +\Delta s = g_A^0\equiv \Delta\Sigma
\nonumber
\eea
the quantity $\Delta\Sigma$ being the total contribution of quarks
and antiquarks to the spin of the proton.
The sum rule (\ref{Gamma1a}) can be now rewritten in the form
\begin{equation}
\Gamma_1^p=\frac{1}{12}A_3 +\frac{1}{36}A_8 +\frac{1}{9}A_0\,.
\label{Gamma1e}
\end{equation}

The NLO QCD corrections modify the above expression as follows
(see~\cite{Lampe00} and references therein):
\begin{equation}
\Gamma_1^p(Q^2)=\left(1-\frac{\al_s(Q^2)}{\pi}\right)
\left[\frac{1}{12}A_3 +\frac{1}{36}A_8 +\frac{1}{9}A_0(Q^2)
\right]
\label{Gamma1f}
\end{equation}
where $\alpha_{s}(Q^{2})$ is the strong coupling constant and the
quantities $A_{i}$ ($i=3,8,0$) are given by the relation (\ref{Ai}).
The quantities $A_{3}$ and $A_{8}$ are determined by the one--nucleon 
matrix elements of the corresponding conserved currents 
(we have assumed SU(3) flavor symmetry).
These quantities  do not depend on $Q^{2}$ and turn out to be 
\begin{equation}
A_{3}= g_{A}\,~~~ A_{8}= g_{A}^{8}\,,
\end{equation}
where the numerical values of the constants $ g_{A}$ and 
and $g_{A}^{8}$ are given by Eqs.~(\ref{gA}) and  (\ref{gA8b}), 
respectively. The quantity $A_{0}$, instead, 
is determined by the matrix element of the {\em non--conserved} singlet
current $\bar\psi \gamma^{\alpha} \gamma_{5}\psi$.
If higher order QCD corrections are taken into account, then this quantity
depends on the renormalization scheme and on the renormalization scale, 
which is usually taken to be equal to $Q^{2}$. 

Two renormalization schemes are commonly employed: the 
\MSbar scheme~\cite{thooft72} and the AB (Adler--Bardeen)
scheme~\cite{Ball96}.
In the \MSbar scheme $A_0(Q^2)$ is determined by the renormalization 
scale--dependent contribution of quarks to the spin of nucleon,
\begin{equation}
A_{0}(Q^{2})=\Delta \Sigma(Q^{2}) = \sum_{q=u,d,s}\Delta q(Q^{2})\,.
\end{equation}
In the AB scheme $A_{0}(Q^{2})$ is determined by the contribution of
quarks and gluons to the spin of the proton
\begin{equation}
A_{0}(Q^{2})= \Delta \Sigma^{AB} - 3 \frac{\alpha_{s}(Q^{2})}{2\pi}
\Delta G(Q^{2})
\label{ABA0}
\end{equation}
where $\Delta \Sigma^{AB}$ does not depend on $Q^{2}$
and all renormalization scale dependence is absorbed by the gluon
contribution $\Delta G(Q^{2})$. The latter,  due to triangle 
anomaly~\cite{Efremo88,Altare88}, behaves as $1/\al_s$ and can give 
a sizable contribution to $A_0(Q^2)$. \footnote{
Let us notice that the relation (\ref{ABA0}) offers the possibility of
explaining the data by the large gluon contribution~\cite{Efremo88,Altare88}.
In fact, $A_0$ can be written in the form
\[A_0=g_A^8+ 3\Delta s -3\frac{\al_s}{2\pi}\Delta G\,.\]
Even if we assume that $\Delta s=0$, the experimental data
can be explained by a large positive $\Delta G$.}

In the  $\overline{\rm{MS}}$ scheme, from Eqs.~(\ref{axialcur}),
(\ref{Aieqs}) and (\ref{Gamma1f}),  for the
matrix element of the axial strange current in NLO approximation we have
\begin{equation}
\frac{1}{2M}\la p|\sbar\gam_{\al}\gam_5 s|p\ra s^{\al} =
-3\left(1-\frac{\al_s(Q^2)}{\pi}\right)^{-1}\Gamma_1^p(Q^2)
+\frac{1}{4}g_A +\frac{5}{12}g_A^8 \,.
\label{melasc}
\end{equation}

Let us stress that the matrix element $\la p|\sbar\gam_{\al}\gam_5 s|p\ra$ 
depends on the renormalization scale. Using the E155 data~\cite{Anthon00}
at $Q^{2} = 5~ \rm{GeV}^{2}$  and the values (\ref{gA}), (\ref{gA8b}) 
of the axial constants $g_{A}$ and $g_{A}^{8}$,
from (\ref{melasc}) we find
\begin{equation}
\frac{1}{2M}\la p|\sbar\gam_{\al}\gam_5 s|p\ra s^{\al} = 0.12\pm 0.03
\end{equation}
Thus, if we take into account higher order QCD corrections,
from the data on deep inelastic scattering of polarized
leptons on polarized protons we can conclude
that the one--nucleon matrix element of the axial strange current
is relatively large. This conclusion does not depend on the
renormalization scheme (matrix elements are measurable quantities).
Similar considerations  can be drawn from the operator product 
expansion (OPE) approach~\cite{Politz74}. 

In this review we will consider possibilities
of obtaining information on the strange vector and axial form factors of
the nucleon from the investigation of neutral current effects.
We will consider in detail the P--odd asymmetry in elastic 
and quasi--elastic scattering of
polarized electrons on nucleons and nuclei (Sections \ref{sec.Poddel},
\ref{sec.Poddelex}, \ref{sec.PoddAel}, \ref{sec.PoddQE}) and
the elastic and quasi--elastic scattering 
of neutrinos (antineutrinos) on nucleons and nuclei
(Sections \ref{sec.nufree}, \ref{sec.freeasymm},
\ref{sec.nuAel}, \ref{sec.nuQE}). 
We will discuss the existing experimental data
and future experiments. Derivations of many 
basic relations will be presented.

%%%%%%%%%%%%%%%%%%%%%%%%%%%%%%%%%%%%%%%%%%%%%%%%%%%%%%%%%%
\section{The Standard Lagrangian of the interaction of 
leptons and quarks with vector bosons}
\label{sec.SL}
 
In the Standard SU(2)$\times$U(1) electroweak Model 
(SM)~\cite{Glashow61,Weinberg67,Salam68}
the Lagrangian of the interaction of the fundamental
fermions (neutrinos, charged leptons and quarks) with vector 
bosons contains three parts: charged current (CC), 
electromagnetic (em) and neutral current (NC) 
interactions~\cite{Cheng84,Weinberg96,Pesk,Quigg,Leader,Bilenky94}.

%%%%%%%%%%%%%%%%%%%%%%%%%%%%%%%%%%%%%%%%%%%%%%%%%%%%%%%%%%
\subsection{The charged current Lagrangian}
\label{sec.SLCC}
The  Lagrangian of the CC interaction of leptons and
quarks with the charged vector bosons  $W^{\pm}$ reads:
\begin{equation}
{\mathcal{L}}^{CC}_I = 
-\frac{g}{2\sqrt{2}}\,j_\al^{CC} W^{\al} + \mathrm{h.c.}
\label{Lcc}
\end{equation}
Here $g$ is a coupling constant which is connected with Fermi constant
$G_{F}$ by the relation 
\begin{equation}
\frac{G_F}{\sqrt{2}} = \frac{g^2}{8 m_\W^2}
\end{equation}
($m_\W$ being the mass of the W--boson) and 
\begin{equation}
j_\al^{CC} = 2\left( j_\al^1 +i j_\al^2\right) \equiv 2 j_\al^{1+i2}
\label{cc1}
\end{equation}
is the charged current. 
In Eq.~(\ref{cc1}) $j_\al^{1,2}$ are components of the isovector 
current:\footnote{We use the Feynman--Bjorken--Drell metric. 
In this metric $g^{00}=1,\, g^{ii}=-1\,\,(i=1,2,3)$, the non--diagonal
 elements of $g^{\al\be}$ being equal to zero.
Thus, the scalar product of vectors $A^\al$ and  $B^\al$ 
is $A\cdot B \equiv A_{\al} B^{\al}=A^0 B^0 -\vec{A}\cdot\vec{B}$. 
Moreover the Dirac matrices
$\gamma^{\al}$ satisfy the commutation relations
$\gamma^{\al}\gamma^{\be}+\gamma^{\be}\gamma^{\al}=2g^{\al\be}$
and we adopt the definition $\gamma_5=i\gamma^0\gamma^1\gamma^2\gamma^3$
for the  matrix $\gamma_5$ and the definition $\epsilon_{0 1 2 3} =1$
for the antisymmetric tensor
$\epsilon_{\al\be\rho\sigma}$.  For the spinors $u(p)$ we 
will use the covariant normalization $\ubar(p)\gam^{\al}u(p)=2p^{\al}$.
In this metric 
${\gamma^{\al}}^\dagger=\gamma^0\gamma^{\al}\gamma^0$.
Notice also that vector of states are normalized in such a way that
$\la p'|p\ra = 2p^0(2\pi)^3\delta^{(3)}(\vec{p'}-\vec{p})$
(see for example~\cite{IZub}). With this choice the normalizing factors
do not appear in the matrix elements of the currents, but only 
in the final expression of the cross sections.
}
\begin{equation}
j_\al^i=\sum_a \psib_{aL}\gam_{\al}\frac{1}{2}\tau^i\psi_{aL}
\label{cc2}
\end{equation}
where  $\psi_{aL}=\half(1-\gam_5)\psi_a$ 
are left--handed doublets of the SU(2)xU(1) gauge group of the 
Standard Model:
\begin{equation}
\begin{array}{lll}
\psi_{eL}=\left(\begin{array}{c} \nu_{eL}\\ e_L\end{array}\right),\quad
&\psi_{\mu L}=\left(\begin{array}{c} \nu_{\mu L}\\ \mu_L\end{array}\right),
\quad
&\psi_{\tau L}=\left(\begin{array}{c} \nu_{\tau L}\\ \tau_L\end{array}
\right)\\
\psi_{1L}=\left(\begin{array}{c} u_{L}'\\ d_L'\end{array}\right),\quad
&\psi_{2L}=\left(\begin{array}{c} c_{L}'\\ s_L'\end{array}\right),\quad
&\psi_{3L}=\left(\begin{array}{c} t_{L}'\\ b_L'\end{array}\right)
\end{array}
\label{doublet}
\end{equation}

In terms of the fields of leptons and quarks with definite masses
the charged current (\ref{cc2}) reads:
\begin{equation}
j_{\al}^{CC}=2\sum_{\ell=e,\mu,\tau}{\nubar_{\ell}}_L\gam_\al\ell_L
+2\left[{\overline{ u}}_L\gam_\al d^{\mathrm{mix}}_L +
{\overline{c}}_L\gam_\al s^{\mathrm{mix}}_L +
{\overline{t}}_L\gam_\al b^{\mathrm{mix}}_L\right]
\end{equation}
where now
\begin{equation}
d^{\mathrm{mix}}_L =\sum_{q=d,s,b} V_{uq}q_L,\quad
s^{\mathrm{mix}}_L =\sum_{q=d,s,b} V_{cq}q_L,\quad
b^{\mathrm{mix}}_L =\sum_{q=d,s,b} V_{tq}q_L,\quad
\end{equation}
and $V$ is the unitary $3\times 3$ Cabibbo--Kobayashi--Maskawa 
mixing matrix.

%%%%%%%%%%%%%%%%%%%%%%%%%%%%%%%%%%%%%%%%%%%%%%%%%%%%%%%%%%
\subsection{ The electromagnetic interaction Lagrangian}
\label{sec.SLem}
The Lagrangian of the electromagnetic interaction has the form:
\begin{equation}
{\mathcal{L}}^{em}_I = -e j_\al^{em} A^{\al}\, ,
\label{Lem}
\end{equation}
$e$ being the charge of the proton and
\begin{equation}
j_{\al}^{em} = \sum_{\ell=e,\mu,\tau} (-1){\overline{\ell}}\gam_{\al}\ell
+ \sum_{q=u,d,\dots} e_q {\overline{q}}\gam_{\al} q
\label{jem}
\end{equation}
the electromagnetic current (with $e_u=2/3$, $e_d=-1/3$, \dots)

\subsection{ The neutral current Lagrangian}
\label{sec.SLNC}
The Lagrangian of the NC interaction
of leptons and quarks with the neutral vector boson $Z^0$ is:
\begin{equation}
{\mathcal{L}}^{NC}_I = -\frac{g}{2\cos\theta_W} j_\al^{NC} Z^{\al}\, ,
\label{Lnc}
\end{equation}
where $\theta_W$ is the weak (Weinberg) angle, the characteristic parameter 
of the electroweak unification, and $j_{\al}^{NC}$
is the neutral current. The structure of the latter in the
Standard Model is determined by the requirements of unification of the
weak and electromagnetic interactions into the unified electroweak
interaction. We have\footnote{
We notice that in the literature different definitions of NC are used.
In particular a frequently used notation differs from (\ref{jnc1}) by a
factor of 2:
\[{\tilde j}_{\al}^{NC} = 2 j_{\al}^{NC}.\] }
\begin{equation}
\label{jnc1}
j_{\al}^{NC}=2 j_{\al}^3 - 2\sin^2\theta_W j_{\al}^{em}\, .
\end{equation}

From Eqs.~(\ref{cc2}), (\ref{doublet}) and (\ref{jnc1}) the neutral current
can be rewritten in the following form:
\bea
\label{jnc2}
j_{\al}^{NC}&&=\sum_{q=u,c,t}\qbar\gam_{\al}(1-\gam_5)\half q
+\sum_{q=d,s,b}\qbar\gam_{\al}(1-\gam_5)\left(-\half\right) q +
\nonumber\\
&&+\sum_{\ell=e,\mu,\tau}\nubar_{\ell}\gam_{\al}(1-\gam_5)\half\nu_{\ell}
+\sum_{\ell=e,\mu,\tau}\lbar\gam_{\al}(1-\gam_5)\left(-\half\right)\ell +
\nonumber\\
&& - 2\sin^2\theta_W j_{\al}^{em}\, .
\eea
In this review we will focus on processes at relatively
small energies (less than a few GeV). Therefore it can be convenient to 
separate, in (\ref{jnc2}), the contribution  of the lightest $u$ and $d$ 
quarks. One obtains:
\begin{equation}
\label{jnc3}
j_{\al}^{NC;q}=v_{\al}^3 -a_{\al}^3 -\half\left(v_{\al}^s-a_{\al}^s\right)
- 2\sin^2\theta_W j_{\al}^{em}\, .
\end{equation}
Here we define:
\bea
v_{\al}^3&&= \ubar\gam_{\al}\half u -\dbar\gam_{\al} \half d
\equiv {\overline{N}}\gam_{\al}\half\tau_3 N,
\nonumber\\
a_{\al}^3&&= \ubar\gam_{\al}\gam_5\half u 
-\dbar\gam_{\al} \gam_5\half d
\equiv {\overline{N}}\gam_{\al}\gam_5\half\tau_3 N,
\nonumber
\eea
where $N=\scriptscriptstyle{\left(\begin{array}{c}u\\d\end{array}\right)}$. 
Indeed the mass difference between  $d$ and $u$ quarks 
($m_{d}-m_{u}\simeq 3$~MeV~\cite{PDG00} is much smaller than the
QCD constant $\Lambda^{\mathrm{QCD}}\simeq 200\div 300$~MeV and can
be neglected: in this case
$N$ is a doublet of the isotopic SU(2) group and the
currents $v^3_{\al}$ and $a^3_{\al}$ are the third components of the
isovectors  
\begin{equation}
\label{isovecq}
v_{\al}^i = \Nbar\gam_{\al}\half\tau^i N,\quad 
a^i_{\al}= \Nbar\gam_{\al}\gam_5\half\tau^i N\, .
\end{equation}
Instead, the currents $v^s_{\al}$ and $a^s_{\al}$ are isoscalars: they
represent the contributions to $j_{\al}^{NC;q}$  of the  $s$, $c$ 
and heavier quarks. Taking into account only $s$--quarks we have 
\begin{equation}
v_{\al}^s=\sbar\gam_{\al}s, \quad
a_{\al}^s=\sbar\gam_{\al}\gam_5 s
\end{equation}

The quark electromagnetic current is given by [see Eq. (\ref{jem})]
\begin{equation}
j_{\al}^{em;q}= \sum_{q=u,d,\dots} e_q \qbar\gam_{\al} q
\end{equation}
Also in this case it is convenient to separate, in the above current, 
the contributions of the lightest  $u$, $d$ quarks. Taking into 
account that $e_q=I_3^q+\scriptstyle{\frac{1}{6}}$ ($q=u,d$) we have:
\begin{equation}
j_{\al}^{em;q}= v_{\al}^3 + v_{\al}^0
\label{jemq}
\end{equation}
where $v_{\al}^0$ is the isoscalar current which, in the $u$, $d$, $s$
approximation, reads:
\begin{equation}
 v_{\al}^0 = \frac{1}{6}\Nbar\gam_{\al} N + 
\left(-\frac{1}{3}\right)\sbar\gam_{\al} s\, .
\label{jemisosca}
\end{equation}
%
%%%%%%%%%%%%%%%%%%%%%%%%%%%%%%%%%%%%%%%%%%%%%%%%%%%%%%%%%%%%%%%%%%%%
\section{ One-nucleon matrix elements of the neutral current}
\label{sec.matrixNC}

We will considerer here in detail the one--nucleon matrix elements of
the neutral current as well as the ones 
 of the electromagnetic current. Let us consider, for example, the
process of the elastic scattering of muon--neutrino on the nucleon:
\begin{equation}
\label{nuscat}
\nu_{\mu} + p \longrightarrow \nu_{\mu} + p
\end{equation}
The amplitude of this process is given by the expression
\bea
&&\la f|S|i\ra =
\label{mat1}\\
&&\quad =
 -i\frac{G_F}{\sqrt{2}}
%{\Ncal}_{k'}{\Ncal}_k
\ubar(k')\gam^{\al}(1-\gam_5)u(k)
\,\out p'|J_{\al}^{NC}(0)|p\in
(2\pi)^4 \delta^{(4)}(p'-p-q)
\nonumber
\eea
where  $k$ and $p$ ($k'$ and $p'$) are the four--momenta of the initial
(final) neutrino and nucleon, respectively, 
$q=k-k'$  and
\begin{equation}
\label{mat2}
\out p'|J_{\al}^{NC}(0)|p\in = \la p'| T\left\{ j_{\al}^{NC}(0)
e^{-i\int\,{\mathcal H}_I^{had}(x) d^4 x}\right\} |p\ra
\end{equation}
is the hadronic matrix element\footnote{The indexes ``in'' and ``out'' 
will be dropped hereafter.}.

In Eq. (\ref{mat2}) ${\mathcal H}_I^{had}(x)$ is the Hamiltonian density 
of the strong interactions, $J_{\al}^{NC}(0)$, $|p\in$ and 
$|p\ra_{\mathrm{out}}$ are the neutral current operator, the initial and 
final nucleon states in the Heisenberg representation.

From (\ref{jnc3}) we get the following expression for the matrix element
of the neutral current:
\bea
\la p'|J_{\al}^{NC}(0)|p\ra &&=
\la p'|\left(V^3_{\al}-A^3_{\al}\right)|p\ra
-\half\la p'|\left(V^s_{\al}-A^s_{\al}\right)|p\ra +
\nonumber\\
&&\qquad
 -2\sin^2\theta_W \la p'|J_{\al}^{em}|p\ra
\label{mat3}
\eea
where $V^{3(s)}_{\al}$, $A^{3(s)}_{\al}$ are the currents in Heisenberg
representation. From the isotopic SU(2) invariance of strong
interactions it follows that $V^3_{\al}$, $A^3_{\al}$ are the third 
components of the isovector currents  $V^i_{\al}$ and $A^i_{\al}$
($i=1,2,3$) while  $V^s_{\al}$, $A^s_{\al}$ are isoscalar currents.

The isotopic invariance of strong interactions allows one to determine
the one--nucleon matrix elements of the current $V_{\al}^3$ from
the one--nucleon matrix elements of the electromagnetic current
$J^{em}_{\al}$. In fact, from (\ref{jemq}) it follows
\begin{equation}
{_{p(n)}\la} p'|J_{\al}^{em}|p\ra_{p(n)} =
{_{p(n)}\la} p'|V_{\al}^{3}|p\ra_{p(n)} +
{_{p(n)}\la} p'|V_{\al}^{0}|p\ra_{p(n)}
\label{jem1}
\end{equation}
where $|p\ra_p$ ($|p\ra_n$) is the state of a proton (neutron) with
momentum  $p$. Furthermore we have 
\begin{equation}
\Ucal V_{\al}^3 \Ucal^{-1}=-V^3_{\al},\quad
\Ucal V_{\al}^0 \Ucal^{-1}=V^0_{\al},
\label{Uop}
\end{equation}
where the charge symmetry operator $\Ucal =\exp\{i\pi I_2\}$ 
(rotation of  $\pi$ around the second axis in the isotopic space)
 transforms proton states into neutron states and viceversa, according to:
\[
\Ucal|p\ra_p=-|p\ra_n,\quad \Ucal|p\ra_n=|p\ra_p\, .
\]
The following relations then hold:
\begin{equation}
\begin{array}{ll}
{_{p}\la} p'|V_{\al}^{3}|p\ra_{p} = -&{_{n}\la} p'|V_{\al}^{3}|p\ra_{n}\,,\\
{_{p}\la} p'|V_{\al}^{0}|p\ra_{p} = +&{_{n}\la} p'|V_{\al}^{0}|p\ra_{n}
\end{array}
\label{iso1}
\end{equation}

From (\ref{jem1}) and (\ref{iso1}) we have then:
\begin{equation}
{_{p}\la} p'|V_{\al}^{3}|p\ra_{p}=
\half \left[{_{p}\la} p'|J_{\al}^{em}|p\ra_{p}-
{_{n}\la} p'|J_{\al}^{em}|p\ra_{n}\right]\,.
\label{vector}
\end{equation}
Moreover
\begin{equation}
{_{p}\la} p'|V_{\al}^{0}|p\ra_{p}=
\half \left[{_{p}\la} p'|J_{\al}^{em}|p\ra_{p}+
{_{n}\la} p'|J_{\al}^{em}|p\ra_{n}\right]\,.
\label{scalar}
\end{equation}

Let us discuss now the one--nucleon matrix elements of the electromagnetic
current. The conservation law of the electromagnetic current, 
$\partial^{\al}J_{\al}^{em}=0$, entails
\begin{equation}
(p'-p)\la p'|J_{\al}^{em}|p\ra =0\, .
\end{equation}
From this relation it follows that the one--nucleon matrix elements of the
electromagnetic current are characterized by two \ffs and have the general
form
\begin{equation}
\label{matem1}
\la p'|J_{\al}^{em}|p\ra =
%\Ncal_{p'}\Ncal_p
\ubar(p')
\left[\gam_{\al} \FVem(Q^2) +\frac{i}{2M}\sigma_{\al\be}q^{\be}\FMem(Q^2)
\right]u(p)
\end{equation}

Here $q=p'-p$ is the four--momentum transfer, $Q^2=-q^2$, 
$\sigma_{\al\be}={\scriptstyle\frac{i}{2}}
(\gam_{\al}\gam_{\be}-\gam_{\be}\gam_{\al})$, 
$\FVem(Q^2)$ and $\FMem(Q^2)$ are the Dirac and Pauli \ffs. 
At $Q^2=0$ we have
\begin{equation}
\FVem(0)=e_N,\quad \FMem(0)=\kappa_N
\end{equation}
where  $e_N$ is the nucleon charge (in  units of the proton charge) 
and  $\kappa_{N}$ is the anomalous magnetic moment of the nucleon (in 
units of the nucleon Bohr magneton). Notice that from the invariance of 
strong interactions under time reversal it follows that the \ffs
are real functions of $Q^2$. 

With the help of the Dirac equation the matrix element (\ref{matem1})
can be rewritten in the form:
\begin{equation}
\label{matem2}
\la p'|J_{\al}^{em}|p\ra =
%\Ncal_{p'}\Ncal_p
\ubar(p')
\left[\gam_{\al} \GMem(Q^2) -n_{\al}\frac{1}{2M}\frac{\GMem(Q^2) -
\GEem(Q^2)}{1+\tau}
\right]u(p)
\end{equation}
Here  $n=p+p'$, $\tau=Q^2/4M^2$ and 
\begin{equation}
\begin{array}{ll}
\GMem(Q^2)=&\FVem(Q^2)+\FMem(Q^2)\\
\GEem(Q^2)=&\FVem(Q^2)-\tau\FMem(Q^2)
\end{array}
\label{Sachs}
\end{equation}
are, correspondingly, the magnetic and electric (charge) Sachs \ffs.
In the $Q^2=0$ limit they yield
\bea
\GMem(0)&&=e_N+\kappa_N=\mu_N
\nonumber\\
\GEem(0)&&=e_N
\nonumber
\eea
$\mu_N$ being the total magnetic moment of the nucleon (in units of the
nucleon Bohr magneton).

Let us notice that the magnetic and electric \ffs  $\GMem$ and 
 $\GEem$  characterize the matrix elements of the operators 
${\vec{J}}^{em}$ and $J_0^{em}$, respectively, in the Breit system
(the system in which ${\vec{n}}={\vec{p}}+{\vec{p'}}=0$). In fact, 
from (\ref{matem2}) it follows that
\begin{equation}
\label{matem3}
\la p'|{\vec{J}}^{em}|p\ra =
%{\Ncal_p}^2
\ubar(p'){\vec\gam}u(p)
\GMem(Q^2)\,.
\end{equation}
Furthermore, in the Breit system $n_0=2p_0$ and we have
\begin{equation}
\la p'|J_0^{em}|p\ra =
%{\Ncal_p}^2
\ubar(p')
\left[\gam_{0} \GMem(Q^2) -\frac{p_0}{M}\frac{\GMem(Q^2) -
\GEem(Q^2)}{1+\tau}
\right]u(p)\,,
\label{matem4}
\end{equation}
while, from the Dirac equation, it follows:
\begin{equation}
\label{matem5}
\ubar(p')({\psla}'+\psla)u(p)=
2p_0\ubar(p')\gam_0 u(p)=
2M\ubar(p')u(p)\,.
\end{equation}

The quantity $p_0^2/M^2$ in the Breit system can be expressed through:
\[
\frac{p_0^2}{M^2} = 1+\frac{Q^2}{4M^2}\equiv 1+\tau\,,
\]
and combining (\ref{matem4}) and (\ref{matem5}) we have
\begin{equation}
\la p'|J_0^{em}|p\ra =
%{\Ncal_p}^2
\ubar(p')\gam_0 u(p)\GEem(Q^2)\,.
\label{matem6}
\end{equation}

Let's consider now the one--nucleon matrix elements of the vector current. 
From the relation (\ref{vector}) it follows: 
\bea
&&{_{p}\la} p'|V_{\al}^{3}|p\ra_{p} = 
- {_{n}\la} p'|V_{\al}^{3}|p\ra_{n} =
\label{matvec1}\\
&&\qquad\quad =
%\Ncal_{p'}\Ncal_p
\ubar(p')
\left[\gam_{\al} \FVV(Q^2) +\frac{i}{2M}\sigma_{\al\be}q^{\be}\FMV(Q^2)
\right]u(p)
\nonumber
\eea
where
\bea
\FVV(Q^2)&=&\half\left({\FVem}_{\!,p}(Q^2)-{\FVem}_{\!,n}(Q^2)\right)\,,
\nonumber\\
\FMV(Q^2)&=&\half\left({\FMem}_{\!,p}(Q^2)-{\FMem}_{\!,n}(Q^2)\right)\,,
\nonumber
\eea
are the isovector Dirac and Pauli \ffs. Alternatively we can use the 
isovector magnetic and electric (charge) \ffs:
\bea
\GMV(Q^2)&=&\half\left({\GMem}_{\!,p}(Q^2)-{\GMem}_{\!,n}(Q^2)\right)\,,
\nonumber\\
\GEV(Q^2)&=&\half\left({\GEem}_{\!,p}(Q^2)-{\GEem}_{\!,n}(Q^2)\right)\,.
\nonumber
\eea

Let us consider now the one--nucleon matrix elements of the operator
$A_{\al}^3$. Information about these matrix elements can be obtained
from the data on the investigation of the quasi--elastic processes
\bea
\nu_{\mu} + n &&\longrightarrow \mu^- + p
\label{numuscat}\\
\nubar_{\mu} + p &&\longrightarrow \mu^+ + n
\label{nubarmuscat}
\eea
In the region $Q^2\ll m_W^2$ we are interested in, the matrix
elements of the processes (\ref{numuscat}) and (\ref{nubarmuscat}) 
have the following form
\bea
\la f|S|i\ra &&= -i\frac{G_F}{\sqrt{2}}\ubar(k')\gam^{\al}(1-\gam_5)u(k)
{_p\la}p'|J_{\al}^{CC}|p\ra_n\,(2\pi)^4\delta^{(4)}(p'-p-q)
\nonumber\\
\la f|S|i\ra &&= -i\frac{G_F}{\sqrt{2}}\ubar(k')\gam^{\al}(1+\gam_5)u(k)
{_n\la}p'|{J_{\al}^{CC}}^{\dagger}|p\ra_p\,(2\pi)^4\delta^{(4)}(p'-p-q)
\nonumber
\eea
Here $k$ is the momentum of the initial $\nu_{\mu}$ ($\bar\nu_{\mu}$),
$k'$ the momentum of the final $\mu^{-}$  ($\mu^{+}$); $p$ and $p'$ are 
the momenta of the initial $n$ ($p$) and of the final $p$ ($n$), respectively.

The quark current which  gives contribution to the matrix element of the
process (\ref{numuscat}) has the form 
\begin{equation}
j_{\al}^{CC} =\ubar\gam_{\al}(1-\gam_5) d\,V_{ud}
\label{jccq1}
\end{equation}
where $V_{ud}$ is an element of the CKM mixing matrix. From the
existing data $|V_{ud}|^2=0.9735\pm0.0008$~\cite{PDG00}
and hereafter we will not take
into account small corrections due to $|V_{ud}|\ne 1$. The current 
(\ref{jccq1}) can be expressed in terms of the above introduced $u$, $d$
quark iso--doublet $N$ as follows:
\begin{equation}
\label{jccq2}
j_{\al}^{CC} = \Nbar\gam_{\al}(1-\gam_5)\half(\tau_1+i\tau2)N
\equiv v_{\al}^{1+i2} - a_{\al}^{1+i2}\,,
\end{equation}
where $v_{\al}^{1+i2}$ and $a_{\al}^{1+i2}$ are the 
``plus--components'' of the isovectors (\ref{isovecq}).

For the Heisenberg currents we have
\begin{equation}
\label{jccHei}
J_{\al}^{CC} = V_{\al}^{1+i2} - A_{\al}^{1+i2}\,,
\end{equation}
where $V_{\al}^{1+i2}$ and $A_{\al}^{1+i2}$
are the ``plus-components'' of the isovectors  $V_{\al}^{i}$
and $A_{\al}^{i}$.

Let us consider now the one--nucleon matrix elements of the axial current.
The charge symmetry operator $\Ucal$ introduced in (\ref{Uop}) transforms
the isovector $A_{\al}^i$ as follows:
\begin{equation}
\begin{array}{lr}
\Ucal A_{\al}^{1,3}\Ucal^{-1}& = -A_{\al}^{1,3}\\
\Ucal A_{\al}^{2}\Ucal^{-1}& = A_{\al}^{2}
\end{array}
\label{csaxial}
\end{equation}
From these relations we have:
\begin{equation}
{_p\la}p'|A_{\al}^{1+i2}|p\ra_n =
{_n\la}p'|A_{\al}^{1-i2}|p\ra_p \,.
\label{dubbio1}
\end{equation}
Eq.~(\ref{dubbio1}) implies:
\begin{equation}
{_p\la}p'|A_{\al}^{1+i2}|p\ra_n =
{_p\la}p|A_{\al}^{1+i2}|p'\ra_n^*
\label{dubbio2}
\end{equation}

The one--nucleon matrix element of the CC axial current has the
following general structure
\bea
&&{_p\la}p'|A_{\al}^{1+i2}|p\ra_n =
%\Ncal_{p'}\Ncal_p
\label{matax1}\\
&&\qquad =
\ubar(p')\left[\gam_{\al}\gam_5\GACC(Q^2) 
+\frac{1}{2M}q_{\al}\gam_5\GPCC(Q^2)
+\frac{1}{2M} n_{\al}\gam_5\GTCC(Q^2)\right]u(p)
\nonumber
\eea
Due to the invariance under time reversal the \ffs $\GACC$, $\GPCC$
and $\GTCC$ are real quantities. Moreover, taking into account the
relation (\ref{dubbio2}), one easily finds that
\begin{equation}
\GTCC(Q^2) = 0\,.
\label{tensor}
\end{equation}

Let us notice that for the quasi--elastic processes 
$\ubar(k')\gam^{\al}(1-\gam_5)u(k)q_{\al}
= - m_{\mu}\ubar(k')(1-\gam_5)u(k)$. Thus the contribution of
the pseudoscalar \ff $\GPCC$ to the matrix element of the
processes (\ref{numuscat}) and (\ref{nubarmuscat}) 
is proportional to the muon mass $m_{\mu}$ and in the region
of neutrino energies $\ge 1$~GeV  can be neglected.

Isotopic invariance of the strong interactions provides 
the relation between the one--nucleon matrix elements of
the operators $A_{\al}^3$ and $A_{\al}^{1+i2}$. In fact for the
isovector $A_{\al}^i$ we have
\begin{equation}
\left[I_k,A_{\al}^j\right] = i\epsilon_{kj\ell}A_{\al}^{\ell}\,,
\label{isocomm}
\end{equation}
$I_{k}$ being the total isotopic spin operator (here $\epsilon_{kj\ell}$
is the totally antisymmetric tensor, with $\epsilon_{123}=1$). 
This relation implies
\begin{equation}
A_{\al}^3=\half\left[A_{\al}^{1+i2},I_{1-i2}\right]
\label{isocom2}
\end{equation}
and taking into account that
\[I_{1-i2}|p\ra_p=|p\ra_n\,,\quad
{_p\la}p'|I_{1-i2}=0\,,\]
from (\ref{isocom2}) the following relation holds
\begin{equation}
{_p\la}p'|A_{\al}^3|p\ra_p=
- {_n\la}p'|A_{\al}^3|p\ra_n=
\half{_p\la}p'|A_{\al}^{1+i2}|p\ra_n\,.
\label{matax2}
\end{equation}

The one--nucleon matrix elements of the axial current $A_{\al}^3$ have
the general structure
\bea
&&{_p\la}p'|A_{\al}^{3}|p\ra_p =
\label{matax3}\\
%\Ncal_{p'}\Ncal_p
&&\qquad =
\ubar(p')\left[\gam_{\al}\gam_5\GAV(Q^2) 
+\frac{1}{2M}q_{\al}\gam_5\GPV(Q^2)
+\frac{1}{2M} n_{\al}\gam_5\GTV(Q^2)\right]u(p)
\nonumber
\eea
where due to the T--invariance of strong interactions the \ffs
$\GAV$, $\GPV$ and $\GTV$ are real. Furthermore
\begin{equation}
{_p\la}p'|A_{\al}^{3}|p\ra_p = {_p\la}p|A_{\al}^{3}|p'\ra_p^*\,.
\label{matax4}
\end{equation}
From (\ref{matax3}) and (\ref{matax4})
it follows that the \ff $\GTV(Q^2)$ vanishes. Moreover the
contribution of the pseudoscalar \ff $\GPV(Q^2)$ to the matrix elements
of the NC--induced processes is proportional to the
lepton mass and can be neglected  (both for neutrino-- and 
electron--induced processes). Finally from (\ref{matax2})
and (\ref{matax3}) we have the following relation
for the axial \ff:
\begin{equation}
\GAV(Q^2)=\half\GACC(Q^2)
\end{equation}

Thus, summarizing, the form factors that characterize the proton matrix
elements of the $u-d$ part of the 
NC are connected with the electromagnetic form factors of proton and neutron
and with the CC axial nucleon form factor by the relations
\bea
G^V_{M,E}(Q^2)&=&\half\left\{ G^{p}_{M,E}(Q^2)-
G^{n}_{M,E}(Q^2)\right\}
\nonumber\\
\GAV(Q^2)&=&\half\GACC(Q^2)
\nonumber
\eea
The matrix elements of proton and neutron are connected by the
charge--symmetry relations
\bea
{_p\la}p'|V_{\al}^3|p\ra_p &=& - {_n\la}p'|V_{\al}^3|p\ra_n
\nonumber\\
{_p\la}p'|A_{\al}^3|p\ra_p &=& - {_n\la}p'|A_{\al}^3|p\ra_n
\nonumber
\eea
%

%%%%%%%%%%%%%%%%%%%%%%%%%%%%%%%%%%%%%%%%%%%%%%%%%%%%%%%%%%%
\section{ Strange form factors of the nucleon}
\label{sec.strangeFF}

In this Section we will consider the strange \ffs of the nucleon.
Let us start by considering the one--nucleon matrix element of
the vector, $v_{\al}^s=\sbar\gam_{\al}s$, and axial,
$a_{\al}^s=\sbar\gam_{\al}\gam_5 s$, strange currents\footnote{
Notice that the present, general discussion is also valid if the
currents of the $c$ and the other heavier (isoscalar) quarks are included.}.
They have the following general structure
\bea
&&\la p'|V_{\al}^s|p\ra =
\nonumber\\
&&\, =
\ubar(p')
\left[\gam_{\al}\FVs(Q^2) +\frac{i}{2M}\sigma_{\al\be}q^{\be}\FMs(Q^2)
+\frac{1}{2M}q_{\al}\FTs(Q^2)\right]u(p)
\label{vstrange1}\\
&&\la p'|A_{\al}^s|p\ra =
\nonumber\\
&&\, =
\ubar(p')
\left[\gam_{\al}\gam_5\GAs(Q^2) +\frac{1}{2M}q_{\al}\gam_5\GPs(Q^2)
+\frac{1}{2M}n_{\al}\gam_5\GTs(Q^2)\right]u(p)
\label{astrange1}
\eea
where, again, $q = p'-p$, $n = p'+p$, the $F_i^s(Q^2)$ ($i=1,2,3$)
are the strange vector \ffs of the nucleon and $G_a^s(Q^2)$ ($a=A,P,T$)
the strange axial ones, respectively.

From the invariance of strong interactions under time reversal it
follows that the \ffs $\FTs(Q^2)$ and $\GTs(Q^2)$ are equal to zero.
In fact, for the axial current T--invariance implies:
\begin{equation}
\la p'|A_{\al}^s|p\ra =  \la p_T|A_{\al}^s|p_T'\ra\eta_{\al}
\label{trev1}
\end{equation}
(repeated indexes, in the r.h.s., are {\it not} summed)
where $\eta=(1,-1,-1,-1)$ and the vector $|p_T\ra$ describes a nucleon
with momentum $p_T=(p_0, -\vec{p})$ and in a spin state
\begin{equation}
u(p_T)= T\ubar^T(p)\,,
\label{trev2}
\end{equation}
the matrix $T$ satisfying the condition
\begin{equation}
\label{trev3}
T\gam_{\al}^T T^{-1}=\gam_{\al}\eta_{\al}
\end{equation}

With the help of (\ref{trev1}) and (\ref{trev3}) it is easy to see that
\[ \GTs(Q^2)=0.\]
Analogously, from T--invariance it follows
\[\FTs(Q^2)=0.\]
Furthermore, from the hermiticity of the neutral currents we have
\begin{equation}
\begin{array}{ll}
\la p'|A_{\al}^s|p\ra &= \la p|A_{\al}^s|p'\ra^*,\\
\la p'|V_{\al}^s|p\ra &= \la p|V_{\al}^s|p'\ra^*.
\end{array}
\label{hermi}
\end{equation}

From (\ref{vstrange1}), (\ref{astrange1}) and (\ref{hermi}), it follows
that the \ffs $F_{1,2}^s(Q^2)$ and $G_{A,P}^s(Q^2)$ are real\footnote{
In general the vector strange current is not conserved. However, due to
T--invariance, the one--nucleon matrix element of the vector strange
current satisfies the condition
\[(p'-p)^{\al}\la p'|V_{\al}^s|p\ra =0\,.\]
}. 
 For the same reasons mentioned above we shall hereafter omit the 
pseudoscalar \ff $G_P^s$.

As an alternative to $\FVs(Q^2)$ and $\FMs(Q^2)$, 
one can define the magnetic and electric strange \ffs
of the nucleon, which are connected with
$F_{1,2}^s(Q^2)$ by the relations
\bea
\GMs(Q^2)&=&\FVs(Q^2) + \FMs(Q^2)
\label{gms}\\
\GEs(Q^2)&=&\FVs(Q^2)-\tau\FMs(Q^2)
\label{ges}
\eea
which, in the $Q^2=0$ limit, assume the values
\bea
\GMs(0)&=& \mu_s
\label{strangemu}\\
\GEs(0)&=&0\, ,
\label{gezero}
\eea
$\mu_s$ being the strange magnetic moment of the nucleon in units of
the nuclear Bohr magneton. Obviously relation (\ref{gezero}) follows 
from the fact that the net strangeness of the nucleon is equal to 
zero\footnote{
In fact, in the Breit system, for the one--nucleon matrix element of the
strangeness operator 
\[ S=\int V_0^s(x) d^3x \]
we have
\bea
&&\la p'|\int V_0^s(x) d^3x|p\ra = (2\pi)^3\delta^{(3)}(\vec{p'}-\vec{p})
\la p'|V_0^s(0)|p\ra =
\nonumber\\
&&\quad
=(2\pi)^3\delta^{(3)}(\vec{p'}-\vec{p})
%\Ncal_p^2
\ubar(p)\gam_0 u(p)\GEs(0)
=(2\pi)^3 2p^0\delta^{(3)}(\vec{p'}-\vec{p})\GEs(0).
\nonumber
\eea
On the other hand, since the net strangeness of the nucleon is equal to zero,
we have
\[\la p'|\int V_0^s(x) d^3x|p\ra = 0.\]
}. 
In the region of small $Q^2$ we have
\begin{equation}
\GEs(Q^2) = -\frac{1}{6}\la r^2_s\ra Q^2
\label{strangera}
\end{equation}
where $\la r^2_s\ra= -6\left(d\GEs/dQ^2\right)_{Q^2=0}$ is a parameter 
which can be interpreted as the mean square strangeness radius of the
nucleon.

As already mentioned in the Introduction, in the framework of the parton 
model the matrix element  $\la p|\qbar\gam_{\al}\gam_5 q|p\ra$ gives the 
contribution of the $q$--quark
and $\qbar$--antiquark to the spin of proton. 
In fact, assuming that
the proton is in a state with momentum $p$ and helicity equal 
to one, we have
\begin{equation}
{_p\la} p|\qbar\gam_{\al}\gam_5 q|p\ra_p = 
%\Ncal_p^2
\ubar(p)\gam_{\al}\gam_5 u(p)g_A^q\, ,
\label{axial1}
\end{equation}
where the spinor $u(p)$ satisfies the equation
\begin{equation}
\gam_5\rlap{/}{s} u(p)= u(p)\,,
\label{axial2}
\end{equation}
and $s_{\alpha}$ is the unit vector which obeys
the condition $s\cdot p=0$. In the rest frame of the nucleon 
$s^{\al}=(0,\vec{\kappa})$ where $\vec{\kappa}$
is the unit vector in the direction of the proton momentum.
From (\ref{axial1}) and (\ref{axial2}) we obtain
\begin{equation}
{_p\la} p|\qbar\gam_{\al}\gam_5 q|p\ra_p
= {\mathrm{Tr}}\gam_{\al}\gam_5\half
\left(1+\gam_5\rlap{/}{s}\right)(\psla+M)g_A^q=2Ms_{\al}g_A^q\,.
\label{axial3}
\end{equation}

Notice that, by combining Eq.~(\ref{axial3}) for $q=s$ with
Eq.~(\ref{melasc}), one obtains the following value for $g_A^s$:
\begin{equation}
\GAs(0)\equiv g_A^s = -0.12 \pm 0.03\,,
\label{expgas}
\end{equation}
which represents the present direct estimate of this parameter 
from deep inelastic scattering experiments.

Let us consider now the matrix element of the axial quark current
in the parton approximation, in the infinite momentum frame. We have
\bea
{_p\la} p|\qbar\gam_{\al}\gam_5 q|p\ra_p &&=
\int_0^1\frac{p^0}{p^0_x}\sum_r \ubar^r(p_x)\gam_{\al}\gam_5 u^r(p_x)
\left(q^r(x)+{\qbar}^r(x)\right)dx
\nonumber\\
&&=\int_0^1\frac{1}{x} 2m_qs_{\al}^q\sum_r r
\left(q^r(x)+{\qbar}^r(x)\right)dx
\label{axial4}
\eea
where $q^r(x)$ (${\qbar}^r(x)$) is the density of $q$--quarks 
($\bar{q}$--antiquarks) with momentum $p_x=xp$ and helicity $r$, 
$x=Q^2/(2p\cdot q)$ is the Bjorken variable 
($0\le x \le 1$) and $m_q$ the mass of the $q$--quark.
Taking into account that
\[ s_{\al}^q = 
%\frac{xp_{\al}}{m_q}= 
x\frac{M}{m_q}s_{\al}\]
from (\ref{axial4}) we obtain 
\begin{equation}
{_p\la} p|\qbar\gam_{\al}\gam_5 q|p\ra_p = 2Ms_{\al}\int_0^1\sum_r
r \left(q^r(x)+\qbar^r(x)\right)dx\, .
\label{axial5}
\end{equation}
Now by comparing (\ref{axial3}) with (\ref{axial5}) one finds that in the
parton approximation
\begin{equation}
g_A^q=\int_0^1 \left[q^{(+)}(x)+\qbar^{(+)}(x)-\{q^{(-)}(x)
+\qbar^{(-)}(x)\}\right]dx\equiv \Delta q\,.
\label{axial6}
\end{equation}
Thus, the constant  $g_A^q\equiv\Delta q$ is the contribution of
$q$-quarks and $\bar{q}$- antiquarks to the spin of the nucleon.

There exists a large number of papers in which the strange magnetic moment 
$\mu_s$ and the strange radius $r_s$ of the nucleon are calculated within
different models (pole models, chiral quark models, soliton models,
Skyrme models, lattice QCD and others). The predicted values of $\mu_s$
and $r_s$ in different models are very different in magnitude
and in sign. It is not our aim here to review these papers and we
recommend the interested reader to refer to the original
literature~\cite{Jaffe89}--\cite{Dubnic01}.

In summarizing the contents of this and of the previous Sections, for
the one--nucleon matrix elements of the vector and axial NC 
of the Standard Model we have
\bea
&&{_{p(n)}\la}p'|V_{\al}^{NC}|p\ra_{p(n)}=
\nonumber \\
&&\qquad = \ubar(p')\left[
\gam_{\al}F_1^{NC;p(n)}(Q^2) +\frac{i}{2M}
\sigma_{\al\be}q^\be F_2^{NC;p(n)}(Q^2)\right]u(p)
\label{vecNClast}\\
&&{_{p(n)}\la}p'|A_{\al}^{NC}|p\ra_{p(n)}= \ubar(p')\gam_{\al}\gam_5 
G_A^{NC;p(n)} u(p)
\label{axNClast}
\eea
where the NC \ffs are given by
\bea
F_{1,2}^{NC;p(n)}(Q^2)&&= \pm\half\left\{ F_{1,2}^p(Q^2) - F_{1,2}^n(Q^2)
\right\}- 
\nonumber\\
&&\qquad
-2\sin^2\theta_W F_{1,2}^{p(n)}(Q^2) -\half F_{1,2}^s(Q^2)
\label{f12NC}\\
G_A^{NC;p(n)}(Q^2)&&=\pm\half \GACC(Q^2) -\half G_A^s(Q^2)
\label{gaNC}
\eea
Equivalently one can consider the NC Sachs \ffs:
\bea
{G_E}^{NC;p(n)}(Q^2) &&= \pm\half\left\{\GEem^p(Q^2)-\GEem^n(Q^2)\right\}-
\nonumber\\
&&\qquad
-2\sin^2\theta_W\GEem^{p(n)}(Q^2) -\half\GEs(Q^2)
\label{geNC}\\
{G_M}^{NC;p(n)}(Q^2) &&= \pm\half\left\{\GMem^p(Q^2)-\GMem^n(Q^2)\right\}-
\nonumber\\
&&\qquad
-2\sin^2\theta_W\GMem^{p(n)}(Q^2) -\half\GMs(Q^2)
\label{gmNC}
\eea

The relations (\ref{f12NC}) [or (\ref{geNC}), (\ref{gmNC})]
and (\ref{gaNC})  are the basic ones.
From these relations it is obvious that the investigation of NC--induced
processes allows one 
to obtain direct information on the strange form factors
of the nucleon providing one can ``a priori'' utilize
information on the value of the parameter
$\sin^2\theta_W$ (which is obtained from the measurement of
different NC processes), information on the electromagnetic \ffs of the
nucleons (which is obtained from the measurement of elastic scattering
of electrons on nucleons) and on the axial \ff of the nucleon (which is
obtained from the measurement of quasi--elastic CC neutrino scattering on
nucleons).

The investigation of NC--induced processes 
in the region $Q^2\ll 1$~GeV$^2$ allows one
to determine the strange magnetic moment of the nucleon
$\mu_s$ and the strange axial constant $g^{s}_{A}$ directly from
experimental data. At larger momentum transfers one could obtain
information on the $Q^2$ behavior of the strange form factors
of the nucleon.
In the next Sections we shall discuss possible experiments from which
direct information on the strange form factors of the nucleon can be
obtained. We will also present the existing experimental data.

%%%%%%%%%%%%%%%%%%%%%%%%%%%%%%%%%%%%%%%%%%%%%%%%%%%%%%%%%%%%%%%
\section{P--odd effects in the elastic scattering
of polarized electrons on the nucleon}
\label{sec.Poddel}

There are two types of NC--induced effects which allow one to obtain
direct information on the strange form factors of the nucleon 
 (see for example Ref.~\cite{Musolf94.rep}):
\begin{enumerate}
\item
The P--odd asymmetry in the elastic scattering of polarized electrons 
on unpolarized nucleons
\item
The NC--induced elastic scattering of neutrinos and antineutrinos on
nucleons.
\end{enumerate}

\begin{figure}
\begin{center}
\mbox{\epsfig{file=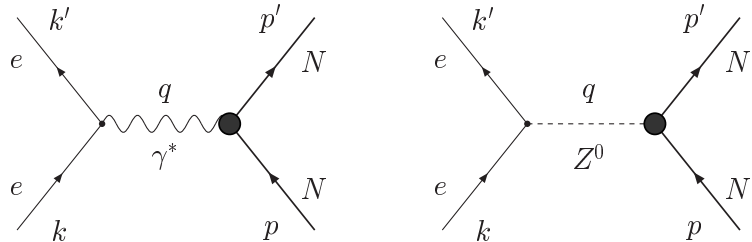,width=0.99\textwidth}}
\end{center}
\caption{Diagrams of the process $\vec{e}+p\to e+p$.}
\label{Fig1}
\end{figure}

In this Section we will discuss the P--odd asymmetry in the process
\begin{equation}
\vec{e} + p\longrightarrow e + p
\label{epolscat}
\end{equation}
The diagrams of the process (\ref{epolscat}) in lowest order in
the constants $e$ and  $g$  are shown in Fig.~\ref{Fig1},
where both the exchange of a photon  and of the vector boson $Z^0$
are considered.

%*************************************************************************
%*************************************************************************

For the matrix element of this process we have the following expression
\bea
\la f|S|i\ra &&= i(2\pi)^4\delta^{(4)}(p'-p-q)\frac{4\pi\al}{Q^2}
\left[ \ubar(k')\gam^{\al}u(k)\la p'|J^{em}_{\al}|p\ra -\right.
\nonumber\\
&& \left. -\frac{G_F Q^2}{2\sqrt{2}\pi\al} \ubar(k')\gam^{\al}
\left(g_V-g_A\gam_5\right)u(k)\la p'|J^{NC}_{\al}|p\ra\right]\,.
\label{polampli}
\eea
Here $k$ and  $k'$ are the momenta of the initial and final electron,
$p$ and  $p'$ the momenta of the initial and final nucleon, $q=k-k'$,
$\al=e^2/4\pi$ and the weak NC vector and axial couplings for the
electron are:
\begin{equation}
\begin{array}{ll}
g_V &= -\half +2\sin^2\theta_W\\
g_A &= -\half\,.
\end{array}
\label{gvga}
\end{equation}

Since we will consider polarized electrons, let us introduce
the density matrix of electrons with momentum  $k$ and polarization $P$.
It is given by:
\begin{equation}
\rho(k) = \half\left(1+\gam_5\Psla\right)\left(\ksla +m\right)
\label{denmat1}
\end{equation}
Here $P^{\al}$ is the four--vector of polarization, satisfying the
condition $P\cdot k=0$. In the electron rest frame we have
$P=(0, \vec{P^0})$, the vector $\vec{P^0}$ being usually written
in the form of the sum of longitudinal and transverse components:
\begin{equation}
\vec{P^0} = P^0_{\parallel}\vec{\kappa} +\vec{P_{\perp}}
\end{equation}
where $\vec{\kappa}$ is the unit vector in the direction of the electron
momentum.

We shall consider scattering of high--energy electrons on nucleons. 
Thus $k_0 \gg m$ and the polarization vector can
be approximated by the expression:
\begin{equation}
P^{\al}=  P^0_{\parallel}\frac{k^{\al}}{m} + P_{\perp}^{\al}
\label{polvec}
\end{equation}
where $P_{\perp}=(0,\vec{P^0_{\perp}})$. 
From (\ref{denmat1}) and (\ref{polvec}) it follows that the density
matrix of ultrarelativistic electrons has the form
\begin{equation}
\rho(k) = \half\left(1+ \lambda\gam_5 +\gam_5\Psla_{\perp}\right)\ksla
\label{denmat2}
\end{equation}

where we have introduced the notation $\lambda= P^0_{\parallel}$. 
Notice that
for the $V-A$ interaction the contribution of the transverse polarization
to the cross section is proportional to the electron mass and
at high energies can be neglected.

The lowest order contribution, to the cross section of the process,
stemming from the second (NC) term of the matrix element (\ref{polampli})
is determined by the quantity
\begin{equation}
\Acal_0\equiv\frac{G_F Q^2}{2\sqrt{2}\pi\al}= 1.798 \times 10^{-4} 
 \frac{Q^2}{\mathrm{GeV}^2}\,,
\label{AA0}
\end{equation}
which is small in the region $Q^2\lesssim M^2$ we are interested in.
Thus, in the calculation of the cross section we shall only
take into account the square of the first (electromagnetic) term
of (\ref{polampli}) 
and the interference of the electromagnetic and NC terms.
Let us notice that the interference of the electromagnetic and P--even
part of the NC term ($v\cdot V$ and $a\cdot A$) gives a very small
correction to the electromagnetic term and can be neglected.

We shall be interested in the pseudoscalar term of the cross section
which is proportional to  $\lambda$  and is due to the interference
of the electromagnetic amplitude and P--odd part of the NC amplitude
($v\cdot A$ and $a\cdot V$). It is obvious that
\begin{equation}
\frac{1}{4}{\mathrm{Tr}}\gam^{\al}\lambda\gam_5\ksla\gam^{\be}
\left(g_V-g_A\gam_5\right)\ksla'=
\lambda\left\{g_V L_5^{\al\be}(k,k') -g_A L^{\al\be}(k,k')\right\}
\label{traces1}
\end{equation}
where
\bea
L^{\al\be}(k,k') &=& k^{\al}{k'}^{\be} +{k'}^{\al}k^{\be}
-g^{\al\be}k\cdot k'\,,
\label{lepten}\\
L_5^{\al\be}(k,k') &=& i\epsilon^{\al\be\rho\sigma}k_{\rho}k'_{\sigma}\,,
\label{lepten5}
\eea
$\epsilon^{\al\be\rho\sigma}$ being the antisymmetric (under the exchange
of any two indexes) tensor, with $\epsilon^{0123}=-\epsilon_{0123}=-1$.

With the help of Eq.~(\ref{traces1}) the cross section of the process
(\ref{epolscat}) can be expressed as follows:
\bea
d\sigma_{\lambda}&& =\frac{4\al^2}{Q^4}\frac{M}{p\cdot k}\half
\left\{ L^{\al\be} W^{em}_{\al\be} +\right.
\label{cross1}\\
&&\, \left.
+\lambda \frac{G_F Q^2}{2\sqrt{2}\pi\al}\left[g_V L_5^{\al\be}
W^I_{\al\be}(A) +g_A L^{\al\be} W_{\al\be}^I(V)\right]\right\}
\frac{d\vec{k'}}{k'_0}\,.
\nonumber
\eea
In the above the hadronic electromagnetic tensor $W^{em}_{\al\be}$
is given by
\begin{equation}
W^{em}_{\al\be}=\frac{1}{2M}\sum\int\la p'|J^{em}_{\al}|p\ra
\la p|J^{em}_{\be}|p'\ra
\delta^{(4)}(p'-p-q)\frac{d\vec{p'}}{2 p'_0}\,,
\label{Wem}
\end{equation}
while the tensor $W^I_{\al\be}(V)$ (pseudotensor $W^I_{\al\be}(A)$)
arises from the interference of the electromagnetic and vector 
(axial) part of the hadronic NC:
\bea
W^{I}_{\al\be}(V)&&=\frac{1}{2M}\sum\int\left\{\la p'|J^{em}_{\al}|p\ra
\la p|V^{NC}_{\be}|p'\ra +\right.
\label{WIV}\\
&&\, \left. +\la p'|V^{NC}_{\al}|p\ra\la p|
J^{em}_{\be}|p'\ra\right\}
\delta^{(4)}(p'-p-q)\frac{d\vec{p'}}{2 p'_0}\,,
\nonumber
\eea
\bea
W^{I}_{\al\be}(A)&&=\frac{1}{2M}\sum\int\left\{\la p'|J^{em}_{\al}|p\ra
\la p|A^{NC}_{\be}|p'\ra +\right.
\label{WIA}\\
&&\, \left. +\la p'|A^{NC}_{\al}|p\ra\la p|
J^{em}_{\be}|p'\ra\right\}
\delta^{(4)}(p'-p-q)\frac{d\vec{p'}}{2 p'_0}\,,
\nonumber
\eea
From Eqs.~(\ref{cross1}), (\ref{WIV}) and (\ref{WIA}) it follows that
information on the one--nucleon matrix elements of NC can be obtained
by investigating the dependence of the cross section of the process
(\ref{epolscat}) on the longitudinal polarization $\lambda$.

The SM values of the constants $g_V$ and $g_A$ are given by 
(\ref{gvga}).
The parameter $\sin^2\theta_W$ is known, at present,
with very high accuracy. Its on--shell value is given by~\cite{PDG00}
\[ \sin^2\theta_W = 0.23117\pm 0.00016\,. \] 
For the constant $g_V$ we have
\[ g_V = -0.0397\pm 0.0003 \,. \]
Thus in the SM $|g_V| \ll |g_A|$. Taking into account this inequality
we can conclude from the general expression for the cross section
(\ref{cross1}) that the main contribution to 
the  $\lambda$--dependent part of the cross section is given by
the interference of the electromagnetic term and the {\it vector} part
of the NC term. The axial part of the NC term can, nevertheless, be not
totally negligible at specific kinematical conditions.

The tensors $W^{em}_{\al\be}$, $W^I_{\al\be}(V)$ and the pseudotensor
$W^I_{\al\be}(A)$ have the following general form
\bea
W^{em}_{\al\be}&=&
-\left(g_{\al\be} -\frac{q_{\al}q_{\be}}{q^2}\right)W_1^{em} +
\frac{1}{4M^2}n_{\al}n_{\be} W_2^{em}
\nonumber\\
W^I_{\al\be}(V)&=&
-\left(g_{\al\be} -\frac{q_{\al}q_{\be}}{q^2}\right)W_1^{I} +
\frac{1}{4M^2}n_{\al}n_{\be} W_2^{I}
\label{tensor1}\\
W^I_{\al\be}(A)&=&
\frac{i}{2M^2}\epsilon_{\al\be\rho\sigma}p^{\rho}q^{\sigma}W^I_3
\nonumber
\eea
where $n=p'+p$.
Calculating the traces in Eqs.~(\ref{Wem}), (\ref{WIV}) and (\ref{WIA}),
one obtains
\bea
W^{em}_1 &&=\tau \GMem^2\delta\left(\nu-\frac{Q^2}{2M}\right)\,,
\nonumber\\
W^{em}_2 &&=\frac{\displaystyle{\GEem^2+\tau\GMem^2}}
{\displaystyle{1+\tau}}
\delta\left(\nu-\frac{Q^2}{2M}\right)
\label{Wem2}
\eea
and
\bea
W^{I}_1 &&=2\tau \GMem\GMNC\delta\left(\nu-\frac{Q^2}{2M}\right)\,,
\nonumber\\
W^{I}_2 &&=2\frac{\displaystyle{\GEem\GENC+\tau\GMem\GMNC}}
{\displaystyle{1+\tau}}
\delta\left(\nu-\frac{Q^2}{2M}\right)\,,
\label{WI2}\\
W^{I}_3 &&=2\GMem\GANC \delta\left(\nu-\frac{Q^2}{2M}\right)
\nonumber
\eea
Here $\nu=p\cdot q/M$ and $\tau=Q^2/4M^2$.

With the help of Eqs.~(\ref{cross1}), (\ref{Wem2}) and (\ref{WI2})
for the cross section of the scattering of electrons with polarization
$\lambda$ on unpolarized nucleons we find the following general
expression:
\begin{equation}
\left(\frac{d\sigma}{d\Omega}\right)_{\lambda}
= \left(\frac{d\sigma}{d\Omega}\right)_0 (1+\lambda\Acal)
\label{cross2}
\end{equation}
where $(d\sigma/d\Omega)_0$ is the cross section for the scattering of
unpolarized electrons on nucleons and is given by the 
Rosenbluth formula
\begin{equation}
\left(\frac{d\sigma}{d\Omega}\right)_0=\sigma_{Mott}
\left\{\frac{\GEem^2+\tau\GMem^2}{1+\tau} +
2\tan^2\frac{\theta}{2}\tau\GMem^2\right\}\,.
\label{sigmaRuth}
\end{equation}
Here $\sigma_{Mott}$ is the Mott cross section
\begin{equation}
\sigma_{Mott}=\frac{\al^2\cos^2(\theta/2)}{\displaystyle
{4E^2\sin^4\frac{\theta}{2}\left(1+\frac{2E}{M}\sin^2\frac{\theta}{2}
\right)}}
\label{sigmaMott}
\end{equation}
where $M$ is the mass of the target nucleon, $E$ and $\theta$ 
are the energy and scattering angle of the electron in the laboratory
system. From Eq.~(\ref{cross2}) it follows that 
the P--odd asymmetry is given by
\begin{equation}
\Acal=\frac{1}{\lambda}\frac{\displaystyle{
\left(\frac{d\sigma}{d\Omega}\right)_{\lambda}
- \left(\frac{d\sigma}{d\Omega}\right)_{-\lambda}}}
{\displaystyle{\left(\frac{d\sigma}{d\Omega}\right)_{\lambda} +
\left(\frac{d\sigma}{d\Omega}\right)_{-\lambda}}}\,.
\label{asymmel1}
\end{equation}
With the help of (\ref{cross1}), (\ref{tensor1}), (\ref{Wem2})
and (\ref{WI2}) we find the following expression for the
 asymmetry $\Acal$ in Born approximation:
\begin{equation}
\Acal = -\Acal_0\,
%\left(-\frac{G_F Q^2}{2\sqrt{2}\pi\al}\right)
\frac{\tau\GMem\GMNC +\varepsilon\GEem\GENC +
(1-4\sin^2\theta_W)\varepsilon' \GMem\GANC}
{\tau\GMem^2 +\varepsilon\GEem^2}\,,
\label{asymmel2}
\end{equation}
where
\[\varepsilon =\frac{1}{1+2(1+\tau)\tan^2(\theta/2)}\,,
\quad
\varepsilon'=\sqrt{\tau(1+\tau)(1-\varepsilon^2)}\,.
\]

We remind the reader 
that the NC vector and axial \ffs [see expressions (\ref{geNC}) 
and (\ref{gmNC})] can be written in the following form 
\bea
G_{M,E}^{NC;p(n)}&&=
\half\left\{\left(1-4\sin^2\theta_W\right)G_{M,E}^{p(n)}
-G_{M,E}^{n(p)}\right\}-\half G_{M,E}^s
\label{gmeNC1}\\
&&\equiv G_{M,E}^{0;p(n)} -\half G_{M,E}^s
\nonumber\\
G_A^{NC;p(n)}&&=\pm \half\GACC -\half\GAs\equiv G_A^{0;p(n)} -\half\GAs
\label{gaNC1}
\eea

Using the expressions (\ref{gmeNC1}) and (\ref{gaNC1}) we can explicitly
separate the terms proportional to strange \ffs in the expression of
the P--odd asymmetry. Indeed the r.h.s. of Eq. (\ref{asymmel2}) can be
split as follows:
\begin{equation}
\Acal= \Acal^{(0)} +\Acal^{(s)}
\label{asymmel3}
\end{equation}
where 
\bea
\Acal^{(0)}&=& -\Acal_0
%\left(-\frac{G_FQ^2}{2\sqrt{2}\pi\al}\right)
\frac{\tau\GMem G_M^0+ \varepsilon\GEem G^0_E +
(1-4\sin^2\theta_W)\varepsilon'\GMem G^0_A}
{\tau\GMem^2+\varepsilon\GEem^2} \,,
\label{A0elec}\\
\Acal^{(s)}&=& -\Acal_0
%\left(-\frac{G_FQ^2}{4\sqrt{2}\pi\al}\right)
\frac{\tau\GMem G_M^s +\varepsilon\GEem G^s_E +
(1-4\sin^2\theta_W)\varepsilon'\GMem G^s_A}
{\tau\GMem^2+\varepsilon\GEem^2}\,.
\label{Aselec}
\eea

According to this separation, the asymmetry $\Acal^{(0)}$ is determined by
the non--strange electromagnetic and axial form factors of
the nucleon and by the electroweak parameter $\sin^2\theta_W$.
The asymmetry  $\Acal^{(s)}$, instead, is the contribution to the P--odd
asymmetry of the strange \ffs. As it is seen from these expressions
the contribution of the axial strange form factor to the asymmetry is 
suppressed by the factor $1-4\sin^2\theta_W\simeq 0.075$.

Let us stress that in order to obtain information on the strange vector
\ffs  of the nucleon from the measurement of the P--odd asymmetry it
is ne\-ces\-sa\-ry to know the nucleonic electromagnetic form factors 
with large
enough accuracy. Due to the isovector nature of $u-d$ part of the
neutral current, even if we limit ourselves to consider the P--odd
asymmetry for the scattering on the proton, this quantity contains the
electromagnetic \ffs of {\it both} proton and neutron.
At present the electromagnetic form factors of the neutron and
particularly its charge form factor is rather poorly known.
New measurements of the electromagnetic \ffs of the
nucleon are under way or in program at the Thomas Jefferson National
Accelerator Laboratory (Jefferson Lab)~\cite{Cebaf}.

%%%%%%%%%%%%%%%%%%%%%%%%%%%%%%%%%%%%%%%%%%%%%%%%%%%%%%%%%%%%%%%%
\section{The experiments on the measurement of P--odd asymmetry 
in elastic $e - p$ scattering}
\label{sec.Poddelex}

We will discuss here the results of recent experiments on
the measurement of the P--odd asymmetry $\Acal$ in elastic
electron--proton scattering.

In the experiment of the HAPPEX collaboration at Jefferson Lab~\cite{Aniol00}
the elastic scattering of electrons with energy of $3.3$~GeV at the
average scattering angle $\theta=12.3^{\circ}$  was measured. Consequently
the average value of the 
square of the momentum transfer was $Q^2=0.477$~GeV$^2$.
The longitudinal polarization of the electrons was in the range
$67\div 76 \%$. A $15$~cm liquid hydrogen target was used. 
In order to select elastic scattering events 
two high--resolution spectrometers were used in the experiment.
Only 0.2 \% of the events were due to background processes.

Electrons with a polarization of about 70\% were obtained 
by irradiation  of  GaAs crystals by circularly polarized
laser light. The polarization of the electron beam was continuously 
monitored by a Compton polarimeter and was also measured by M{\o}ller
scattering. The combined asymmetry obtained from the results of the
1998 and 1999 data taking is equal to
\begin{equation}
\Acal(Q^2=0.477) =\left\{ -15.05\pm 0.98 (\mathrm{stat})\pm
0.56 (\mathrm{syst})\right\}\times 10^{-6}\,.
\label{asymHap}
\end{equation}

The measured value of the asymmetry allows one to obtain information on
the following combination of the strange form factors
(at $Q^2=0.477$~GeV$^2$):
\begin{equation}
 \GEs + 0.392\GMs\,.
 \label{strangecomb}
 \end{equation}
From (\ref{Aselec}) and (\ref{asymHap}) it was found
\begin{equation}
\frac{\GEs + 0.392\GMs}{\GMem^p/\mu_p} =
0.069\pm 0.056\pm 0.039
\label{strangeHap}
\end{equation}
where the first error is the combined (in quadrature) statistical and
systematic errors and the second error is determined by the
uncertainties on the electromagnetic form factors. For the HAPPEX
kinematics $\left( \GMem^p/\mu_p \right) \simeq 0.36$.
In accordance with the existing data, the ratios of the electromagnetic \ffs
of proton and neutron to the magnetic form factor of the proton 
were taken to be
\bea
\frac{\GEem^p}{(\GMem^p/\mu_p)} &=& 0.99\pm 0.02\,;
\nonumber\\
\frac{\GEem^n}{(\GMem^p/\mu_p)} &=& 0.16\pm 0.03\,;
\label{ffratio}\\
\frac{\GMem^n/\mu_n}{(\GMem^p/\mu_p)} &=& 1.05\pm 0.02\,;
\nonumber
\eea
The estimated contribution to the asymmetry of the axial form factor
$\GANC$ was
\begin{equation}
\Acal_A=(0.56\pm 0.23)\times 10^{-6}\,,
\label{axasym}
\end{equation}
where the main uncertainty is due to radiative 
corrections~\cite{Musolf94.rep,MusHol90,Zhu00}.

Taking into account (\ref{gezero}) one can put
\[ \frac{\GEs}{\GMem^p/\mu_p} = \tau\rho_s\]
where $\rho_s$ is a constant. Furthermore in Ref.~\cite{Aniol00}
the same $Q^2$-dependence of $\GMem^p(Q^2)$ was
assumed for the
strange magnetic form factor; 
in this case, from (\ref{strangeHap}) one finds:
\begin{equation}
\rho_s+2.9\mu_s = 0.51\pm 0.41\pm 0.29\,,
\label{strangeHap2}
\end{equation}
where $\mu_s$ is the strange magnetic moment of the nucleon
[see Eq. (\ref{strangemu})].

\begin{figure}
\begin{center}
\mbox{\epsfig{file=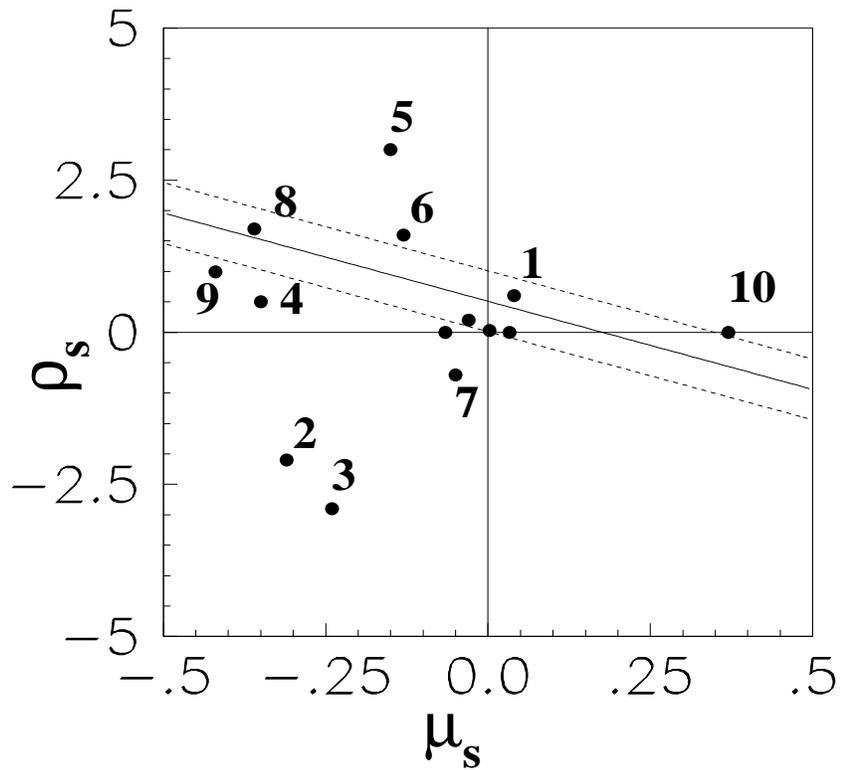,width=0.9\textwidth,height=0.6\textheight}}
%\mbox{\epsfig{file=fig4_aniol.eps,width=0.9\textwidth,height=0.6\textheight}}
\end{center}
\vskip -1cm
\caption{Band: allowed region from the results of Ref.~\cite{Aniol00} with
the assumptions discussed in the text. Points are the theoretical
estimates from  various models. 
The numbers refer to the list of references for the models as 
they appear in~\cite{Aniol00} (the corresponding reference numbers in
the present work are listed in~\cite{list}). 
(Taken from Ref.~\cite{Aniol00})
}
\label{Fig2}
\end{figure}

The allowed region of values of the parameters $\rho_s$ and $\mu_s$,
obtained from (\ref{strangeHap2}), is shown in Fig.~\ref{Fig2}. Points 
are the predictions of different models~\cite{list}.
%
%**************************************************************************
%**************************************************************************
%

The main uncertainties in the determination of the quantity
(\ref{strangecomb}) are connected with the
electromagnetic form factors of the neutron, which are not known, at present,
with accuracy large enough. The value (\ref{strangeHap}) was
obtained~\cite{Anklin98} by
using (\ref{ffratio}) for the magnetic form factor of the neutron.
If, instead, we take the value~\cite{Bruins95}
\begin{equation}
\frac{\GMem^n/\mu_n}{\GMem^p/\mu_p}= 1.12\pm 0.04
\end{equation}
then
\begin{equation}
\frac{\GEs + 0.392\GMs}{\GMem^p/\mu_p} =
0.122\pm 0.056\pm 0.047\,.
\label{strangeHap3}
\end{equation}
Thus the new measurements of electromagnetic form factors
of the nucleon which
are in progress at Jefferson Lab  will have an important impact on
the possibility of obtaining a more precise information
on the strange form factors of the nucleon from future measurements
of the P--odd asymmetry.

An extension of the HAPPEX experiment (HAPPEX2)~\cite{happex2}
is planned at Jefferson Lab: it will measure the P-odd asymmetry
at a scattering angle $\theta\simeq 6 ^\circ$, corresponding to
$Q^2\simeq 0.1$ GeV$^2$ thus smaller than in the HAPPEX measurement.
The motivation for this extension is to explore the possibility
that the strange form factors can be large at small $Q^2$ but then
fall off significantly at the current HAPPEX kinematics.

Another experiment on the measurement of the P--odd asymmetry in 
elastic $e-p$ scattering was carried out by the SAMPLE collaboration at
the MIT/Bates Linear Accelerator Center~\cite{Spayde00}. In this experiment
longitudinally polarized electrons with energy of $200$~MeV were
scattered in backward direction, at scattering angles
$130^{\circ} \le \theta\le 170^{\circ}$. The average value of the
momentum transfer squared was $\langle Q^2\rangle =0.1$~GeV$^2$. A liquid
hydrogen target was used in the experiment. The scattered
electrons were detected by air \u{C}herenkov counters. The average
polarization of the electron beam was equal to $36.3 \pm 1.4 \%$.
In the latest measurements the following value of the P--odd
asymmetry was obtained:
\begin{equation}
\Acal_p(Q^2=0.1) =\left\{ -4.92\pm 0.61 (\mathrm{stat})\pm
0.73 (\mathrm{syst})\right\}\times 10^{-6}\,.
\label{asymSam}
\end{equation}

When electrons are scattered in the backward direction, the
parameter $\varepsilon$ in the expressions (\ref{A0elec}) and (\ref{Aselec})
is small and the contribution to the asymmetry of the electric strange form 
factor $\GEs$ is suppressed. The measurement of the P--odd asymmetry 
allows one in this case to
obtain information on the strange magnetic form factor of the nucleon,
$\GMs$. From Eq.~(\ref{asymmel2}) for $\theta=\pi$ we obtain the
following expression for the asymmetry:
\begin{equation}
\Acal_p = -\frac{G_F Q^2}{2\sqrt{2}\pi\al}
\left[ \frac{\GMNC}{\GMem} +
(1-4\sin^2\theta_W)\sqrt{1+\frac{1}{\tau}}\frac{\GANC}{\GMem}\right]\,.
\label{asymmel4}
\end{equation}
The last, axial, term in the above expression is multiplied by the factor
 $(1-4\sin^2\theta_W)$ which is small ($\simeq 0.07$). 
 However, in the SAMPLE  experiment
the value of $\tau$ is small ($\tau \simeq 0.03$): hence the contribution
of the axial form factor turns out to be kinematically enhanced. 

In Eq.~(\ref{asymmel4}) the weak axial form factor of the proton is
given, at tree level in the Standard Model, by:
\begin{equation}
\GANC=\half\left(\GACC-\GAs\right)\,.
\label{ganc2}
\end{equation}
However, as it was pointed out in  Ref.~\cite{MusHol90}, 
the contribution to the P--odd asymmetry of the radiative
corrections can be large. Taking the latter into account, the expression
(\ref{ganc2}) can be written in the form
\begin{equation}
\label{gancRC}
\GANC=\half\left[\left(1+R_A^1\right)\GACC - R_A^0 -\GAs\right]\,,
\end{equation}
where $R_A^1$ and $R_A^0$  are the radiative corrections to the
isovector and isoscalar parts of the matrix element. They were
calculated to be~\cite{MusHol90}:
\begin{equation}
R_A^1= -0.34\pm 0.28;\qquad R_A^0= -0.12\pm 0.12
\label{radcor}
\end{equation}

The electroweak corrections to the
nucleon vertex induce the following anapole axial term in the matrix
element of the electromagnetic current:
\begin{equation}
\la p'|J^{em}_{\al}|p\ra = e\frac{a(Q^2) Q^2}{M^2}
\ubar(p')\left(\gam_{\al}-\frac{\qsla q_{\al}}{q^2}\right)\gam_5 u(p)\,.
\label{anapole1}
\end{equation}
Here $a(0)$ is the anapole moment of the nucleon~\cite{Zeldovich61}.
We recall that the anapole moment of Cs nuclei was measured in a recent
experiment~\cite{Wood97}. In Ref.~\cite{Zhu00}
the contribution to the P--odd asymmetry of the anapole moment of the
nucleon  has been calculated in the framework of chiral perturbation theory,
both for the isovector [$(R_A^1)_a$] and isoscalar [$(R_A^0)_a$] terms. 
They are given by~\cite{Zhu00}:
\begin{equation}
\left(R_A^I\right)_a = -\frac{8\sqrt{2}\pi\al}{G_F\Lambda_{\chi}^2}
\frac{1}{(1-4\sin^2\theta_W)}\frac{a_I}{\GACC}\,,\quad(I=0,1)
\label{anapole2}
\end{equation}
where $\Lambda_{\chi}$ is the scale of chiral symmetry breaking.
In Ref.~\cite{Zhu00} for the contribution of the anapole moments to
$R_A^1$ and $R_A^0$ it was found:
\begin{equation}
\left(R_A^1\right)_a = -0.06\pm 0.24\,,\qquad
\left(R_A^0\right)_a = 0.01\pm 0.14\,.
\label{anapole3}
\end{equation}
and for the total radiative corrections to the axial form factor $\GANC$
the following values were obtained: 
\begin{equation}
R_A^1= -0.41\pm 0.24\,;\qquad
R_A^0= 0.06\pm 0.14\,.
\label{radcor2}
\end{equation}

The SAMPLE data for the proton were first studied by
assuming the values (\ref{radcor}) for the radiative corrections
and the value $g_A^s=-0.1$ for the axial strange form factor
(in agreement with the data of the experiments on the deep inelastic
scattering of polarized leptons on polarized protons). Under these 
assumptions the following value of the strange magnetic form factor
at $Q^2=0.1$~GeV$^2$ was obtained~\cite{Spayde00}:
\begin{equation}
\GMs = 0.61 \pm 0.17\pm 0.21\pm 0.19\,,
\label{Samgms}
\end{equation}
where the last error is due to uncertainties in the radiative
corrections. 

Recently the SAMPLE collaboration has published the first results
of the experiment on the measurement of the P-odd asymmetry
in the quasi-elastic scattering of polarized electrons on 
deuterium~\cite{Beise99,Sample01} 
in the same kinematical region as in the proton case.
The P--odd asymmetry in $\vec{e}-d$ scattering is given by the 
following expression~\cite{Sample01}:
\bea 
\Acal_d&=&\left(-7.27 + 1.78 G_{A}^{e}(T=1) + 0.75G_M^s
\right) \times 10^{-6}\,,
\label{eq.deutasy} 
\eea
where the term
\bea
 G_{A}^{e}(T=1) = - G_{A}( 1 + R_{A}^1 ) 
\eea
includes the axial form factor and the isovector part of the radiative 
corrections. The (small) isoscalar part of the radiative corrections and 
the contribution of $G_{A}^s$ are included in the constant term in 
Eq.~(\ref{eq.deutasy}).

The P-odd asymmetry in the scattering of polarized electrons
on protons can be  expressed as follows~\cite{Sample01}:
\begin{equation} 
\Acal_p=\left(-5.72 + 1.55 G_{A}^{e} + 3.49G_M^s
\right) \times 10^{-6}\,.
\label{eq.pasy} 
\end{equation}

\begin{figure}
\begin{center}
\mbox{\epsfig{file=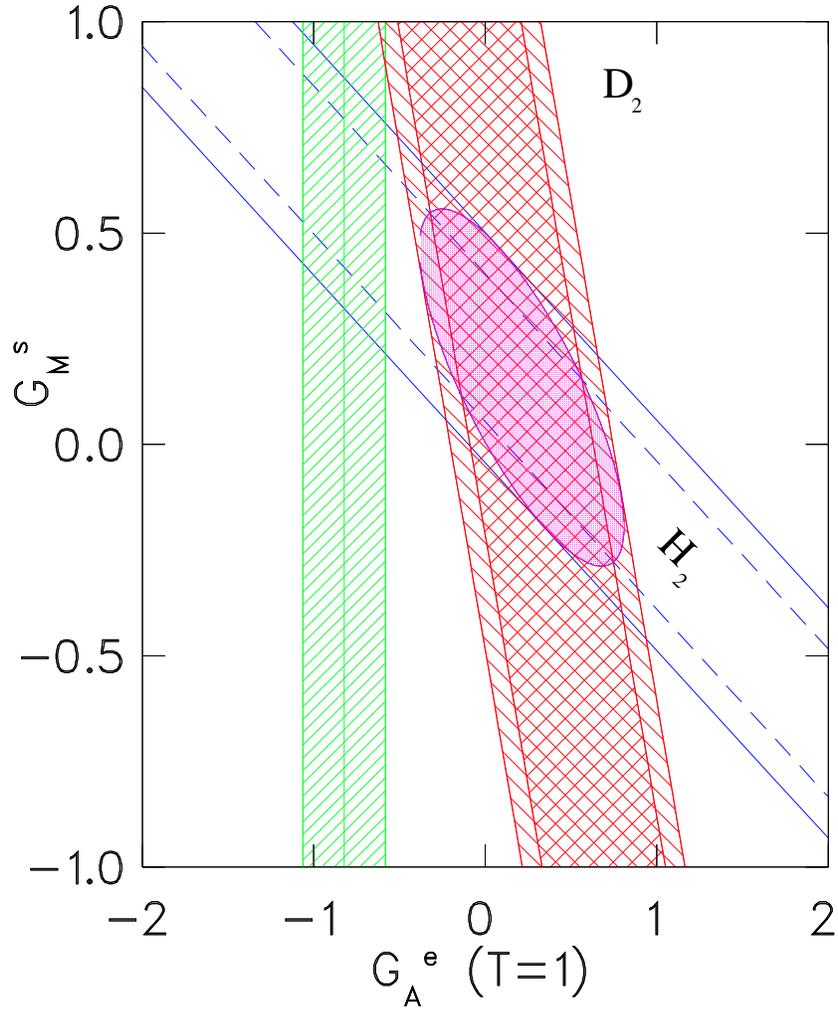,width=0.8\textwidth}}
\end{center}
\vspace{1cm}
\caption{
Allowed regions for the form factors $G_M^s$ and 
$-(1+R_A^1)G_A \equiv G^e_A$ (in the notation of the figure),
corresponding to the measurements of Refs.~\cite{Spayde00,Sample01} 
%Refs.~\protect\cite{Spayde00,Sample01} 
of the P--odd asymmetries for the proton ($H_2$, unshaded region)
and for the deuteron ($D_2$, hatched region), respectively. 
The inner regions include statistical errors only, 
the outer ones include statistic and
systematic uncertainties added in quadrature. The vertical shaded band
corresponds to the calculated value of $-(1+R_A^1)G_A$ 
using the theoretical estimate of Ref.~\cite{Zhu00}
%Ref.~\protect\cite{Zhu00}.
The isoscalar corrections $R_A^0$ are assumed to be the ones
calculated in Ref.~\cite{Zhu00}.
%Ref.~\protect\cite{Zhu00}.
(Taken from Ref.~\cite{Sample01})
%Ref.~\protect\cite{Sample01})
}
\label{Fig3}
\end{figure}

The measured value of the asymmetry in the SAMPLE $\vec{e}-p$ 
experiment~\cite{Spayde00} is given by (\ref{asymSam}), while the 
P--odd asymmetry measured in $\vec{e}-d$ scattering turned out to 
be~\cite{Sample01}:
\begin{equation}
\Acal_{d}(Q^2=0.1) =\left\{ -6.79\pm 0.64 (\mathrm{stat})\pm
0.55 (\mathrm{syst})\right\}\times 10^{-6}\,.
\label{Sam1}
\end{equation}
By combining Eqs.~(\ref{eq.deutasy}) and (\ref{eq.pasy}) with the 
corresponding experimental values, the authors of Ref.~\cite{Sample01}
obtained two bands in the $(G_A^e,G_M^s)$ plane, which are shown in 
Fig.~\ref{Fig3}.
The inner parts of the bands include only statistical errors
while the outer bounds take into account statistical and systematic 
errors combined in quadrature.
The shaded ellipse in Fig.~\ref{Fig3} corresponds to the $1\sigma$
allowed region for both quantities.

%*************************************************************************
%*************************************************************************

The best--fit values of the form factors  $ G_{M}^{s}$ and
 $ G_{A}^{e}$ at $Q^{2}=0.1$~GeV$^2$ are given by:
\bea
G_M^s &&= 0.14\pm 0.29 \,{\mathrm{stat}} 
\pm 0.31\, {\mathrm{syst}}\,,
\label{eq.gms-samplenew}\\
G_A^{e}(T=1) &&= 0.22\pm 0.45 \,{\mathrm{stat}} 
\pm 0.39\, {\mathrm{syst}}\;.
\label{eq.sam2}
\eea

In order to obtain from Eq.(\ref{eq.gms-samplenew}) the strange 
magnetic moment of the nucleon it is necessary to assume a 
$Q^{2}$--dependence of the strange form factor. In Ref.~\cite{Sample01} 
a model proposed by Hemmert {\it et al.}~\cite{Hemmert98},
based on heavy baryon chiral perturbation theory, was used.
For the strange magnetic moment the following value was then obtained 
\begin{equation}
\mu_s= \left[0.01 \pm 0.29 \,{\mathrm{stat}}
\pm 0.31 \,{\mathrm{syst}} \pm 0.07 \,{\mathrm{theor}}\right] \mu_{N}\,,
\label{eq.musnew}
\end{equation}
where the third error takes into account the theoretical uncertainty
coming from different theoretical predictions.
Thus, from the latest SAMPLE data it is impossible to draw any definite
conclusion on the value of the strange magnetic moment of the nucleon.

Let us discuss now the value (\ref{eq.sam2}) of the axial constant.
At $Q^{2}=0.1$~GeV$^{2}$, in Born approximation and assuming the usual
axial dipole form factor,  we have $G_A^{e}(T=1)= -1.071\pm 0.005$.  
Taking into account the radiative corrections calculated in Ref.~\cite{Zhu00},
the following value of the form factor $G_A^{e}(T=1)$ was 
obtained~\cite{Sample01}:
\begin{equation}
G_A^{e}(T=1)= -0.83\pm 0.26
\label{eq.samoil3}
\end{equation}
This value corresponds to the vertical band in Fig.~\ref{Fig3}.
Thus, the predicted value of  $G_A^{e}(T=1)$ differs considerably from 
the experimental value, Eq.~(\ref{eq.sam2}). One possible origin of this 
disagreement could be connected with a large anapole moment of the 
nucleon~\cite{Sample01}.

The surprising results which have been obtained in the SAMPLE
experiments on the measurement of the P--odd asymmetry in $\vec{e}-p$ and 
$\vec{e}-d$ experiments require further theoretical efforts in 
the calculations of the radiative corrections and further experiments,
which  will allow to check these results (see also the recent review 
\cite{Holst01}). At present the SAMPLE collaboration has proposed a new 
experiment \cite{BATES,Ito01} on the measurement of the P--odd asymmetry 
in the scattering on deuterium of polarized electrons with energy of
$120$~MeV (thus lower than in the previous run). 
At this energy the asymmetry will be smaller, but
the cross section will be significantly larger.

One final remark about the measurement of strange magnetic moment of
the nucleon is in order: with the help of expression (\ref{jem})
for the electromagnetic current, we can present the Pauli form factors of
proton and neutron in the following form:
\begin{equation}
\begin{array}{l}
F_2^p= \frac{2}{3}F_2^u +\left(-\frac{1}{3}\right)F_2^d +
\left(-\frac{1}{3}\right)F_2^s\\
F_2^n= \frac{2}{3}F_2^d +\left(-\frac{1}{3}\right)F_2^u +
\left(-\frac{1}{3}\right)F_2^s
\end{array}
\label{Pauliff}
\end{equation}
where $F_2^q$ is the contribution of the $q$--quark ($q=u,d,s$)
to the Pauli form factor of the proton. We have used in Eqs.~(\ref{Pauliff})
the isotopic SU(2) symmetry, from which it follows 
\[ \left(F_2^{u,d}\right)_p = \left(F_2^{d,u}\right)_n\,.\]
If we set $Q^2=0$ and take into account that
$F_2^p(0)\equiv\kappa_p=1.79$, $F_2^n(0)\equiv\kappa_n=-1.91$,
we obtain:
\bea
1.79&=& \frac{2}{3}F_2^u(0) +\left(-\frac{1}{3}\right)F_2^d(0) +
\left(-\frac{1}{3}\right)\mu_s\,,
\nonumber\\
-1.91&=& \frac{2}{3}F_2^d(0) +\left(-\frac{1}{3}\right)F_2^u(0) +
\left(-\frac{1}{3}\right)\mu_s
\nonumber
\eea
These relations can be combined to give:
\begin{equation}
\begin{array}{rcl}
F_2^u(0) &=& \mu_s + 1.67\\
F_2^d(0) &=& \mu_s -2.03
\end{array}
\label{F2ud}
\end{equation}

Thus, the measurement of the strange magnetic moment of the nucleon
will allow one to determine the contribution of the $u$ and $d$ quarks
to the magnetic moments of proton and neutron.
From (\ref{F2ud}) it follows that if $\mu_s>0.18$ then
$F_2^u(0)\ge |F_2^d(0)|$,
while in the case that $\mu_s<0.18$ the opposite inequality holds,
$F_2^u(0)\le |F_2^d(0)|$.

A new experiment on the measurement of P--odd asymmetry in elastic
electron--proton scattering is going on at the Mainz Microtron
Facility~\cite{Maas99}. The energy of the electron beam
in this experiment is of $855$~MeV. The scattered electrons are
detected at a scattering angle of $35^{\circ}$ ($ Q^2 = 0.227$~GeV$^2 $).
The polarization of the  electron beam is $80\%$.
It is expected that
the P--odd asymmetry will be measured with statistical accuracy of $ 3\%$
and systematic error of $4\%$.
The  combination of strange Dirac and Pauli \ffs $F_1^s+0.13 F_2^s$ at
$Q^2=0.227$~GeV$^2$ will be determined from this experiment
with accuracy  of 0.02.

Finally,  the
G0 collaboration is measuring now the  P--odd asymmetry in elastic
electron--proton scattering at Jefferson Lab~\cite{Beck99}. It is expected 
that from the data of this experiment the
strange form factors will be determined with a few \% accuracy at different
values of $ Q^2 $
in the interval $ 0.1 \le Q^{2} \le 1$~GeV$^2$.

Thus, in the nearest years we will have new information on the
 strange vector form factors of nucleon, an information which could
 have an important impact on our understanding of the nucleon structure
 and of strong interactions.

%%%%%%%%%%%%%%%%%%%%%%%%%%%%%%%%%%%%%%%%%%%%%%%%%%%%%%%%%%%%%%%%%%%%%%%
\section{P--odd asymmetry in the elastic scattering of electrons
on nuclei with $S=0$ $T=0$ }
\label{sec.PoddAel}

In this Section we shall consider the elastic scattering of polarized
electrons on nuclei with spin and isotopic spin equal to zero 
(like $^4$He, $^{12}$C, etc.). Interest for this case was raised in
 past works~\cite{Feinb75,Walec77,DonnPeccei,Beck89}.
\begin{equation}
\vec{e} +A\longrightarrow e + A
\label{reaction1}
\end{equation}
It is evident from the SU(2) isotopic invariance of strong
interactions that in this case the matrix elements of the isovector
currents $V^3_{\al}$ and  $A^3_{\al}$ are equal to zero.
Also the matrix element of the axial strange current 
$A^s_{\alpha}$ is equal to zero\footnote{
In fact, the matrix element $\la p'|A^s_{\al}|p\ra$ is a pseudovector
and depends only on $p$ and $p'$ ($p$ and $p'$ being the momenta of the
initial and final nucleus). It is obvious that from two vectors
it is impossible to build pseudovector.}.

The matrix element of the process (\ref{reaction1}) is given by the
general expression (\ref{polampli}) in which $p$ ($p'$) refers now to
the initial (final) nucleus and, as before, 
 $q=p'-p$ and  $Q^2= - q^2$. The process (\ref{reaction1})
can be represented by the same diagrams of Fig.~\ref{Fig1}, with
the exchange of $\gamma$ and $Z^0$ between the electron and the nucleus.

For the matrix element of the electromagnetic current  $J^{em}_{\al}$
we have now
\begin{equation}
\la p'|J^{em}_{\al}|p\ra =\la p'|V^0_{\al}|p\ra =
(p+p')_{\al} F(Q^2)
\label{emamplnuc}
\end{equation}
where  $V^{0}_{\al}$ is the isoscalar component of the
electromagnetic current and $F(Q^2)$ is the electromagnetic 
form factor of the nucleus (there is only one \ff in the case
of a spin zero nucleus). 

Similarly the nuclear matrix element of the NC reads
\begin{equation}
\la p'|J^{NC}_{\al}|p\ra =\la p'|\left(-2\sin^2\theta_W J^{em}_{\al}
-\half V^s_{\al}\right)|p\ra =
(p+p')_{\al} F^{NC}(Q^2)
\label{NCamplnuc}
\end{equation}
where
\begin{equation}
\label{weakff}
F^{NC}(Q^2)= -2\sin^2\theta_W F(Q^2) -\half F^s(Q^2)
\end{equation}
and $F^s(Q^2)$ is the strange form factor of the nucleus\footnote{
Let us notice that the current 
 $V^{s}_{\alpha}$ is not conserved and the matrix element of the strange
vector current has the following general form
\[ \la p'|V^s_{\al}|p\ra =
(p'+p)_{\al} F^s(Q^2)+ (p'-p)_{\al}G^s(Q^2)\,.\]
However from T--invariance of the strong interactions it follows that
the \ff $G^s$  is equal to zero.}.

At  $Q^2=0$ the form factor $F(Q^2)$ is equal to the total charge of the
nucleus
\[ F(Q^2=0) = Z \]
while  for the strange form factor we have, in the same limit,
\[ F^s(Q^2=0) =0 \]
(the  net strangeness of the nucleus is equal to zero).

At small  $Q^2$ we can expand the two form factors as follows:
\bea
F(Q^2) &=& Z\left(1-\frac{1}{6}\la r^2\ra Q^2 +\dots\right)
\nonumber\\
F^s(Q^2) &=& -\frac{1}{6}\la r_s^2\ra Q^2 +\dots
\nonumber
\eea
where $\la r^2\ra$ is the mean square of the electromagnetic radius
of the nucleus and $\la r^2_s\ra$ is the mean square of the
nuclear ``strangeness radius''.
Let us notice that in the impulse approximation, for nuclei with $N=Z$,
we have
\begin{equation}
\frac{F^s(Q^2)}{F(Q^2)} =
\frac{2G_E^s(Q^2)}{\GEem^p(Q^2) + \GEem^n(Q^2)}\,.
\end{equation}

The general expression for the cross section of the scattering
of electrons with longitudinal polarization $\lambda$ on nuclei
with zero spin can be obtained from Eq.~(\ref{cross1}) by setting
$W^I_{\al\be}(A)=0$. One gets then:
\begin{equation}
d\sigma_{\lambda} =\frac{4\al^2}{Q^4}\frac{M_A}{p\cdot k}\half
\left\{ L^{\al\be} W^{em}_{\al\be} +
\lambda \frac{G_F Q^2}{2\sqrt{2}\pi\al}
g_A L^{\al\be} W_{\al\be}^I(V)\right\}
\frac{d\vec{k'}}{{k'}^0}\,,
\label{crossnucl}
\end{equation}
where $M_{A}$ is the mass of the nucleus and the tensors $W^{em}_{\al\be}$
and $W^I_{\al\be}(V)$ are given by Eqs.~(\ref{Wem}) and (\ref{WIV}).

For a spin zero nucleus the tensors $W^{em}_{\al\be}$ and
$W^I_{\al\be}(V)$ have the following general form
\begin{equation}
\begin{array}{l}
W^{em}_{\al\be} = \frac{\displaystyle{n_{\al}n_{\be}}}
{\displaystyle{4M_A^2}}\,  W^{em}_2\\
W^I_{\al\be}(V) = \frac{\displaystyle{n_{\al}n_{\be}}}
{\displaystyle{4M_A^2}}\, W^I_2
\end{array}
\label{nuctensor}
\end{equation}
where $n=p+p'$. It is then easy to show that
\begin{equation}
\begin{array}{l}
W^{em}_2 = F^2(Q^2)\, \delta\left(\nu-\frac{\displaystyle{Q^2}}
{\displaystyle{2M_A}}\right)\,,\\
W^I_2 = 2F^{NC}(Q^2)F(Q^2)\,\delta\left(\nu-\frac{\displaystyle{Q^2}}
{\displaystyle{2M_A}}\right)
\end{array}
\label{nucW2}
\end{equation}
Here $\nu=E-E'=p\cdot q/M_A$ is the energy transferred to the
nucleus in the laboratory system.

By inserting (\ref{nuctensor}) and (\ref{nucW2}) into (\ref{crossnucl}),
we obtain the following expression for the cross section of the
scattering  of electrons with longitudinal polarization $\lambda$ on
spin zero nuclei:
\begin{equation}
\left(\frac{d\sigma}{d\Omega}\right)_{\lambda} =
\left(\frac{d\sigma}{d\Omega}\right)_0 (1+\lambda\Acal)\,.
\label{crossnuc2}
\end{equation}
Here
\begin{equation}
\left(\frac{d\sigma}{d\Omega}\right)_0=\sigma_{Mott} F^2(Q^2)
\label{sigma0}
\end{equation}
is the cross section for the scattering of unpolarized electrons on
nuclei, $\sigma_{Mott}$ being the Mott cross section for a target nucleus
of mass $M_A$:
\begin{equation}
\sigma_{Mott}=\frac{\al^2\cos^2(\theta/2)}
{\displaystyle{4E^2\sin^4\frac{\theta}{2}\left(1+\frac{2E}{M_A}
\sin^2\frac{\theta}{2}
\right)}}\,.
\label{sigmaMott2}
\end{equation}
In the above $\theta$ is the scattering angle and $E$ the initial energy
of the electron in the laboratory system.

The P--odd asymmetry $\Acal$ is then given by~\cite{Bernabeu92}
\bea
\Acal(Q^2) &=& - \frac{G_F Q^2}{2\sqrt{2}\pi\al}
\frac{F^{NC}(Q^2)}{F(Q^2)} =
\nonumber\\
&=&  \frac{G_F Q^2}{2\sqrt{2}\pi\al}
\left(2\sin^2\theta_W +
\frac{F^s(Q^2)}{2 F(Q^2)}\right)
\label{nuclasym}
\eea
As it is clearly seen from (\ref{nuclasym}), the measured value of the
asymmetry can be different from
\begin{equation}
\Acal_{(0)}(Q^2) =\frac{G_F Q^2}{\sqrt{2}\pi\al}\sin^2\theta_W =
8.309\times 10^{-5} \frac{Q^2}{\mathrm{GeV}^2}
\end{equation}
only if the strange form factor $F^s(Q^2)$ is different
from zero. Important information on the strange  form factor of the
nucleus can be obtained from the investigation of the $Q^2$ dependence
of the asymmetry: if the quantity $\Acal(Q^2)/Q^2$ depends on $Q^2$,
it would be the proof that the strange nuclear form factor is different
from zero. Finally, should it occur that the P--odd asymmetry
(\ref{nuclasym}) is negative, it would imply 
that the strange form factor of the nucleus is large and negative.

From (\ref{nuclasym}) it follows that the strange form factor of a
nucleus with S=0 and T=0 is determined by quantities that can be
experimentally measured. In fact we have
\begin{equation}
F^s(Q^2) = 2\sqrt{\frac{1}{\sigma_{Mott}}
\left(\frac{d\sigma}{d\Omega}\right)_0}
\left[\frac{2\sqrt{2}\pi\al}{G_F Q^2}\Acal(Q^2) -2\sin^2\theta_W
\right]\,.
\label{sffnucl1}
\end{equation}

An experiment on the measurement of the P--odd asymmetry in the
elastic scattering of polarized electrons on $^4$He is under
preparation at the Jefferson Lab~\cite{Beise00}. The square of the
momentum transfer in this experiment is expected to be $Q^2=0.6$~GeV$^2$.
Two high resolution spectrometers will be employed. The target will be
a circulating  $^4$He gas system. Thus this experiment will be able to
measure the above discussed strange form factor of a spin zero nucleus.

We like to mention, here, that the P--odd asymmetry in the elastic
scattering of polarized electrons on nuclei represents an almost direct
measurement of the Fourier transform of the neutron density, since the
$Z^0$--boson preferentially couples to neutrons. Indeed for $0^+\to 0^+$
transitions, it is easy to show that the P--odd asymmetry can be expressed
in the following form~\cite{AMParis}:
\begin{equation}
\Acal=-\Acal_0\frac{1}{2}\left\{(1-4\sin^2\theta_W)-
\frac{\int d\vec{r}j_0(qr)\rho_n(\vec{r})}
{\int d\vec{r}j_0(qr)\rho_p(\vec{r})}\right\}
\label{neutronden}
\end{equation}
which is valid both for isospin symmetric ($Z=N$) and asymmetric 
($Z\ne N$) nuclei. In Eq.~(\ref{neutronden}) $\rho_n$ ($\rho_p$) is
the neutron (proton) density and $j_0(x)$ the spherical Bessel function
of order zero. 
Taking into account the value of $\sin^2\theta_W$, the last term in the
right hand side of Eq.~(\ref{neutronden}) dominates the asymmetry and 
directly gives information on the neutron distribution. In fact the 
denominator coincides with the form factor $F(Q^2)$, which can be 
measured independently [see Eq.~(\ref{sigma0})].
The Parity Radius Experiment (PREX) at the Jefferson Laboratory plans
to measure the neutron radius $R_n$ in $^{208}$Pb through parity violating 
electron scattering~\cite{PREX}. The measurement of the neutron skin in a 
heavy nucleus ($R_n$ is generally assumed to be a few \% larger that the
proton radius) will have important implications on our knowledge of the
structure of neutron stars, which are expected to have a solid, 
neutron--rich crust~\cite{HorPie01}.

%%%%%%%%%%%%%%%%%%%%%%%%%%%%%%%%%%%%%%%%%%%%%%%%%%%%%%%%%%%%%%%%%%%%%%%
\section{Inelastic Parity Violating (PV) electron scattering on nuclei}
\label{sec.PoddQE}

In addition to PV elastic electron scattering on proton, deuterium and
$S=0, T=0$ nuclei, the  P--odd asymmetry can be considered in the 
process of inelastic scattering of polarized electrons on nuclei. 
Several basic ideas motivate this investigation: the scattering on 
the single proton is not sufficient to determine the various unknown
form factors which enter into the PV hadronic response; one is thus 
immediately led to consider also neutrons (namely deuterium) and more
generally nuclei~\cite{Musolf94.rep,DonnPeccei,Mus-Don1}.

 As we have discussed in the previous Section, the special case
of elastic scattering on spin--zero, isospin-zero targets offers an
unambiguous possibility to measure the strange form factor of the nucleus.
However this type of investigation is confined to modest momentum transfers
since the elastic nuclear form factors rapidly fall off with increasing
momentum, with the exception of the very light nuclei.
Therefore one would like to have additional, complementary information
from other electron scattering measurements. One possibility is the 
inelastic excitation of discrete
states in nuclei, but most probably the corresponding cross sections are
not large enough to permit high precision information to be extracted. 

A more promising case is the quasi--elastic (QE) scattering namely the 
inelastic scattering of electrons in the region of the so--called 
quasi--elastic peak~\cite{Bertozzi92}. 
This process roughly corresponds to ``knocking'' 
individual nucleons out of the nucleus without too much complication in the
final nuclear state, in particular from final state interactions. 
QE scattering occurs for a given three--momentum transfer $q\equiv|\qv\,|$,
approximately at energy transfer $\omega=Q^2/(2M)$\footnote{
Here we adopt the customary notation $q_0=\omega$ for the energy 
transferred to the nucleus; hence $Q^2=-q^2={\qv}^2-\omega^2$.}
The width of the peak is characterized by the Fermi momentum $p_F$ 
of the specific nucleus under study. In this kinematical region the 
cross sections are generally proportional to the number of nucleons 
in the nucleus, and thus are prominent features in the inelastic spectrum.
One might then hope to perform high precision studies, which would complement
work on parity--violating elastic electron 
scattering~\cite{Musolf94.rep,Mus-Don93}.

The focuses of this investigation are multiple: on the one hand one 
wishes to understand the role played by the various single--nucleon \ffs
in the total asymmetry. By changing the kinematics ($q$, $\omega$ and the 
scattering angle $\theta$) and by adjusting $N$ and $Z$ through the choice of
different targets, one can hope to alter the sensitivity of the asymmetry
to the underlying form factors. 
Of course a precise study of nucleonic \ffs from the
scattering on nuclei is possible only if nuclear model uncertainties are
well under control. On the other hand the measurement of the asymmetry
in PV QE electron scattering brings into play new aspects of the nuclear
many--body physics, namely the ones related to the nuclear response 
functions to NC probes. This might involve a sensitivity of the cross
sections to specific dynamical aspects which cannot be revealed with the
customary reactions employed in nuclear structure studies. 

\begin{figure}
\begin{center}
\mbox{\epsfig{file=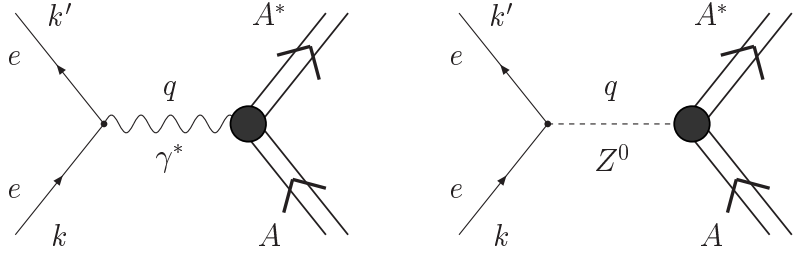,width=0.99\textwidth}}
\end{center}
\caption{Single photon exchange and $Z^0$ exchange diagrams for 
 electron scattering from nuclei.}
\label{fig.PVscat}
\end{figure}

These issues were extensively discussed in Ref.~\cite{DMABDM}
where PV quasi--elastic electron scattering was studied within the context 
of the relativistic Fermi gas (RFG). Let us  consider the inclusive 
process in which a polarized electron with four--momentum $k$ and 
longitudinal polarization $\lambda$ is scattered through an angle 
$\theta$ to four--momentum $k'$, exchanging a photon or a $Z^0$ to the
target nucleus:
\begin{equation}
\vec{e} + A\longrightarrow e +A^*
\label{einescatt}
\end{equation}
We generically indicate with $A^*$  the final nucleus in an excited state,
in which  one (or more) nucleons are ejected. The leading order 
diagrams contributing to the amplitude of the process (\ref{einescatt}) 
are illustrated in Fig.~\ref{fig.PVscat}.
Only the final electron is detected and fixes the 
kinematics of the process. 

%*************************************************************************
%*************************************************************************

The inclusive cross section for the scattering of polarized 
electrons on unpolarized nuclei can be written as:
\bea
d\sigma_{\lambda} && =\frac{4\al^2}{Q^4}\frac{1}{2k^0}
\left\{ L^{\al\be} W^{em}_{\al\be} +\right.
\label{crossepolA}\\
&&\, \left.
+\lambda \frac{G_F Q^2}{2\sqrt{2}\pi\al}\left[g_V L_5^{\al\be}
W^I_{\al\be}(A) +g_A L^{\al\be} W_{\al\be}^I(V)\right]\right\}
\frac{d\vec{k'}}{{k'}_0}\,,
\nonumber
\eea
where the leptonic tensor and pseudotensor, 
$L^{\alpha \beta}$ and $L_5^{\alpha \beta}$, are given in 
Eqs.~(\ref{lepten}) and (\ref{lepten5}). The
hadronic (electromagnetic and interference) tensors are 
defined as:
\bea
W_{\al\be}^{em}&&= \frac{3 {\Ncal} M^2}{4\pi p_F^3}\int
\frac{d\vec{p}}{E_p}\frac{d\vec{ p'}}{E_{p'}}\delta^{(4)}(p'-p-q)\times
\nonumber\\
&&\qquad\times 
\theta(p_F-|{\vec{p}}\,|)\theta(|{\vec{p}}^{\,\prime}|-p_F)
\left(W_{\al\be}^{em}\right)_{s.n.}\,,
\label{hadrtenem}
\eea
and
\bea
W_{\al\be}^{I}(V,A)&&= \frac{3 {\Ncal} M^2}{4\pi p_F^3}\int
\frac{d\vec{p}}{E_p}\frac{d\vec{ p'}}{E_{p'}}\,
\delta^{(4)}(p'-p-q)\times
\nonumber\\
&&\qquad\times 
\theta(p_F-|{\vec{p}}\,|)\theta(|{\vec{p}}^{\,\prime}|-p_F)
\left[W_{\al\be}^{I}(V,A)\right]_{s.n.}\,.
\label{hadrtenI}
\eea
In Eqs.~(\ref{hadrtenem}) and (\ref{hadrtenI}) $E_p=\sqrt{\vec{p}^2+M^2}$
and the single--nucleon (s.n.) tensors are:
\bea
&&\left[W_{\al\be}^{em}\right]_{s.n.}=-\tau\GMem^2
\left(g_{\al\be} -\frac{q_{\al}q_{\be}}{q^2}\right)+
\frac{X_{\al}X_{\be}}{M^2}
\frac{\displaystyle{\GEem^2+\tau\GMem^2}}
{\displaystyle{1+\tau}}
\label{Wpvem}\\
&&\left[W_{\al\be}^{I}(V)\right]_{s.n.}= -2\tau\GMem\GMNC
\left(g_{\al\be} -\frac{q_{\al}q_{\be}}{q^2}\right)+
\nonumber\\
&&\qquad\qquad\qquad\qquad +2\frac{X_{\al}X_{\be}}{M^2}
\frac{\displaystyle{\GEem\GENC+\tau\GMem\GMNC}}
{\displaystyle{1+\tau}}
\label{WpvNCV}\\
&&\left[W_{\al\be}^{I}(A)\right]_{s.n.}=
\frac{i}{2M^2}\epsilon_{\al\be\rho\sigma}p^{\rho}q^{\sigma}
4\GMem\GANC\,,
\label{WpvNCA}
\eea
where $X_{\al}=p_{\al}-q_{\al}(q\cdot p)/q^2$. The expressions 
(\ref{hadrtenem}) and (\ref{hadrtenI}) for the nuclear hadronic tensors 
are obtained in the Impulse Approximation (IA), which amounts to 
consider the electron--nucleus interaction as an incoherent 
superposition of electron--nucleon scattering processes. Moreover
the nucleus is described as a gas of non--interacting, relativistic 
nucleons, with momentum distribution 
$3{\Ncal}/(4\pi p_F^3)\theta(p_F-|\vec{p}\,|)$, $p_F$ being the 
Fermi momentum. In Eqs.~(\ref{hadrtenem}) and (\ref{hadrtenI}) 
${\Ncal}=Z,N$ is the number of protons or neutrons in the nucleus 
and the function $\theta(|\vec{p'}|-p_F)$ ensures that the final 
nucleon is excited above the Fermi level (Pauli blocking effect).
We also notice that the total cross section is obtained from the sum of 
the contributions from protons and neutrons, each of them being 
calculated by using the pertinent nucleonic \ffs in the single--nucleon 
tensors (\ref{Wpvem})--(\ref{WpvNCA}).

From the above equations one can derive the expression for 
the  double--differential (with respect to the energy, 
$k_0'\equiv\epsilon'$, and scattering angle, $\Omega$, 
of the final electron)
inclusive  cross section for the inelastic scattering of 
polarized electrons on nuclei. The  sum of the  cross sections for 
electrons with opposite polarization
\begin{equation}
\left(\frac{d^2\sigma}{d\Omega d\epsilon'}\right)^{pc} =
 \frac{d^2\sigma_{+}}{d\Omega d\epsilon'} +
\frac{d^2\sigma_{-}}{d\Omega d\epsilon'} =
\sigma_{Mott}\left\{v_L R^L(q,\omega) +v_T R^T(q,\omega)\right\}
\label{emcross1}
\end{equation}
coincides with the inclusive, parity--conserving cross section for 
unpolarized electrons, which is obtained from the electromagnetic 
hadronic tensor only.  Their difference, instead,
\bea
&&\left(\frac{d^2\sigma}{d\Omega d\epsilon'}\right)^{pv} =
 \frac{d^2\sigma_{+}}{d\Omega d\epsilon'} -
\frac{d^2\sigma_{-}}{d\Omega d\epsilon'}=
\nonumber\\
&&\, =-\sigma_{Mott}\frac{G_F Q^2}{2\sqrt{2}\pi\al}\left\{
v_L R^L_{AV}(q,\omega) +v_T R^T_{AV}(q,\omega)+
v_{T'} R^{T'}_{VA}(q,\omega)\right\}
\label{pvcross1}
\eea
denotes the parity--violating inclusive cross section, which is
obtained from the interference hadronic tensors $ W_{\al\be}^I(V,A)$.
It corresponds to the interference between the matrix elements for the
exchange of one photon and the one for the exchange of a $Z^0$ boson.
In Eqs.(\ref{emcross1}), (\ref{pvcross1}) 
$\sigma_{Mott}$ is the Mott cross section and
\bea
v_L&&=\left(\frac{Q^2}{\vec{q}^2}\right)^2,\qquad\quad 
v_T=\half\frac{Q^2}{\vec{q}^2}+\tan^2\half\theta
\label{vlvt}\\
v_{T'}&&=\sqrt{\frac{Q^2}{\vec{q}^2}+\tan^2\half\theta}\,\tan\half\theta
\label{vtprime}
\eea
are lepton kinematical factors.

The functions $R^{L(T)}(q,\omega)$ ($q=|\vec{q}|$) are the longitudinal
(transverse) electromagnetic nuclear response functions, which are
given by:
\bea
R^L(q,\omega)&&=W_{00}^{em},
\label{nuclRL}\\
R^T(q,\omega)&&=W_{11}^{em}+W_{22}^{em},
\label{nuclRT}
\eea
the direction of the three--momentum transfer $\vec{q}$ being assumed 
as $z$--axis. The corresponding parity--violating nuclear response 
functions are defined as:
\bea
R^L_{AV}(q,\omega)&&=g_A W_{00}^{I}(V)\,,
\label{nuclRLAV}\\
R^T_{AV}(q,\omega)&&=g_A\left[W_{11}^{I}(V)+W_{22}^{I}(V)\right]\,,
\label{nuclRTAV}\\
R^{T'}_{VA}(q,\omega)&&=ig_V W_{12}^{I}(A)\,.
\label{nuclRTVA}
\eea

By measuring the cross sections for the scattering of electrons with
both polarizations one can determine the asymmetry:
\bea
{\Acal}=&&
%\displaystyle{\frac
\left({\displaystyle{\frac{d^2\sigma_{+}}{d\Omega d\epsilon'} -
\frac{d^2\sigma_{-}}{d\Omega d\epsilon'}}}\right)/
\left({\displaystyle{\frac{d^2\sigma_{+}}{d\Omega d\epsilon'} +
\frac{d^2\sigma_{-}}{d\Omega d\epsilon'}}}\right)
\nonumber\\
&&=\Acal_0\, \frac{v_L R^L_{AV}(q,\omega) +v_T R^T_{AV}(q,\omega)+
v_{T'} R^{T'}_{VA}(q,\omega)}
{v_L R^L(q,\omega) +v_T R^T(q,\omega)}
\label{nuclasym1}
\eea
where $\Acal_0$ is defined in Eq.~(\ref{AA0}).

Within the RFG model the above defined nuclear response functions 
can be analytically evaluated. By performing the integrals 
over $\vec{p}$ one obtains:
\bea
R^{L,T}(q,\omega)&&=R_0(q,\omega)U^{L,T}(q,\omega)
\label{nuclRem1}\\
R^{L,T}_{AV}(q,\omega)&&=R_0(q,\omega) g_A U^{L,T}_{AV}(q,\omega)
\label{nuclRLTAV1}\\
R^{T'}_{VA}(q,\omega)&&=R_0(q,\omega) g_V U^{T'}_{VA}(q,\omega)
\label{nuclRTVA1}
\eea
where
\begin{equation}
R_0(q,\omega)=\frac{3{\Ncal}M^2}{2q p_F^3}\left(E_F-\Gamma\right)
\theta\left(E_F-\Gamma\right)\,.
\label{nuclR0}
\end{equation}
In Eq.~(\ref{nuclR0}) $E_F=\sqrt{p_F^2+M^2}$ is the Fermi energy and
\[ \Gamma(q,\omega)=\max \left\{(E_F-\omega),\,\, 
\half\left(q\sqrt{1+\frac{1}{\tau}}-\omega\right)\right\}\,.
\]
Two regimes exist: (i) $q<2p_F$, where $\Gamma=E_F-\omega$ and Pauli
blocking occurs; and (ii) $q\ge 2p_F$, where 
$\Gamma=\half\left(q\sqrt{1+\frac{1}{\tau}}-\omega\right)$ and the 
responses are not Pauli blocked. The remaining dependence on $q$ and 
$\omega$ in Eqs.~(\ref{nuclRem1})--(\ref{nuclRTVA1}) is contained in
the reduced responses:
\bea
U^L(q,\omega)&=&\frac{q^2}{Q^2}\left\{\GEem^2+\frac{1}{1+\tau}
\left(\GEem^2+\tau\GMem^2\right)\Delta\right\}
\nonumber\\
U^T(q,\omega)&=&2\,\tau\GMem^2+\frac{1}{1+\tau}
\left(\GEem^2+\tau\GMem^2\right)\Delta
\nonumber\\
U^L_{AV}(q,\omega)&=&2\,\frac{q^2}{Q^2}\left\{\GEem\GENC+
\frac{1}{1+\tau}\left(\GEem\GENC+
\tau\GMem\GMNC\right)\Delta\right\}
\label{redresp}\\
U^T_{AV}(q,\omega)&=&4\,\tau\GMem\GMNC+\frac{2}{1+\tau}
\left(\GEem\GENC+ \tau\GMem\GMNC\right)\Delta
\nonumber\\
U^{T'}_{VA}(q,\omega)&=&
\sqrt{\tau(1+\tau)}\,4\,\GMem\GANC(1+\Delta')\,.
\nonumber
\eea
Here the following functions of $q$, $\omega$ and $p_F$ have been 
introduced:
\bea
&&\Delta = \frac{Q^2}{q^2}\left\{
\frac{1}{3M^2}\left(E_F^2+E_F\Gamma+\Gamma^2\right)+
\frac{\omega}{2M^2}\left(E_F+\Gamma\right)+\frac{\omega^2}{4M^2}\right\}+
\nonumber\\
&&\qquad\qquad\qquad
-(1+\tau)
\label{Delta}\\
&&\Delta' =\frac{1}{2q}\sqrt{\frac{\tau}{1+\tau}}\left\{
E_F+\Gamma+\omega\right\}-1
\label{Delta1}
\eea

In most kinematical situations the quantities $\Delta$ and  $\Delta'$ are
small and their effect on the P--odd asymmetry is negligible, below the 
percent. It is interesting to notice that by setting  $\Delta=\Delta'=0$
the presence of the nuclear medium in the response functions 
(\ref{nuclRem1})--(\ref{nuclRTVA1}) is felt through the function 
$R_0(q,\omega)$ only. The latter obviously cancels in the expression 
(\ref{nuclasym1}) of the asymmetry, thus leading to the same combination of
\ffs which was obtained in the case of elastic electron--proton scattering.
This fact endures the possibility of using the measurements of quasi--elastic 
cross sections in the scattering of polarized electrons on nuclei to extract
information on the strange \ffs of the nucleon. Indeed, as discussed in 
Ref.~\cite{DMABDM}, the nuclear physics dependence of the P--odd 
asymmetry which emerges from the calculations in the RFG model is rather 
weak. 

\begin{figure}
\begin{center}
\mbox{\epsfig{file=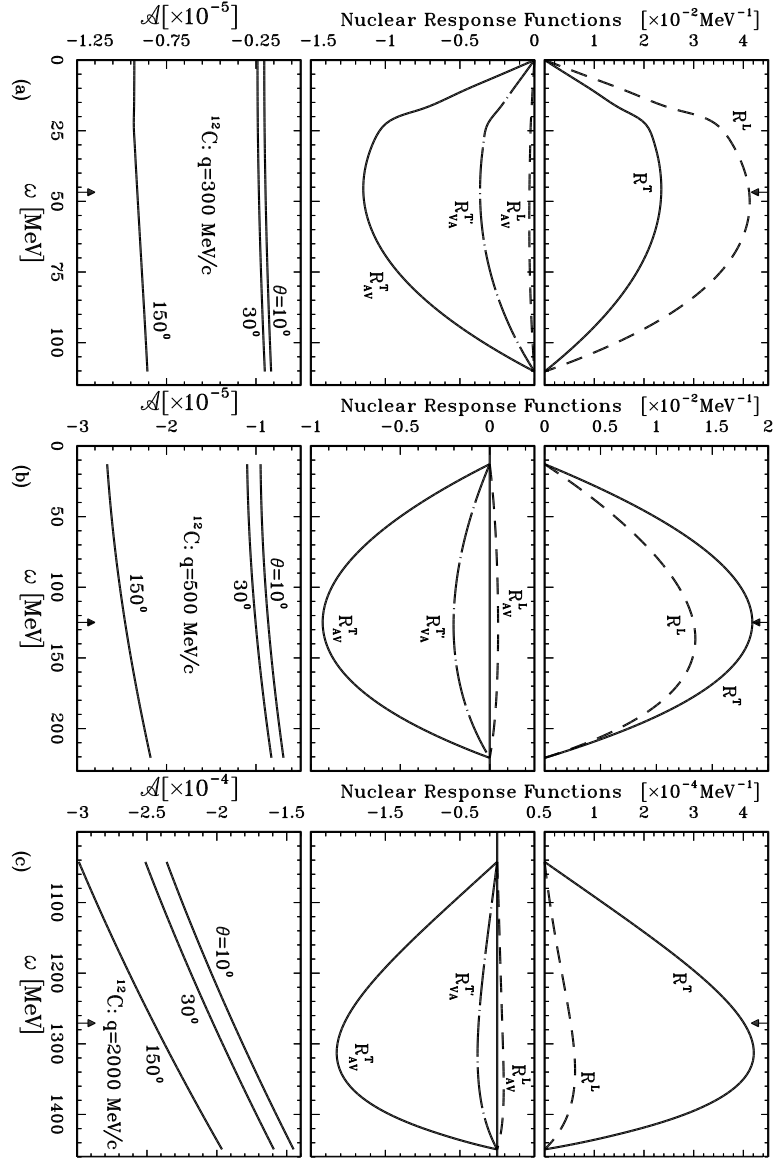,height=0.8\textheight,angle=-180}}
\end{center}
\vskip 3.7cm
\caption{Relativistic Fermi gas response functions and asymmetry for $^{12}$C
at different values of $q$, shown as a function of $\omega$. The location
of $\omega=Q^2/(2M)$ is indicated by the arrows. Further explanations in the 
text (Taken from Ref.~\cite{DMABDM}).
}
\label{fig.pvrfg.1}
\end{figure}

Typical results for $^{12}$C are shown in Fig.~\ref{fig.pvrfg.1}. The Fermi
momentum is taken to be $p_F=225$~MeV and the strange \ffs are set to zero.
In the upper panels the two electromagnetic response functions ($R^L$, $R^T$)
are displayed as a function of energy transfer $\omega$, for three 
typical 3--momentum transfers, $q=0.3,\, 0.5$ and $2$~GeV. The intermediate 
panels show the corresponding parity violating responses $R^L_{AV}$, $R^T_{AV}$
and $R^{T'}_{VA}$. Quasi--elastic scattering allows one 
to explore these quantities
(as well as the P--odd asymmetry) in different kinematical ranges. The $Q^2$
values corresponding to the peak of the responses ($\omega=Q^2/2M$) are 
0.09, 0.24 and 2.4 GeV$^2$, respectively, but for each case a range of 
different $Q^2$ is explored (for example $1.9\le Q^2\le 2.9$~GeV$^2$ in the
right panels), thus showing one of the advantages of this investigation.
Once these response functions are multiplied by the (angle dependent) lepton
kinematical factors (\ref{vlvt}), (\ref{vtprime}) and combined as in 
Eq.~(\ref{nuclasym1}), one obtains the asymmetry shown in the lower panels
of Fig.~\ref{fig.pvrfg.1}. It ranges from a few $\times 10^{-6}$ at forward
angle and low $q$ to a few  $\times 10^{-4}$ for a broad range of angles 
at $q=2$~GeV.

%*************************************************************************
%
%\vskip -1.5cm
%
%*************************************************************************

Further, one can examine the sensitivity of the P--odd asymmetry to the
nucleon strange and non--strange \ffs. At backward scattering angles
$v_L/v_T\to 0$, $v_{T'}/v_T\to 1$ and the terms containing magnetic and
axial \ffs dominate (in spite of the $g_V$ factor which penalizes the
interference with the axial nuclear current). In Ref.~\cite{DMABDM} the
electromagnetic \ffs of the proton were parameterized with the usual
dipole form, with a cutoff mass $M_V=843$~MeV; for the neutron the
following \ffs were used:
\bea
\GMem^n(Q^2)&&=\frac{\rho_{M_n}\mu_n}
{\displaystyle{\left(1+\frac{Q^2}{M_V^2}\right)^2}},\qquad
\label{neutronMff}\\
\GEem^n(Q^2)&&= \frac{-\mu_n\tau}
{\displaystyle{\left(1+\frac{Q^2}{M_V^2}\right)^2(1+\lambda_n\tau)}}\,,
\label{neutronEff}
\eea
where the Galster~\cite{Galster,Platch90} parameterization for 
$\GEem^n$ was assumed,
with $\mu_n=-1.1913$ and $\lambda_n=5.6$.
The standard value of the parameter $\rho_{M_n}$ is unity; it accounts for
possible deviations as in Eq.~(\ref{ffratio}). The axial isovector
\ff was parameterized as
\begin{equation}
\GACC(Q^2)=\frac{g_A}
{\displaystyle{\left(1+\frac{Q^2}{M_A^2}\right)^2}}\,,
\label{GA1}
\end{equation}
with $M_A= 1$~GeV. 
For the strange \ffs the following parameterization was adopted:
\bea
\GEs(Q^2)&=&\frac{\rho_s\tau}
{\displaystyle{\left(1+\frac{Q^2}{M_V^2}\right)^2(1+\lambda_E^s\tau)}}\,,
\label{GEsDonn}\\
\GMs(Q^2)&=&\frac{\mu_s}
{\displaystyle{\left(1+\frac{Q^2}{M_V^2}\right)^2(1+\lambda_M^s\tau)}}\,,
\label{GMsDonn}\\
\GAs(Q^2)&=&\frac{g_A^s\tau}
{\displaystyle{\left(1+\frac{Q^2}{M_A^2}\right)^2(1+\lambda_A^s\tau)}}\,,
\label{GAsDonn}
\eea
where the second factor in the denominators accounts for possible 
deviations of the high--$\tau$ dipole fall--off. In Ref.~\cite{DMABDM}
the values $\lambda_E^s=\lambda_n$, $\lambda_M^s=\lambda_A^s=0$ were 
 used. 

\begin{figure}
\begin{center}
\mbox{\epsfig{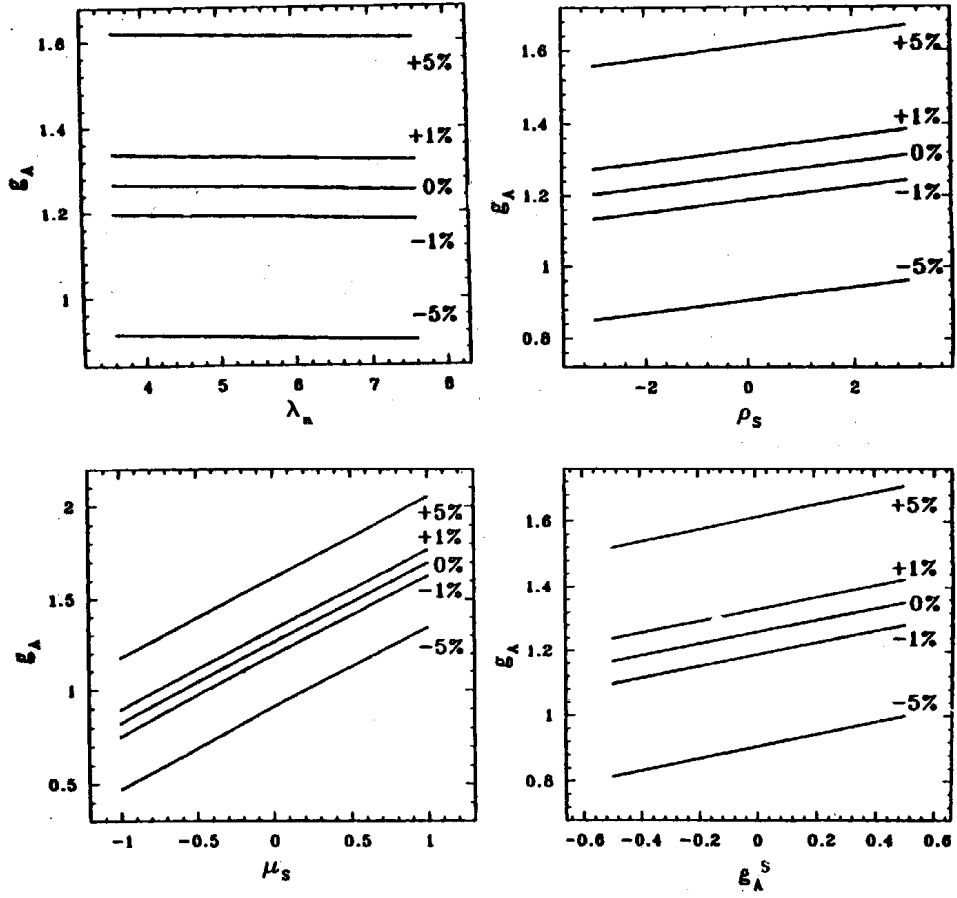}}
\end{center}
%\vskip -0.8cm
\caption{Correlation plots for  $^{12}$C at $q=500$~MeV, 
$\omega=Q^2/(2M)$ and $\theta=150^{\circ}$. The asymmetry is constant 
for any pairs of parameters corresponding to the line marked $0\%$. For lines
marked $\pm 1\%$ ($\pm 5\%$) the asymmetry remains constant at values $1\%$ 
($5\%$) larger or smaller than that obtained with the canonical choice of
parameters.  The panels show correlations of $g_A$ 
 with $\lambda_n$, $\rho_s$, $\mu_s$ and $g_A^s$.  
Further explanations in the text 
(Taken from Ref.~\cite{DMABDM}).
}
\label{fig.pvrfg-3}
\end{figure}

The correlations between different parameters used in modeling the \ffs 
were examined by looking at the dependence of one parameter from a 
second one (all the remaining ones being fixed) keeping the 
P--odd asymmetry  constant. In Fig.~\ref{fig.pvrfg-3} a few examples of 
these correlation plots are shown for $^{12}$C at $q=500$~MeV, 
$\omega=Q^2/(2M)$ and $\theta=150^{\circ}$: 
the lines marked $0\%$ correspond to 
a (constant) value of $\Acal$ which is obtained starting with the 
``standard'' values (e.g. $g_A=1.26$, $\rho_s=0$ in the upper right panel).
Lines marked $+1\%$ ($-1\%$) have asymmetries
$1.01\Acal$ ($0.99\Acal$), with a corresponding meaning for the lines 
marked $\pm5\%$. From these curves one observes, for example, that at this
particular choice of kinematics, a $\pm 1\%$ determination of the asymmetry
would permit a $\pm 5.6\%$ determination of $g_A$ if everything else 
were known (we refer the reader to the above discussion on the uncertainties
on this parameter due, e.g., to radiative corrections). A $\pm 5\%$ 
determination of $\Acal$ likewise would translate into a $\pm 28\%$ 
uncertainty in $g_A$. In fact, there are uncertainties in the other
parameters which enter into the problem and in the nuclear model itself. 
%*************************************************************************
%
%*************************************************************************

The parameter $\lambda_n$ characterizes the high--$\tau$ fall--off of 
the electric form factor of the neutron, $\GEem^n$: we see from the left 
upper panel
of Fig.~\ref{fig.pvrfg-3} that, if the latter will be 
determined in future experiments to $\pm 10\%$,  
this only translates into a $\pm 0.6\%$ 
uncertainty in $g_A$. Obviously this relatively minor effect is due 
the backward kinematics, which suppresses the longitudinal contributions.
The left lower panel of Fig.~\ref{fig.pvrfg-3} shows the correlation between
 $g_A$ and the strength of $\GMs$, $\mu_s$: if this parameter goes 
from 0 to $-1$, (from 0 to 1) then  $g_A$ would decrease (increase) 
by $34.7\%$. This 
correlation is rather important: for example if  $g_A$ were known to
$\pm 10\%$, then $\mu_s$ would be constrained to $\pm 0.29$. Finally the
right lower panel shows a non--negligible correlation between $g_A$ and
$g_A^s$.

One should also notice, here, the potentialities offered by the use of 
different nuclear targets. Indeed the measurement of $\Acal$ in inelastic 
electron--nucleus scattering can give information not only on the strange 
nucleon form factors, but also on the non--strange parts of the weak neutral
\ffs of the nucleon.
This can be achieved by filtering the latter with a suitable
choice of $Z$ and $N$, such that, for example, it enhances the isovector
contributions to the nuclear response and eventually cancels the isoscalar
ones. 

\begin{figure}
\begin{center}
\mbox{\epsfig{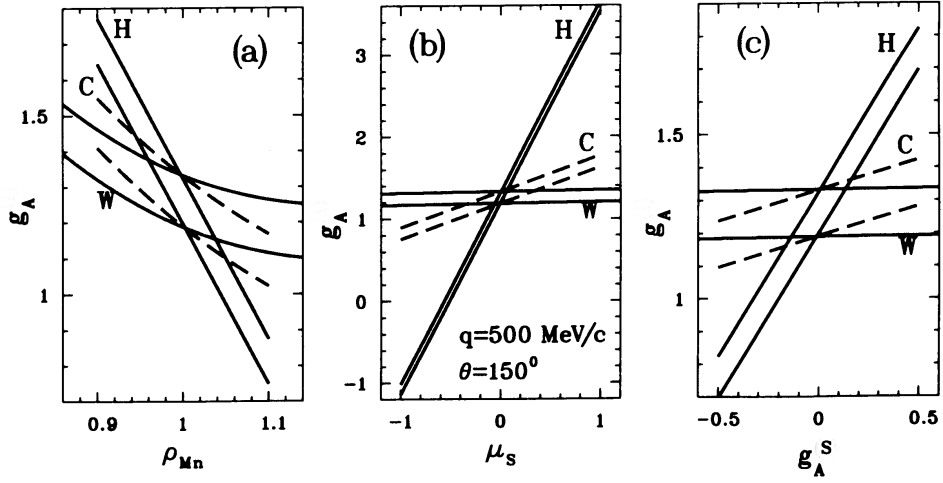}}
\end{center}
%\vskip -0.8cm
\caption{Correlation plots as in Fig.~\ref{fig.pvrfg-3}, showing  $g_A$
%($g_A^{(1)}$ in the figure) 
correlated with (a) $\rho_{M_n}$, (b) $\mu_s$ and (c) $g_A^s$, for
%($g_A^{(s)}$ in the figure), for
$q=500$~MeV and $\theta=150^{\circ}$. Three cases are shown: 
elastic scattering from hydrogen (labelled H) and quasi--elastic scattering 
from carbon (C) and tungsten (W). Only the $\pm 1\%$ contour lines are shown.
For tungsten two different Fermi momenta are used: $p_F^p=250$~MeV for 
protons and $p_F^n=285$~MeV for neutrons.  (Taken from Ref.~\cite{DMABDM}).
}
\label{fig.pvrfg-5}
\end{figure}

As an illustration, let us consider the product $\GMem\GMNC$ in $U^T_{AV}$
or $\GMem\GANC$ in $U^{T'}_{VA}$, in Eqs.~(\ref{redresp}): in a $(Z,N)$ 
nucleus both $\GMs$ and $\GAs$ enter $\Acal$ multiplied by the combination
$Z\GMem^p+N\GMem^n$, whereas the non--strange contributions (e.g. the 
isovector part of $\GANC$) enter with $Z\GMem^p-N\GMem^n$. Hence the ratio
of the strange to the non--strange transverse pieces is 
$(Z\GMem^p+N\GMem^n)/(Z\GMem^p-N\GMem^n)$. This ratio is 1 for the proton, 
0.187 for a $Z=N$ nucleus and can be further reduced to be nearly zero 
by choosing a nucleus whose $N/Z$ ratio is very close to $-\mu_p/\mu_n$.
The case of tungsten ($^{184}$W), having advantages for high luminosity 
experiments, gives a ratio of $-0.009$. In Fig.~\ref{fig.pvrfg-5} 
correlation plots are shown for $q=500$~MeV and $\theta=150^{\circ}$; the 
results for $^{12}$C and $^{184}$W are obtained by integrating over the 
entire quasi--elastic response region, although there is no significative 
difference between the peak asymmetry and the integrated result.
In (a) a significative interplay appears between $g_A$ and $\rho_{M_n}$,
more pronounced in H than in C and W, but not negligible even in the last case.
In (b) and (c) instead, it is clearly shown that no correlation exists in W
between $g_A$ and the strangeness parameters $\mu_s$ and $g_A^s$:
thus an $N>Z$ nucleus such as tungsten appears to have advantages for the
determination of $\GACC$ in backward--angle scattering.

%
%*************************************************************************
%*************************************************************************

A special case with $N=Z$ is  the deuteron, which we have already 
mentioned in Section \ref{sec.Poddelex}, discussing the SAMPLE experiment 
at MIT/Bates. In the kinematic region around the QE peak the P-odd asymmetry
for the deuteron can be evaluated in the so--called ``static'' approximation,
in which the contributions to the cross sections of 
protons and neutrons at rest
are summed incoherently~\cite{Hadji}:
\bea
&&\left( 
\frac{\mathcal{A}}{\mathcal{A}_0}
\right)_d
=
2\,\left\{  
g_A 
\left[
v_L\frac{q^2}{2 Q^2}
\left(
\GEem^p G_{E}^{NC;p} +\GEem^n G_{E}^{NC;n}
\right)
\right.\right. +
\nonumber\\
&& \left.
+ v_T \; \tau
\left( 
\GMem^p G_{M}^{NC;p} +\GMem^n G_{M}^{NC;n}
\right)
\right] +
\nonumber \\
&& +  g_V
\left. \left[
v_T \sqrt{\tau (1+ \tau)} 
\left( 
\GMem^p G_{A}^{NC;p} +\GMem^n G_{A}^{NC;n}
\right)\right]
\right\}\times
\nonumber \\
&& \times 
\left[ 
v_L\frac{q^2}{2 Q^2}
\left(
\left(\GEem^p\right)^2 +\left(\GEem^n\right)^2
\right)
+ v_T \left[ \tau
\left( 
(\GMem^p)^2 +(\GMem^n)^2
\right)
\right] \right]^{-1}\;.
\label{eq.deuteron}
\eea
This formula can be obtained from the RFG , setting the terms
$\Delta$ and $\Delta'$ in Eqs.~(\ref{redresp}) 
to zero and taking $N=Z=1$.
The dependence of the P-odd asymmetry on the deuteron structure
was studied in Ref.~\cite{Hadji}, under different conditions. 
The authors concluded that in the kinematical region
around the QE peak deviations from the static model are within 
1 or 2 \%. A more recent study of the deuteron structure effects
in $A_d$ for the specific kinematic
conditions of the SAMPLE experiment can be found in \cite{Diaconescu}. 
To get a flavor of the general sensitivity of $\mathcal{A}_d$
to the single nucleon form factors one can consider 
the transverse contributions in
Eq.~(\ref{eq.deuteron}), which are dominant at large scattering angles, 
as in the SAMPLE
experiment. As already stated for the general case $N=Z$, 
the strange form factors $G_M^s$ and $G_A^s$ enter the transverse
contributions to the asymmetry multiplied by the combination
$\GMem^p + \GMem^n$ and are suppressed by
a factor $\simeq 0.187$ with respect to the contribution
 coming from the isovector axial form factor $G_A$, 
which is multiplied by $\GMem^p - \GMem^n$.

\begin{figure}
\begin{center}
%\mbox{\epsfig{file=fig11c.eps2,width=\textwidth}}
\mbox{\epsfig{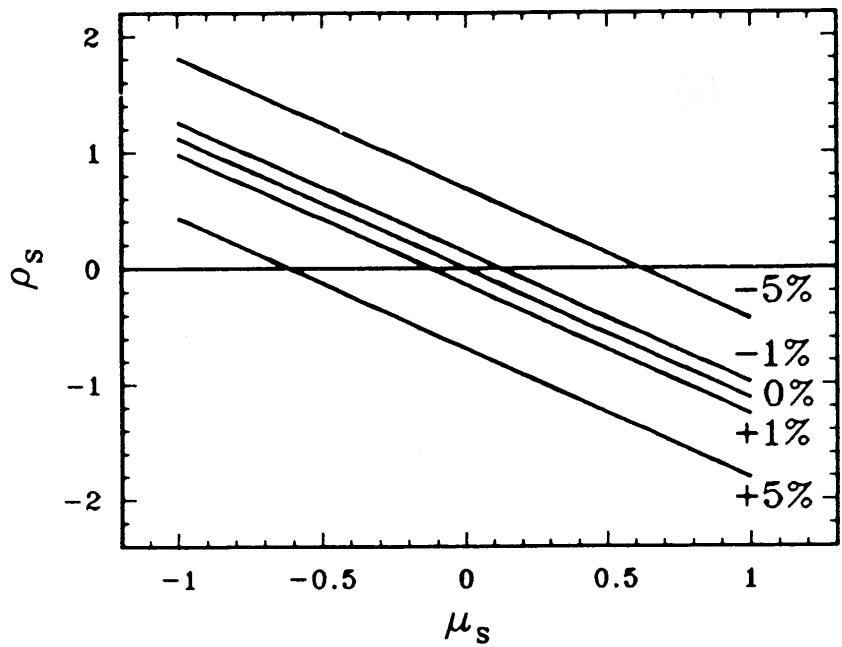}}
\end{center}
%\vskip -0.8cm
\caption{Correlation plots as in Fig.~\ref{fig.pvrfg-3}, showing  $\rho_s$
correlated with  $\mu_s$, for $q=500$~MeV, $\omega=Q^2/(2M)$ and 
$\theta=10^{\circ}$. (Taken from Ref.~\cite{DMABDM}).
}
\label{fig.pvrfg-6}
\end{figure}

As a final issue in this Section let us consider forward--angle scattering:
at small $\theta$ and fixed $q$, $\omega$ it is $v_L/v_T\to 2Q^2/q^2$ and
$v_{T'}/v_T\to 0$, so that the contribution of the response 
$R_{T'}(q,\omega)$ is suppressed. In this situation considerable 
sensitivity of the asymmetry to the electric strange \ff can be achieved.
In Fig.~\ref{fig.pvrfg-6} the correlation plot of $\rho_s$ versus $\mu_s$
is shown for $^{12}$C at $q=500$~MeV and $\theta=10^{\circ}$.

%
%*************************************************************************
%*************************************************************************

From all the above consideration, quasi--elastic scattering of polarized 
electrons on nuclei can be considered, with appropriate choices of the
kinematical conditions, as a useful tool to determine the 
strange \ffs of the nucleon; it can also provide important information 
on the non--strange components of various nucleonic \ffs, which are 
still waiting for a precise determination. Although it goes outside the
scopes of the present review, we also mention that the measurement of
parity--violating nuclear response functions would open new and 
interesting possibilities to explore the nuclear dynamics as viewed by 
weak neutral probes and to test nuclear models in the domain of 
medium--high excitation energy~\cite{Barbaro94,Barbaro96}.

%\vspace{1cm}
%%%%%%%%%%%%%%%%%%%%%%%%%%%%%%%%%%%%%%%%%%%%%%%%%%%%%%%%%%%%%%%%%%%%%%%
\section{Elastic NC scattering of neutrinos (antineutrinos) on the
nucleon}
\label{sec.nufree}

Direct information on the strange form factors of the nucleon can be
obtained from the investigation of the NC processes~\cite{Kap,Ellis}
\begin{equation}
\nu_{\mu} (\nubar_{\mu}) + N \longrightarrow
\nu_{\mu} (\nubar_{\mu}) + N
\label{nuelas}
\end{equation}
The amplitude for the process of neutrino (antineutrino) scattering is
given by the following expression
\begin{equation}
\la f|S|i\ra = \mp \frac{G_F}{\sqrt{2}}
\ubar(k')\gam^{\al}\left(1\mp\gam_5\right)u(k)\la p'|J^{NC}_{\al}|p\ra
(2\pi)^8\delta^{(4)}(p'-p-q)
\label{nuampl}
\end{equation}
where $k$ and $k'$ are the momenta of the initial and final neutrino
(antineutrino), $p$ and $p'$ the momenta of the initial and final nucleon,
$q=k-k'$ and
\[J_{\al}^{NC}= V_{\al}^{NC} -A_{\al}^{NC} \]
is the hadronic neutral current in the Heisenberg representation.

The matrix elements of the vector and axial NC 
are given  in the Standard Model by the expressions (\ref{vecNClast}) and
(\ref{axNClast}). The cross sections of the processes (\ref{nuelas}) turn
 out to be
\begin{equation}
d\sigma_{\nu(\nubar)}=\frac{G_F^2}{(2\pi)^2}\frac{M}{p\cdot k}
\left\{ L^{\al\be}(k,k')\mp L_5^{\al\be}(k,k')\right\}
W^{NC}_{\al\be}(p,q)\frac{d\vec{k'}}{k'_0}\,,
\label{nucross}
\end{equation}
where
\begin{equation}
W_{\al\be}^{NC}(p,q) = \frac{1}{2M}\sum\int\la p'|J^{NC}_{\al}|p\ra
\la p|J^{NC}_{\be}|p'\ra\delta^{(4)}(p'-p-q)\frac{d\vec{p'}}{2 p'_0}
\label{nutensor1}
\end{equation}
while the tensor $L^{\al\be}(k,k')$ and pseudotensor $L^{\al\be}_5(k,k')$
are given by (\ref{lepten}) and (\ref{lepten5}), respectively.

It is evident that $W_{\al\be}^{NC}$ has the following structure
\begin{equation}
W^{NC}_{\al\be} = W_{\al\be}^{VV+AA} - W_{\al\be}^{VA}
\label{nutensor2}
\end{equation}
where, with obvious notation, the tensor$W_{\al\be}^{VV+AA}$  is due to
the contribution of the vector--vector and axial--axial NC,
whereas the pseudotensor $W_{\al\be}^{VA}$ is due to the
interference of the vector and axial NC.

After performing the traces over spin states,
they become ($\tau=Q^2/4M^2$, $n=p+p'$):
\bea
&&W_{\al\be}^{VV+AA}= \left\{ -\left[\tau\left(\GMNC\right)^2
+\left(1+\tau\right)\left(\GANC\right)^2\right]\left(g_{\al\be} -
\frac{q_{\al}q_{\be}}{q^2}\right) \right. +
\nonumber\\
&&\qquad\quad
 +\left[\frac{\left(\GENC\right)^2 +\tau\left(\GMNC\right)^2}
{1+\tau} + \left(\GANC\right)^2\right]\frac{n_{\al}n_{\be}}{4M^2}
\nonumber\\
&&\qquad\qquad \left.
-\frac{q_{\al}q_{\be}}{q^2}\left(\GANC\right)^2\right\}
\delta\left(\nu-\frac{Q^2}{2M}\right)
\label{nutensor3}
\eea
and
\begin{equation}
W^{VA}_{\al\be}=
\frac{i}{M^2}\epsilon_{\al\be\rho\sigma}p^{\rho}{p'}^{\sigma}
\GMNC\GANC \delta\left(\nu-\frac{Q^2}{2M}\right) \,,
\label{nutensor5}
\end{equation}
respectively.

Taking into account that
\begin{equation}
\frac{d\vec{k'}}{{k'}_0} =\pi\frac{M}{p\cdot k}\,dQ^2d\nu\,,
\end{equation}
from Eqs.~(\ref{nucross}), (\ref{nutensor3}) and (\ref{nutensor5})
we obtain, correspondingly, the following expressions for the differential
cross sections of the processes (\ref{nuelas}):
\bea
&&\left(\frac{d\sigma}{dQ^2}\right)^{NC}_{\nu(\nubar)} = 
\nonumber\\
&&\qquad =
\frac{G_F^2}{2\pi}\left[\half y^2(\GMNC)^2 +\left(1-y-\frac{M}{2E}y\right)
\frac{\displaystyle{(\GENC)^2+\frac{E}{2M}y(\GMNC)^2}}
{\displaystyle{1+\frac{E}{2M}y}}\right.
\nonumber\\
&&\qquad\left.
+\left(\half y^2+1-y+\frac{M}{2E}y\right)(\GANC)^2
\pm 2y\left(1-\half y\right)\GMNC\GANC\right]\,.
\label{nucross2}
\eea
Here
\begin{equation}
y=\frac{p\cdot q}{p\cdot k} =\frac{Q^2}{2p\cdot k}
\label{ydef}
\end{equation}
and $E$ is the energy of neutrino (antineutrino) 
in the laboratory system.

In order to obtain information on the strange form factors of the
nucleon from the investigation of the processes (\ref{nuelas}) it is
necessary to know the axial CC form factor $\GACC$
[see relation (\ref{gaNC})]. The latter can be determined by
investigating the quasi--elastic processes
\begin{equation}
\begin{array}{l}
\nu_{\mu}+n\longrightarrow \mu^-+p\,,\\
\nubar_{\mu}+p\longrightarrow \mu^+ +n\,.
\end{array}
\label{nuquasiel}
\end{equation}
The amplitudes 
of these processes are given, respectively, by the expressions
\bea
\la f|S|i\ra &=& -i\frac{G_F}{\sqrt{2}}\ubar(k')\gam^{\al}(1-\gam_5)u(k)
\la p'|J_{\al}^{CC}|p\ra (2\pi)^4\delta^{(4)}(p'-p-q)
\label{nunmup}\\
\la f|S|i\ra &=& i\frac{G_F}{\sqrt{2}}\ubar(k')\gam^{\al}(1+\gam_5)u(k)
\la p'|{J_{\al}^{CC}}^\dagger|p\ra (2\pi)^4\delta^{(4)}(p'-p-q)
\label{nupmun}
\eea
where $k$ and $k'$ are the momenta of the initial $\nu_{\mu}$
($\bar\nu_{\mu}$) and final $\mu^-$ ($\mu^+$) lepton, $p$ is the
momentum of the initial $n$ ($p$) and $p'$ the momentum of the final
$p$ ($n$).
The Heisenberg vector and axial charged currents are the 
``plus''--components of the isovectors $V_{\al}^i$ and $A_{\al}^i$
[see Eq.~(\ref{jccHei})].

In Section \ref{sec.matrixNC} we have  considered the one--nucleon matrix
elements of the axial current $A_{\al}^{1+i2}$. Let us discuss now
the matrix element of the vector current $V_{\al}^{1+i2}$.
Due to isotopic invariance of the strong interactions the vector current
$V_{\al}^i$ is conserved (Conserved Vector Current, CVC)~\cite{CVCbook}:
\[\partial^{\al}V_{\al}^i =0.\]
Thus the matrix element of the vector current satisfies the condition
\begin{equation}
\left(p'-p\right)^{\al}\,{_p\la}p'|V^{1+i2}_{\al}|p\ra_n =0
\end{equation}
and has the following general form
\begin{equation}
{_p\la}p'|V^{1+i2}_{\al}|p\ra_n = \ubar(p')\left[\gam_{\al} F_1^{CC}(Q^2)
+\frac{i}{2M}\sigma_{\al\be}q^{\be} F_2^{CC}(Q^2)\right] u(p)
\label{matelplus}
\end{equation}
where $F_{1,2}^{CC}(Q^2)$ are CC form factors. The corresponding Sachs CC
form factors are given by
\bea
\GMCC(Q^2)&=& F_1^{CC}(Q^2) +F_2^{CC}(Q^2)
\label{gmcc}\\
\GECC(Q^2)&=& F_1^{CC}(Q^2) -\frac{Q^2}{4M^2}F_2^{CC}(Q^2)
\label{gecc}
\eea

An important property of the isovector current $V_{\al}^i$ is
given by its commutation relation with the isospin operator
\begin{equation}
\left[ I_k, V^j_{\al}\right] = i\epsilon_{kj\ell}V_{\al}^{\ell}
\label{isocommv}
\end{equation}
where $I_k$ is the total isotopic spin operator.
From Eq.~(\ref{isocommv}) it follows that
\begin{equation}
V_{\al}^{1+i2} = \left[ V_{\al}^3, I_{1+i2}\right]\,.
\label{isocommv2}
\end{equation}

Taking into account the charge symmetry of strong
interactions, from (\ref{isocommv2}) the following relations hold:
\[
{_p\la}p'|V^{1+i2}_{\al}|p\ra_n = {_n\la}p'|V^{1-i2}_{\al}|p\ra_p 
={_p\la}p'|J^{em}_{\al}|p\ra_p - {_n\la}p'|J^{em}_{\al}|p\ra_n\,.
\]
Let us notice that in the derivation of these relations we have used 
the expression (\ref{jemq}) for the e.m. current. 
Thus the  CC vector form factors are connected with the
electromagnetic \ffs of proton and neutron by:
\bea
\GMCC(Q^2) &=& \GMem^p(Q^2) - \GMem^n(Q^2)
\label{gmcc1}\\
\GECC(Q^2) &=& \GEem^p(Q^2) - \GEem^n(Q^2)
\label{gecc1}
\eea

The cross sections of the processes (\ref{nuquasiel}) are
given by the expression (\ref{nucross}) in which $W_{\al\be}^{NC}$
have to be replaced by 
\begin{equation}
W_{\al\be}^{CC}(p,q) = \frac{1}{2M}\sum\int\la p'|J^{CC}_{\al}|p\ra
\la p|{J^{CC}_{\be}}^{\dagger}|p'\ra
\delta^{(4)}(p'-p-q)\frac{d\vec{p'}}{2 p'_0}
\label{nutensorc1}
\end{equation}

It is obvious that in order to obtain the cross sections of the
quasi--elastic processes (\ref{nuquasiel})
it is necessary to replace the NC \ffs in the expression
(\ref{nucross2}) by the CC ones (we are neglecting the muon mass).
One gets:
\bea
&&\left(\frac{d\sigma}{dQ^2}\right)^{CC}_{\nu(\nubar)} =
\label{nucrossc3}\\
&&\qquad=
\frac{G_F^2}{2\pi}\left[\half y^2(\GMCC)^2 +\left(1-y-\frac{M}{2E}y\right)
\frac{\displaystyle{(\GECC)^2+\frac{E}{2M}y(\GMCC)^2}}
{\displaystyle{1+\frac{E}{2M}y}}\right.+
\nonumber\\
&&\qquad\qquad\left.
+\left(\half y^2+1-y+\frac{M}{2E}y\right)(\GACC)^2
\pm 2y\left(1-\half y\right)\GMCC\GACC\right]\,.
\nonumber
\eea

The most detailed study of the elastic NC scattering of neutrinos
(antineutrinos) on protons was done in the experiment 734 at BNL in
1987~\cite{Ahrens87}.
A 170~ton  high resolution liquid--scintillator target--detector
was used in this experiment. The liquid--scintillator cells were
segmented by proportional drift tubes. About $79 \%$ of the target
protons were bound in Carbon and Aluminum nuclei and about $21 \%$
were free protons.

The neutrino beam was a horn--focused wide band beam. 
The average energy of
neutrinos was 1.3~GeV and the average energy of antineutrinos was 1.2~GeV.
The spectrum of neutrinos and antineutrinos was determined from the
detection of quasi--elastic $\nu_{\mu}+n\to\mu^-+p$ and
$\nubar_{\mu}+p\to \mu^++n$ events.

The angle between the momenta of the recoil protons and 
of the incident neutrinos as well as the range and energy loss
were measured. The measurement of the range and energy loss provided an
effective particle identification and the determination of the kinetic
energy of the recoil protons.

The background from the neutrons entering into the detector was
eliminated by restricting the fiducial volume down to about $19 \%$ of
the total volume of the detector. After all cuts, 951 neutrino events
and 776 antineutrino events were selected.

\begin{figure}
\begin{center}
\mbox{\epsfig{file=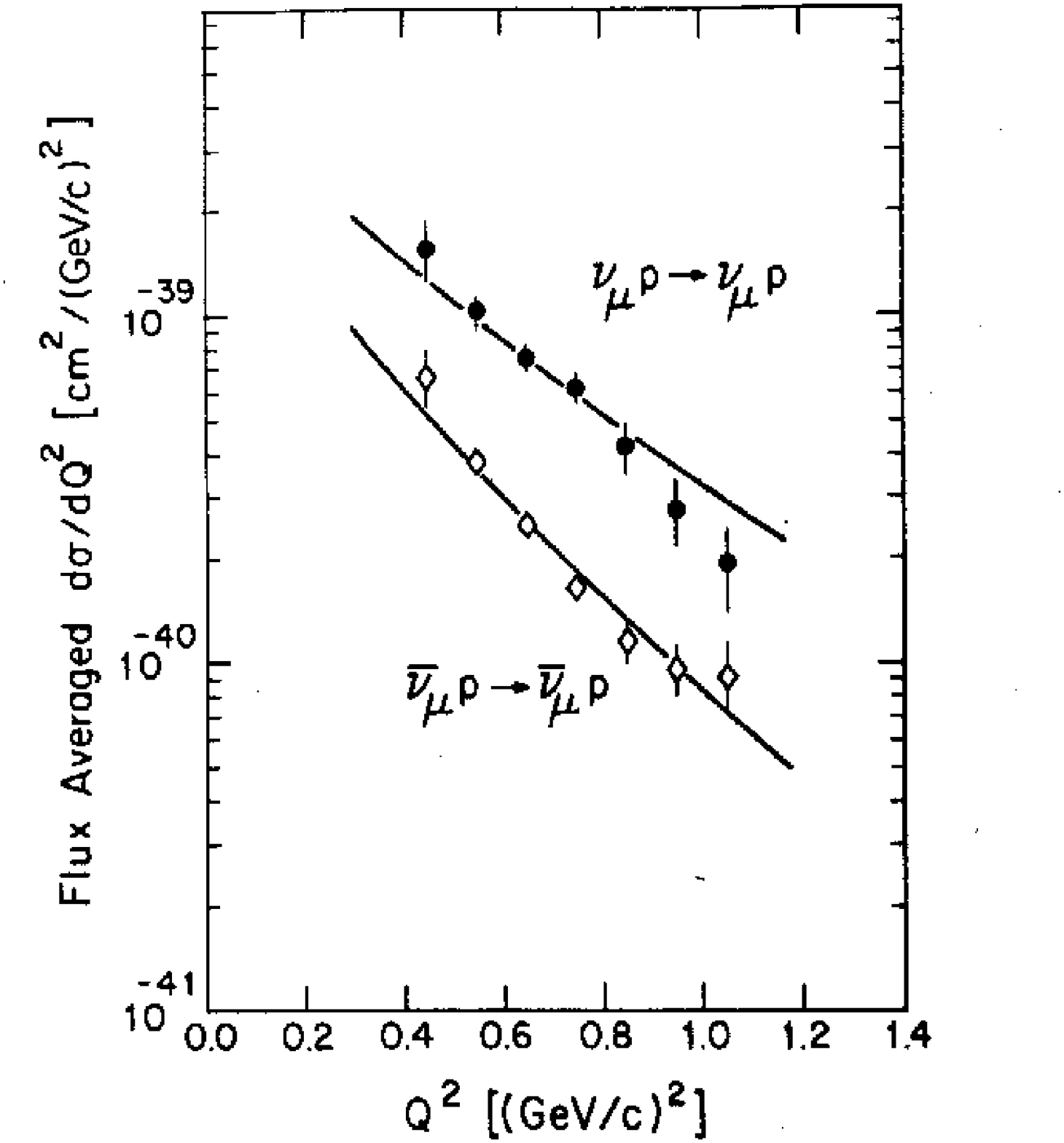,width=0.9\textwidth}}
\end{center}
%\vspace 0.8cm
\caption{The data points are the measured flux--averaged 
differential cross sections for $\nu_{\mu}p\to\nu_{\mu}p$ and 
$\nubar_{\mu}p\to\nubar_{\mu}p$ measured in the experiment of 
Ref.~\cite{Ahrens87}. The solid curves are best fits to the combined 
data with the values $M_A=1.06$~GeV and $\sin^2\theta_W=0.220$.
The error bars represent statistical error and also include 
$Q^2$--dependent systematic errors.
(Taken from  Ref.~\cite{Ahrens87})}
\label{Fig4}
\end{figure}

The differential cross sections
\begin{equation}
\langle\frac{ d\sigma}{d Q^2}
\rangle^{NC}_{\nu (\nubar)} =
\frac{ \displaystyle{
\int d 
E_{\nu (\overline{\nu})} 
\left(d\sigma/dQ^2\right)^{NC}_{\nu (\nubar)} 
\Phi_{\nu(\overline{\nu})} 
\left( 
E_{\nu (\overline{\nu})} 
\right)
} }
{ \displaystyle{
\int d E_{\nu(\overline{\nu})}
\Phi_{\nu(\overline{\nu})} 
\left( 
E_{\nu (\overline{\nu})} 
\right)
} }\;,
\label{eq.crossfold}
\end{equation}
obtained by folding the cross sections (\ref{nucross2}) with the
BNL neutrino and antineutrino spectra
$\Phi_{\nu(\overline{\nu})} \left( E_{\nu (\overline{\nu})} \right)$,
were determined from the data of the experiment~\cite{Ahrens87}.
Their values  are presented in Fig.~\ref{Fig4}
by points. For the ratios of (flux averaged) total elastic
and quasi--elastic cross sections, for  $Q^2$ in the interval 
 $0.5\le Q^2\le 1$~GeV$^2$, the following values were obtained:
\bea
R^{BNL}_{\nu}&=& 
\frac{\langle\sigma(\nu_{\mu}p\to\nu_{\mu}p)\rangle}
{\langle\sigma(\nu_{\mu}n\to\mu^- p)\rangle}=0.153\pm 0.007\, (\mathrm{stat})\,
\pm 0.017\,(\mathrm{syst})
\label{Rnu}\\
R^{BNL}_{\nubar}&=& \frac{\langle\sigma(\nubar_{\mu}p\to\nubar_{\mu}p)
\rangle}
{\langle\sigma(\nubar_{\mu}p\to\mu^+ n)\rangle}=0.218\pm 0.012\,
 (\mathrm{stat})\,
\pm 0.023\,(\mathrm{syst})
\label{Rnubar}
\\
R^{BNL}&=& \frac{\langle\sigma(\nubar_{\mu}p\to\nubar_{\mu}p)\rangle}
{\langle\sigma(\nu_{\mu}p\to\nu_{\mu}p)
\rangle} =0.302\pm 0.019\, (\mathrm{stat})\,
\pm 0.037\,(\mathrm{syst})\,.
\label{RBNL}
\eea
%

%*************************************************************************
%*************************************************************************

The fit of the data presented in Ref.~\cite{Ahrens87} was done
under the assumption that the contribution of the strange form factors 
of the nucleon can be neglected and that the axial CC form factor is 
given by the dipole formula
\begin{equation}
\GANC(Q^2)=\half \GACC(Q^2)= \half\frac{\GACC(0)}{\displaystyle{
\left(1+ \frac{Q^2}{M_A^2}\right)^2}}
\end{equation}
with $\GACC(0)=1.26$. The parameters $M_A$ and $\sin^2\theta_W$ were 
considered as free parameters. From the simultaneous fit of the 
neutrino and antineutrino data the following values
\bea
&&\sin^2\theta_W = 0.218^{+0.039}_{-0.047}
\nonumber\\
&&M_A = 1.06\pm 0.05\, \mathrm{GeV}
\nonumber
\eea
were found (with  $\chi^2 = 15.8$ at 14 DOF).
The value of the axial cutoff  $M_A$ was in a good agreement with the 
existing (at that time) world--average value
\begin{equation}
M_A= 1.032\pm 0.036\,  \mathrm{GeV}
\label{waMA}
\end{equation}
which was found from the data of the experiments on quasi--elastic 
neutrino and antineutrino scattering.
The solid curves in Fig.~\ref{Fig4} were obtained with the above 
best--fit values of the parameters.

\begin{figure}
\begin{center}
%\mbox{\epsfig{file=BNL2.eps2,width=0.75\textwidth,height=0.4\textheight}}
\mbox{\epsfig{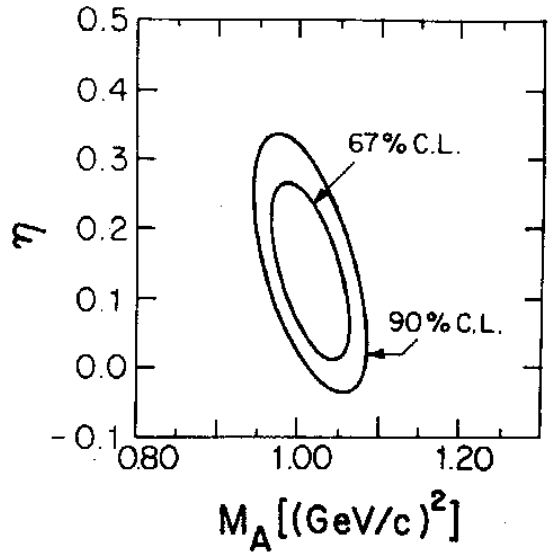}}
%\vskip 2cm
\end{center}
%\vspace{0.8cm}
\caption{Simultaneous fit of $d\sigma/dQ^2$ for the neutrino and 
antineutrino elastic scattering samples in the ($M_A,\eta$)--plane
($\eta\equiv -\GAs(0)$), with $\sin^2\theta_W$ fixed at $0.220$. In the
fit $M_A$ has been constrained at the world--average value 
$M_A=1.032\pm 0.036$~GeV. (Taken from  Ref~\cite{Ahrens87}).
}
\label{Fig5}
\end{figure}

In Ref.~\cite{Ahrens87} it was also reported the result of the fit of
the data on NC elastic neutrino (antineutrino)--proton scattering 
under the assumption that the contribution of the strange
vector form factors can be neglected and the axial strange form factor 
has the same  $Q^2$ dependence as the CC axial form factor
\begin{equation}
\GANC(Q^2)= \half \frac{\left[\GACC(0)-\GAs(0)\right]}
{\left(1+Q^2/M_A^2\right)^2}\,.
\end{equation}
For the parameter $\sin^2\theta_W$  the value $0.22$ was taken.
The parameter $M_A$ was constrained to the world--averaged value 
(\ref{waMA}).
From this fit it was found
\begin{equation}
\GAs(0)= -0.12\pm 0.07\,.
\end{equation}
Thus, from the results of the fit we described above, it follows that
$-0.25\le \GAs(0)\le 0$ at $90 \%$ CL.
It is necessary to stress, however, that there is a strong correlation
between the values of the parameters $\GAs(0)$ and  $M_A$ 
(see Fig.~\ref{Fig5})

%*************************************************************************
%*************************************************************************

The data of the BNL experiment  were re--analyzed
in Ref.~\cite{Garvey93a}. In this work not only strange {\em axial} 
but also strange {\em vector} form factors, $\FVs(Q^2)$ and $\FMs(Q^2)$,
were taken into account. It was assumed that all non--strange 
form factors have the
same dipole $Q^2$--dependence, with $M_{V} = 0.843$~GeV. 
The parameter $M_{A}$ was considered as a free parameter. 
It was also assumed that the electric form factor of the neutron 
is given by Eq.~(\ref{neutronEff}).
These authors made several fits of the BNL data under different 
assumptions on the values of the parameters that characterize the strange
form factors. For the latter the following parameterizations were chosen:
\bea
\FMs(Q^2)&&=\frac{\mu_s}{(1+\tau)
\left(\displaystyle{1+\frac{Q^2}{M_V^2}}\right)^2}\, ,
\qquad
\FVs(Q^2)=\frac{\FVs Q^2}{(1+\tau)
\left(\displaystyle{1+\frac{Q^2}{M_V^2}}\right)^2}\, ,
\nonumber\\
\GAs(Q^2)&&=\frac{\GAs(0)}
{\left(\displaystyle{1+\frac{Q^2}{M_A^2}}\right)^2}\,,
\label{Garveyff}
\eea
which have the same (dominant) $Q^2$--dependence as the non--strange
form factors.\footnote{We notice that some authors, both in the study of
PV electron scattering and of neutrino scattering, have assumed 
parameterizations similar to the ones of the non--strange 
form factors directly for
the Sachs form factors $G_{E,M}^s$. The parameterizations used here
for $F_{1,2}^s$ correspond to:
\bea
G_E^s(Q^2) &&= 
\frac{\left(4 M^2 F_1^s - \mu_s \right) \tau}{1 + \tau}
\frac{1}{\left(
1+\frac{Q^2}{M_V^2}\right)^2}
\nonumber \\
G_M^s(Q^2) &&= 
\frac{\left(4 M^2 F_1^s \tau + \mu_s \right)}{1 + \tau}
\frac{1}{\left(1+\frac{Q^2}{M_V^2}\right)^2}
\nonumber 
\eea
}
For the value of the parameter $\sin^2\theta_W$ the world average 
 value  $\sin^2\theta_W=0.2325$ was taken.

If we neglect the contribution of all strange form factors 
and keep as the only variable parameter $M_{A}$, then an acceptable fit
to the data can be found with 
\[ M_A=1.086\pm 0.015\,{\mathrm{GeV}} \]
($\chi^2 = 14.12 $ at 14 DOF).
This value of $M_{A}$ is in a good agreement with the world--average value
\[ M_A=1.061\pm 0.026\,{\mathrm{GeV}}\,. \]

If we neglect the contribution of the vector strange form factors  only, 
the best fit to the data is obtained with:
\[ \GAs(0)= -0.15\pm 0.07;\qquad
 M_A=1.049\pm 0.019\,{\mathrm{GeV}} \]
($\chi^2 = 9.73 $ at 13 DOF).

Hence there is a strong correlation between the values of parameters
 $M_{A}$ and $\GAs(0)$. This correlation is connected with the
fact that both negative values of $\GAs(0)$
and larger values of $M_{A}$  increase the cross
sections of neutrino-- and antineutrino--proton scattering.

Finally, if we consider all strange form factors 
and  $M_{A}$ to be variable parameters, then from 
the fit of the BNL data we get
\bea
\mu_s&=-0.39\pm 0.70,\qquad
\FVs &= 0.49\pm 0.70\,{\mathrm{GeV}}^{-2}
\nonumber\\
\GAs(0)&= -0.13\pm 0.09,\qquad
 M_A&=1.049\pm 0.023\,{\mathrm{GeV}}
\nonumber
\eea
($\chi^2 = 9.28 $ at 11 DOF).

The authors of Ref.~\cite{Garvey93a} concluded that from the data of the BNL
experiment it is not possible to make firm conclusions on the values of
the strange form factors of the nucleon. 
The result of the fit strongly depends on the value of the parameter
 $M_{A}$ which determines the $Q^2$ behavior of the axial CC form factor.
Satisfactory fits were obtained
for values of the strange axial form factor $\GAs(0)$ in the range
from 0 to $-0.15\pm 0.07$,
depending on the value of  $M_{A}$. There are no doubts that new
experiments on the measurement of NC elastic neutrino (antineutrino)
proton scattering are necessary.

Information on the strange axial constant of the nucleon
$\GAs(0)\equiv g_{A}^s$ can be obtained from the measurement of the ratio
$R$ of the total cross sections for the production
of protons and neutrons in quasi--elastic scattering of neutrinos 
(antineutrinos) on nuclei with isotopic spin equal to zero.
 
A detailed calculation was done for the nucleus $^{12}$C 
and for the neutrino beam of the Los Alamos Meson Physics Facility (LAMPF) 
(with  neutrino energies less than 200 MeV)~\cite{Garvey92,Garvey93b}. 
The nuclear structure
effects were taken into account in the framework of the random phase
approximation; the final state interaction of the ejected nucleon was
included through a finite--range force derived from the Bonn potential.
Although this issue goes somewhat beyond the subject of this Section,
we present it here since the original suggestion was founded on the 
hypothesis that nuclear structure and/or final state interaction effects
appreciably cancel in the ratio of quasi--elastic $\nu (\nubar)$--nucleus
cross sections.

In order to illustrate the sensitivity of the method proposed 
in~\cite{Garvey92,Garvey93b}, let us notice that the main contribution 
to the cross sections of quasi--elastic neutrino scattering comes 
from the axial NC form factor:
\begin{equation}
\sigma^{p,n}_{\nu}\propto \left| \GANC\right|^2
\simeq \frac{1}{4}{\GACC}^2\left(1\mp 2\frac{\GAs}{\GACC}\right)
\end{equation}
For the ratio $R$ we have then\footnote{We address the reader 
to a possible source of confusion
existing in the literature: this approximate expression for the ratio $R$
suggests a ``reference'' value, without strangeness effects, $R=1$.
In fact, even if the axial NC form factor contribution is dominant,
also the vector NC form factors (which are different for protons and
neutrons also in the absence of strange form factors) contribute to the
neutrino scattering cross section, producing a deviation from 1 of
the reference value. The size of this deviation depends on the
kinematical conditions and on the specific kinematical cuts applied
in calculating the cross sections.}:
\begin{equation}
R=\frac{\sigma^p_{\nu}}{\sigma^n_{\nu}}\simeq
1-4\frac{\GAs}{\GACC}\simeq 1-\frac{16}{5}\GAs\,.
\end{equation}
Thus, in the ratio of the cross sections the effect of the strange
form factor is more than doubled.
An important advantage of measuring the ratio $R$ is connected with
the fact that this quantity is not affected by the absolute 
normalization of the cross sections.

A measurement of the above ratio is planned at the LSND detector at
Los Alamos. The detector is made of mineral oil (CH$_2$) and reveals
NC neutrino induced knockout reactions by measuring the energy deposition
of the recoiling nucleons, with an energy resolution of 5\% for nucleon
kinetic energies $T_N > 50$~MeV.
The available neutrino beam is produced by pion in flight decay and
is composed of about 80\% neutrinos and 20\% antineutrinos. 
The yields of protons and neutrons can be measured as a function of
the recoiling nucleon kinetic energy, with an average over the nucleon
angle, as the detector has no angular resolution.
The total cross sections are then obtained
by integrating the differential yields over the nucleon kinetic energy.
While all neutron events come obviously from quasi--elastic scattering,
the proton events can be produced both in free scattering on protons
and in QE scattering on $^{12}$C, therefore a kinematical cut has to be
applied in order to exclude free proton contributions, when forming
the proton over neutron ratio. The maximum kinetic energy that can be
transferred by the neutrinos available at Los Alamos to a free proton,
at rest in the laboratory frame, is about 60~MeV; therefore the lower
limit $T_N> 60$~MeV is adopted in calculating the cross sections.

In Ref.~\cite{Garvey92} it was shown that in the range of neutrinos energies
of LAMPF, the ratio $R$ practically does not depend on the neutrino
energy and consequently on the uncertainties in the neutrino spectrum.
In particular the ratio calculated averaging the cross sections over
the expected neutrino spectrum was found to be practically the same as
the one obtained for fixed $E_{\nu} = 200$~MeV, a value which has been
used in further studies (see Section \ref{sec.nuQE}).

It was also shown that the ratio of the cross sections for the scattering
of neutrinos on free protons and neutrons  differs
from the ratio of the cross sections of the quasi--elastic knockout
of protons and neutrons from $^{12}$C, calculated within the random
phase approximation, by no more than $10 \%$.
\footnote{More specifically for the case of free nucleons
the following limits on the outgoing nucleon kinetic energy
were assumed in the calculation of the total
cross sections: $50 \le T_N \le 59.7$ MeV. 
With this choice the neutrino p/n ratio was found to be:
\[
R = \frac{0.66 -0.84 G_A^s(0) + 0.25 \left(G_A^s(0)\right)^2 
-0.254 F_2^s(0) + 0.2G_A^s(0)F_2^s(0)}
{0.75 + 0.92 G_A^s(0) + 0.25 \left(G_A^s(0)\right)^2 
+0.254 F_2^s(0) + 0.2G_A^s(0)F_2^s(0)}\,.
\]
}
This fact was presented
as a  reason for using the free expression as a qualitative guidance
in understanding the role of the different contributions to the ratio.
This small difference can also be taken as a first indication that the
ratio $R$  weakly depends on nuclear effects.
This argument will be further developed in Section \ref{sec.nuQE}.

Beside the axial form factor $G_A^s$ the effects of the strange form
factor $F_2^s$ were also studied, while $F_1^s$ was not considered since
its effects are expected to be small at the energies of interest for the
Los Alamos experiment. 

Since the Los Alamos beam is in fact a mixture of neutrinos and
antineutrinos, the experimentally measured quantity will be obtained
as the ratio of a ``weighted average'' of both types of cross sections,
namely
\begin{equation}
\overline{R}_{LAMPF} =
\frac{\langle\sigma_{\nu}^p \rangle +
  \langle\sigma_{\overline{\nu}}^p \rangle}
{\langle\sigma_{\nu}^n \rangle +  \langle\sigma_{\overline{\nu}}^n
\rangle}  
\label{rbar}
\end{equation}
$\langle\sigma\rangle = \int \Phi(E)\, \sigma(E) \,dE$
being the cross section averaged over the neutrino or antineutrino
flux $\Phi_{\nu(\nubar)}(E)$.

\begin{figure}
\begin{center}
\mbox{\epsfig{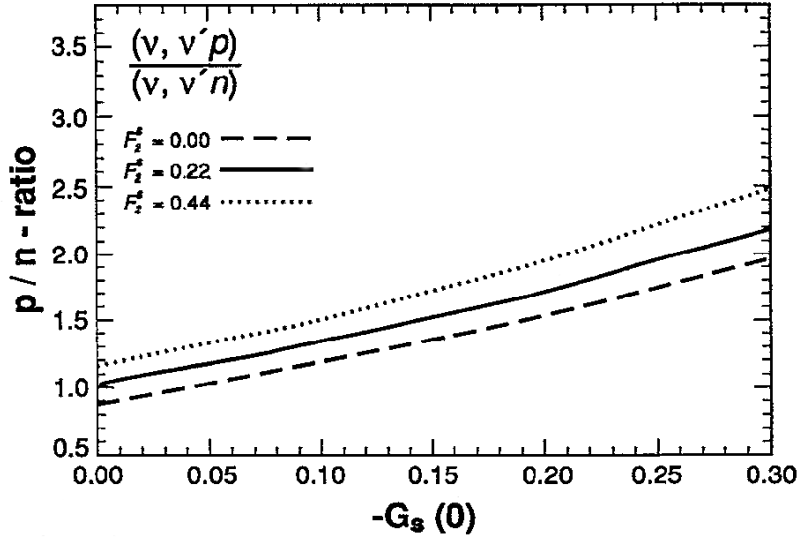}}
\end{center}
\caption{Ratio of the integrated proton-- to neutron yield for the 
quasi--elastic neutrino induced reactions on $^{12}$C as a function of
$-\GAs(0)\equiv -G_s(0)$ in the conditions of the LAMPF decay--in--flight
neutrino beam. The various curves correspond to different values of 
$\FMs(0)\equiv\mu_s$. (Taken from  Ref.~\cite{Garvey93b})
}
\label{Fig6}
\end{figure}

The effects of $F_2^s$ are of opposite sign for neutrinos and
antineutrinos, hence the sensitivity of the experimental ratio 
to $F_2^s$ can be reduced in the measured quantity (\ref{rbar}).
Due to the different rate of change of the separate
$\nu$ and $\overline{\nu}$ cross sections 
with the strange axial form factor, the effects of $g_A^s$ on the
experimental ratio can also be altered.
In Ref.~\cite{Garvey93b} it was found that assuming a spectrum
made of 80\% neutrinos and 20\% antineutrinos the sensitivity
of (\ref{rbar}) to $g_A^s$ was increased with respect to the
pure neutrino ratio, while the sensitivity to $F_2^s$ was decreased.
The corresponding results are presented in  Fig.~\ref{Fig6}, where
 the dependence of the ratio $\overline{R}$ on
the axial strange constant $-\GAs(0)$ is shown.

Notice, however, that
a subsequent study~\cite{Kolbe97}, which used an updated spectrum for the
antineutrinos, showed that the sensitivity of $\overline{R}$
to both $g_A^s$ and $F_2^s$
is close to the one of the pure neutrino ratio.
An analysis of the experiment on the measurement of the ratio 
$R$ is going on at LAMPF~\cite{Louis73}. It is planned that in this 
experiment the ratio $R$ should be measured with an accuracy of $10 \%$.
If  $-G_{A}(0)< -0.2 $  an effect of  $G_{A}(0)$ will be seen.

%
%*************************************************************************
%*************************************************************************

%%%%%%%%%%%%%%%%%%%%%%%%%%%%%%%%%%%%%%%%%%%%%%%%%%%%%%%%%%%%%%%%%%%%%%%
\section{ Neutrino--antineutrino asymmetry in elastic neutrino
(anti\-neu\-tri\-no)--nucleon scattering.}
\label{sec.freeasymm}

As we have seen before, the precise measurement of the cross sections of
the NC neutrino (antineutrino) scattering on nucleons allows one 
to obtain direct information on  the strange
form factors of the nucleon {\em if} the electromagnetic form factors of
proton and neutron and the CC axial form factor are known.
At present there exist detailed information on the electromagnetic
form factors of proton and neutron, which was obtained from the data of 
numerous experiments on elastic electron--proton and
electron--deuteron scattering~\cite{Bosted95,Jager99}. 
New information on the electromagnetic form factors of the 
proton was recently obtained by the
Jefferson Lab Hall A Collaboration~\cite{Jones00}, which
 measured the polarization of recoil protons in
the scattering of polarized electrons on unpolarized protons.
The axial form factor of the proton is rather poorly known and, as it
was pointed out in the previous Section,
 the main problems in extracting information on the strange 
form factors from the existing neutrino (antineutrino)--proton data are 
connected with the uncertainty on the axial form factor.

In Ref.~\cite{ABGM} a method  was proposed, which
allows one to obtain information on the strange form factors of the 
nucleon in a model independent way.

Let us consider the NC processes:
\bea
\nu_{\mu} + N &&\longrightarrow \nu_{\mu}+N
\label{numu}\\
\nubar_{\mu} + N &&\longrightarrow \nubar_{\mu}+N\,.
\label{nubarmu}
\eea
The difference of the cross sections of the processes (\ref{numu}) and
(\ref{nubarmu}) are given by [see Eq.~(\ref{nucross2})]:
\begin{equation}
\left(\frac{d\sigma}{dQ^2}\right)^{NC}_{\nu} -
\left(\frac{d\sigma}{dQ^2}\right)^{NC}_{\nubar} =
\frac{2G_F^2}{\pi} y\left(1-\half y\right)\GMNC\GANC \,,
\label{nucrossdif}
\end{equation}
$y$ being defined as in Eq.~(\ref{ydef}).
Hence, the above difference of the neutrino and antineutrino cross sections
is determined only by the magnetic and axial NC form factors.
The axial and magnetic NC form factors of the nucleon are given by
Eqs.~(\ref{gaNC}) and (\ref{gmNC}), respectively. The latter can be
rewritten in the form:
\begin{equation}
{G_M}^{NC;p(n)}= \pm\GMem^3 -2\sin^2\theta_W\GMem^{p(n)} -\half\GMs
\label{gmNC2}
\end{equation}
where $\GMem^3=\half(\GMem^p-\GMem^n)$ is the isovector form factor
of the nucleon.

Let us consider now the CC processes
\bea
\nu_{\mu}+n&&\longrightarrow \mu^-+p\,,
\label{numun}\\
\nubar_{\mu}+p&&\longrightarrow \mu^+ +n\,.
\label{nubarmup}
\eea
From Eq.~(\ref{nucrossc3}) it follows that the difference
of the cross sections of the reactions (\ref{numun}) and
(\ref{nubarmup}) is given by
\begin{equation}
\left(\frac{d\sigma}{dQ^2}\right)^{CC}_{\nu} -
\left(\frac{d\sigma}{dQ^2}\right)^{CC}_{\nubar} =
\frac{4G_F^2}{\pi} y\left(1-\half y\right)\GMem^3\GACC\, .
\label{cccrossdif}
\end{equation}
Thus, this difference is determined by the isovector electromagnetic
form factor and the CC axial form factor, which are the $u-d$ part of
the magnetic and axial NC form factors of the nucleon.

Let us now define the following neutrino--antineutrino asymmetry:
\begin{equation}
\Acal(Q^2)= \frac{\displaystyle{\left(\frac{d\sigma}{dQ^2}\right)^{NC}_{\nu} -
\left(\frac{d\sigma}{dQ^2}\right)^{NC}_{\nubar}}}
{\displaystyle{\left(\frac{d\sigma}{dQ^2}\right)^{CC}_{\nu} -
\left(\frac{d\sigma}{dQ^2}\right)^{CC}_{\nubar}}}\,.
\label{nuasymm}
\end{equation}
From Eqs.~(\ref{nucrossdif}) and (\ref{cccrossdif}), we have\footnote{
More precisely, in Eq.~(\ref{nuasymm2}) enters $\Acal |V_{ud}|^2$; 
however, we do not write explicitly the CKM matrix element since we 
neglect the small deviation of $|V_{ud}|$ from unity.}
\begin{equation}
\Acal_{p(n)}=\frac{1}{4}\left(\pm 1-\frac{\GAs}{\GACC}\right)
\left(\pm 1-2\sin^2\theta_W\frac{{\GMem}^{p(n)}}{\GMem^3} 
-\half\frac{\GMs}{\GMem^3}\right)\,.
\label{nuasymm2}
\end{equation}
Thus, in the asymmetry $\Acal$ the strange axial and vector form 
factors enter in the form of the ratios ${\GAs}/{\GACC}$ and
${\GMs}/{\GMem^3}$.
Taking into account only terms which depend linearly on the strange
form factors we can rewrite Eq.~(\ref{nuasymm2}) in the following form:
\begin{equation}
\Acal_{p(n)}=\Acal_{p(n)}^0 \mp\frac{1}{8}\frac{\GMs}{\GMem^3} 
\mp \frac{\GAs}{\GACC}\Acal_{p(n)}^0
\label{nuasymm3}
\end{equation}
where
\begin{equation}
\Acal_{p(n)}^0= \frac{1}{4}\left(1\mp 
2\sin^2\theta_W\frac{{\GMem}^{p(n)}}{\GMem^3} \right)
\label{nuasymm0}
\end{equation}
is the expected asymmetry in the case that all strange form factors 
are equal to zero.
Thus, should it turn out that the measured asymmetry is different
from $\Acal^0$, it would be a model independent proof that the
strange form factors are not negligible 
with respect to the non--strange ones.

Usually in neutrino experiments nuclear targets are used. 
Accordingly we can average  the asymmetry (\ref{nuasymm}) 
over the protons and neutrons; then we obtain (assuming, e.g., 
an isospin symmetrical nucleus) the following expression:
\begin{equation}
\la\Acal\ra = \la\Acal^0\ra +\half\sin^2\theta_W
\frac{\GMem^0}{\GMem^3}\frac{\GAs}{\GACC}\,.
\label{aveasymm}
\end{equation}
Here
\[ \la\Acal^0\ra = \frac{1}{4}\left(1 -
2\sin^2\theta_W \right) 
\]
and ${\GMem^0}=(\GMem^p+\GMem^n)/2$ is the isoscalar magnetic form
factor of the nucleon. In the expression for the averaged asymmetry
$\la\Acal\ra$ only the axial strange form factor enters: in fact
the interference between the (isoscalar) strange vector form factor
 and the isovector axial form factor vanishes after
 averaging over $p$ and $n$.

We notice that the electromagnetic \ffs of the nucleon enter into the
expression for the asymmetry $\Acal_{p(n)}(Q^2)$ in the form of the 
ratio $\GMem^{p(n)}/\GMem^3$. It is well known that the
electromagnetic form factors satisfy the approximate scaling relation
\begin{equation}
\frac{\GMem^p(Q^2)}{\GMem^n(Q^2)} =\frac{\mu_p}{\mu_n}
\label{scaling}
\end{equation}
where $\mu_p$ and  $\mu_n$
are the total magnetic moments of proton and neutron.
Using the values
\[
\mu_p=2.79,\qquad \mu_n=-1.91,\qquad \sin^2\theta_W=0.23,
\]
we obtain the following expressions for the asymmetries in the 
scaling approximation 
\bea
\Acal_p&=& 0.12 -0.12\,\frac{\GAs}{\GACC} -0.13\,\frac{\GMs}{\GMem^3}
\nonumber\\
\Acal_n&=& 0.16 +0.16\,\frac{\GAs}{\GACC} +0.13\,\frac{\GMs}{\GMem^3}
\nonumber\\
\la\Acal\ra&=& 0.14 +0.02\,\frac{\GAs}{\GACC}\,. 
\nonumber
\eea
Thus, the asymmetries $\Acal_p$ and $\Acal_n$ are rather sensitive
to both axial and magnetic strange form factors.
The contribution of the axial strange form factor to the averaged
asymmetry $\la\Acal\ra$, instead,  is suppressed due to the smallness 
of the isoscalar magnetic moment of the nucleon with respect to the 
isovector one.

The neutrino--antineutrino asymmetry depends on the ratio of the magnetic
form factors of the proton and neutron.
There exists a rather detailed information on the  magnetic
form factor of the proton in a wide range of $Q^2$ up to 
$30$~GeV$^2$~\cite{Andiva94,Arnold86,Bosted90,Kirk73,Krupa84,Bartel73}.
The experimental data show that the
behavior of the form factor at high $Q^2$ can not be described by
the standard dipole formula.

The magnetic form factor of the neutron is less known than the one of
the proton. A large part of the information on the neutron form factors 
has been obtained from experiments on the measurement of quasi--elastic
scattering of electrons on deuterium and other nuclei~\cite{Lung93,Rock82}.
In order to extract from these data the electromagnetic form factors of the
neutron it is necessary to take into account the neutron binding, 
final state interaction, contribution of meson exchange currents and
other effects which rely on theoretical models.

In order to calculate the expected neutrino--antineutrino asymmetry
with an error band connected with the uncertainties of the electromagnetic
form factors, in Ref.~\cite{ABGM} a fit of the data on the magnetic
form factors of proton and neutron was done.
The range $0.5~ \mathrm{GeV}^{2} \le Q^{2}\le 10~ \mathrm{GeV}^{2}$
was considered and the following two--poles formula was adopted for
the magnetic \ff of the proton:
\begin{equation}
\frac{\GMem^p}{\mu_p}=\frac{a_1}{1+a_2 Q^2} +\frac{1-a_1}{1+a_3 Q^2}\,.
\label{ourffp}
\end{equation}
It generalizes the dipole formula and was previously proposed in
Ref.~\cite{Bilenkaya80}. For the magnetic form factor of the
neutron the expression
\begin{equation}
\frac{\GMem^n}{\mu_n} = \frac{\GMem^p}{\mu_p}\left(1+a_4 Q^2\right)\,,
\label{ourffn}
\end{equation}
was taken, 
where the parameter $a_4$ takes into account the deviation from the
scaling relation (\ref{scaling}). From the fit of the data the
following values for the parameters were found:
\begin{equation}
\begin{array}{l}
a_1=-0.50\pm 0.04,\\
a_2=0.71\pm 0.02,\\
a_3=2.20\pm 0.04,\\
a_4=-0.019\pm 0.004\,.
\end{array}
\label{ourpar}
\end{equation}
In the calculation of the expected asymmetry the parameterizations proposed
in Ref.~\cite{Bosted95,Watanabe95} were also used. 

\begin{figure}
%\vspace{6cm}
\mbox{\epsfig{file=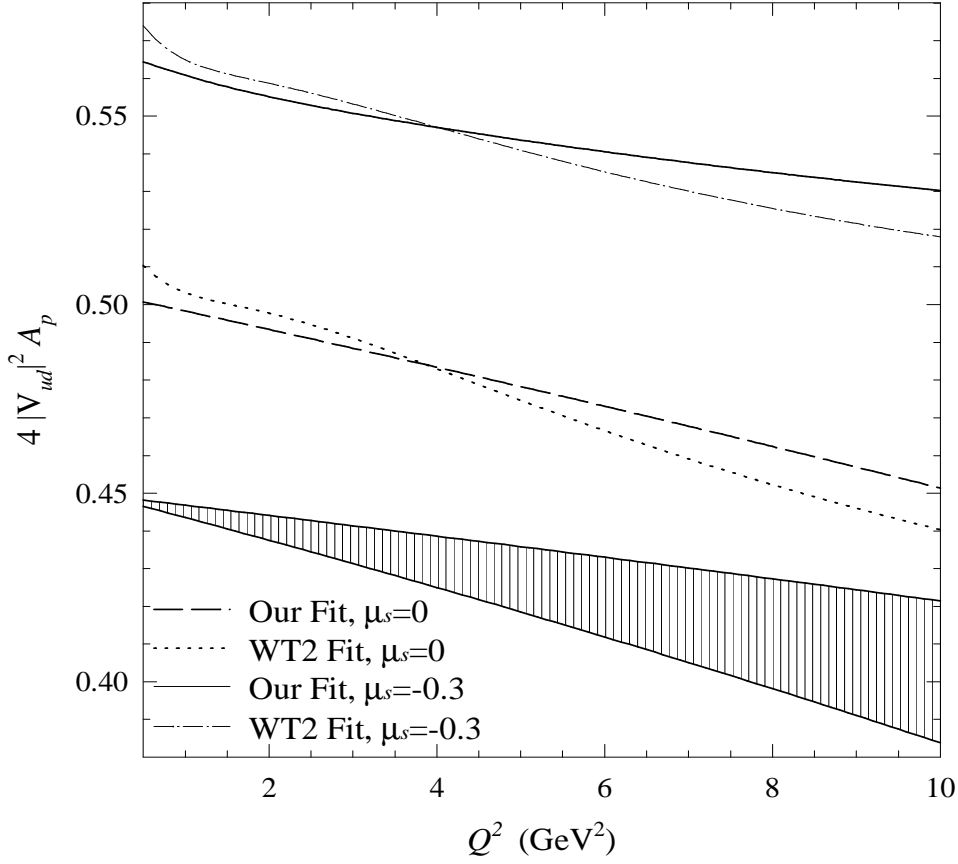,width=0.9\textwidth,height=0.6\textheight}}
\vspace{0.8cm}
\caption{Plot of
$4|V_{ud}|^2{\Acal}_p$ as a function of $Q^2$.
The shadowed area corresponds to the uncertainty induced by the 
errors in the magnetic \ffs in the absence of strange contributions.
All the curves were obtained using a dipole form for
$ F_A^s(Q^2) $ with $g_A^s=-0.15$.
The dashed (dotted) curve was obtained with $ G_M^S(Q^2) = 0 $
utilizing the fit of Eqs.~(\ref{ourffp}) and (\ref{ourffn}) 
(respectively, the WT2 fit~\cite{Watanabe95})
for the magnetic form factors of the nucleon.
The solid (dot--dashed) line was obtained using a dipole form for
$ G_M^S(Q^2)$ with $\mu_s=-0.3$ and
utilizing the fit of Eqs.~(\ref{ourffp}) and (\ref{ourffn}) 
(respectively, the WT2 fit)
for the magnetic form factors of the nucleon.
(Taken from  Ref.~\cite{ABGM})
}
\label{Fig7}
\end{figure}

The  $Q^2$--behavior of the strange form factors is unknown.
Different para\-me\-te\-ri\-za\-tions of the strange form factors were 
thus considered in Ref.~\cite{ABGM}. For the sake of 
illustration in Fig.~\ref{Fig7} the result of the calculation of the
asymmetry in the elastic neutrino (antineutrino)--proton scattering 
is shown.

%*************************************************************************
%*************************************************************************

Here it was assumed that the strange form factors are given by the
dipole formulas
\begin{equation}
\GAs(Q^2)=\frac{g_A^s}{\displaystyle{\left(1+\frac{Q^2}{{M_A^s}^2}
\right)}}\,,
\qquad
\GMs(Q^2)=\frac{\mu_s}{\displaystyle{\left(1+\frac{Q^2}{{M_V^s}^2}
\right)}}\,,
\label{dipstrangff}
\end{equation}
in which the following values were used for $M_A^s$ and $M_V^s$:
\begin{equation}
M_A^s=M_A=1.032~\mathrm{GeV}, M_V^s=M_V=0.84~\mathrm{GeV}\,.
\label{MAMV}
\end{equation}

The dashed area was obtained under the assumption that $g_A^s=\mu_s=0$:
it corresponds to the error band associated (to a confidence level of
$90\%$) to the uncertainties (\ref{ourpar}) in the parameterization
of the magnetic proton and neutron \ffs.
The dashed and dotted curves display the effect of the axial strange
form factor (with $g_A^s=-0.15$) and correspond to different 
parameterizations (\cite{ABGM} and \cite{Watanabe95}, respectively)
of the electromagnetic form factors. These curves were obtained
under the assumption that $\mu_s=0$.
The solid and dot--dashed curves demonstrate the effect of both strange
axial and vector form factors. They were obtained
with $g_A^s=-0.15$ and $\mu_s=-0.3$.
It is worth noticing that the slow variation of the asymmetry with 
$Q^2$ is due to the deviation of the magnetic \ffs from the scaling 
relation (\ref{scaling}). 

Thus the measurement of neutrino--antineutrino
asymmetry could allow one to resolve the contribution of the strange
form factors. Let us notice that the combined effect of the axial 
and vector strange form factors depends on the relative sign of 
$g_A^s$ and $\mu_s$. 
If the sign are the same (as it is assumed in Fig.~\ref{Fig7}) the 
contribution of both form factors to the asymmetry sum up. 
Should the signs be opposite (which could be the case if 
$\mu_s>0$~\cite{Spayde00})
the contributions to the asymmetry of the vector form factor 
would tend to compensate the effect of the axial one.

We have assumed in Eq.~(\ref{dipstrangff}) that strange and non--strange
form factors have the same $Q^2$ behavior. 
According to the  asymptotic quark counting 
rule~\cite{Brodsky76a,Brodsky76b}, it would be 
natural to expect that the strange \ffs decrease with $Q^2$ more 
rapidly than the non--strange ones.
In this case the contribution of the strange form factors to the
asymmetry will disappear at high $Q^2$. Calculations which consider this
situation were made in Ref.~\cite{ABGM} and show that
the region $Q^2\simeq 1\div 2$~GeV is probably the optimal one to look 
for the effects of strangeness in neutrino--nucleon scattering.

%%%%%%%%%%%%%%%%%%%%%%%%%%%%%%%%%%%%%%%%%%%%%%%%%%%%%%%%%%%%%%%%%%%%%%%
\section{ The elastic scattering of neutrinos (antineutrinos) on nuclei
with S=0 and T=0}
\label{sec.nuAel}

In this Section we will consider the processes of the elastic NC
scattering of neutrinos (antineutrinos) on nuclei with S=0 
T=0~\cite{DonnPeccei,Bernabeu92,Suzuki90,Henley91}.
\begin{equation}
\nu~(\nubar) + A \longrightarrow \nu~(\nubar) + A 
\label{nuAscat}
\end{equation}

The matrix element of the process for the scattering of neutrinos
(antineutrinos) is given by the expression
\begin{equation}
\la f|S|i\ra = \mp \frac{G_F}{\sqrt{2}} \ubar(k')\gam^{\al}
\left(1\mp\gam_5\right)u(k)\la p'|J_{\al}^{NC}|p\ra
(2\pi)^4\delta^{(4)}(p'-p-q)
\label{nuAmat1}
\end{equation}
Here $k$ and  $k'$ are the momenta of the initial and final neutrino 
(antineutrino), $q=k-k'$, $p$ and $p'$ are the momenta of the initial and
final nucleus. The cross section of the processes (\ref{nuAscat}) are 
given by the general expression (\ref{nucross}). 

The axial current, $A_{\al}^{NC}$, and the isovector part of the vector
 NC, $V_{\al}^{3}(1-2\sin^2\theta_W)$,
do not give contribution to the matrix element of the processes 
(\ref{nuAscat}).
For the isoscalar part of the vector NC we have, in general,
\begin{equation}
\la p'|V_{\al}^{NC}|p\ra = n_{\al} F^{NC}(Q^2) +q_{\al} G^{NC}(Q^2)\,.
\label{genvec1}
\end{equation}
Here $n=p+p'$, $q=p'-p$ and $Q^2 = 2M_{A}T$
($T$ is the kinetic energy of the final nucleus).

The form factor $G^{NC}(Q^2)$ is equal to zero due to T--invariance of
the strong interactions. The remaining \ff $F^{NC}(Q^2)$ can be written
in the form
\begin{equation}
F^{NC}(Q^2)= -2\sin^2\theta_W F(Q^2) -\half F^s(Q^2)
\end{equation}
where $F(Q^2)$ is the isoscalar electromagnetic form factor
 and $F^s(Q^2)$ the strange form factor of the nucleus.
From Eqs.~(\ref{genvec1}) and (\ref{nutensor1}) we have
\begin{equation}
W_{\al\be}^{NC}= \frac{1}{4M^2}\, n_{\al}n_{\be}
\left[F^{NC}(Q^2)\right]^2\delta\left(\nu-\frac{Q^2}{2M_A^2}\right)\,.
\label{nutensor00}
\end{equation}

Then, from Eqs.~(\ref{nucross}) and (\ref{nutensor00}) it follows that 
the cross sections of the scattering of neutrinos and antineutrinos 
on a nucleus with $S=0$, $T=0$ are equal to each other and are given 
by the expression
\begin{equation}
\frac{d\sigma_{\nu}}{dQ^2}=\frac{d\sigma_{\nubar}}{dQ^2}=
\frac{G_F^2}{2\pi}\left(1-\frac{p\cdot q}{M_A E} -\frac{Q^2}{4E^2}\right)
\left[F^{NC}(Q^2)\right]^2
\label{nucross00}
\end{equation}
where $E$ is the neutrino energy in the laboratory system.

Thus, the strange form factor of the nucleus can be determined 
from the measurement of the cross section of the process (\ref{nuAscat}) 
if the electromagnetic form factor of the nucleus $F(Q^2)$ is known.
The latter can be determined 
from the measurement of the cross section of the elastic scattering
of unpolarized electrons on the nucleus, which is given by the expression
\begin{equation}
\frac{d\sigma_e}{dQ^2}=\frac{4\pi\al^2}{Q^4}
\left(1-\frac{p\cdot q}{M_A E} -\frac{Q^2}{4E^2}\right)
\left[F(Q^2)\right]^2\,.
\label{eAelas}
\end{equation}
From Eqs.~(\ref{nucross00}) and (\ref{eAelas}) we can find the relation 
connecting the strange form factor of the nucleus with the
corresponding (measurable) cross sections:
\begin{equation}
F^s(Q^2)= \pm 2 F(Q^2)\left\{\left(\frac{2\sqrt{2}\pi\al}{G_F Q^2}\right)
\sqrt{\frac{(d\sigma_{\nu}/dQ^2)}{(d\sigma_e/dQ^2)}} \mp 2\sin^2\theta_W
\right\}
\label{sffnucl2}
\end{equation}
or, equivalently, by extracting the elastic \ff from (\ref{eAelas}):
\begin{equation}
F^s(Q^2)= \pm 2\frac{1}{
%\displaystyle
{\sqrt{1-\frac{p\cdot q}{M_A E} -\frac{Q^2}{4E^2}}}}
\left\{ \sqrt{\frac{2\pi}{G_F^2}\frac{d\sigma_{\nu}}{dQ^2}}
\mp 2\sin^2\theta_W\sqrt{
\frac{Q^4}{4\pi\al^2}\frac{d\sigma_e}{dQ^2}} \right\}.
\label{sffnucl3}
\end{equation}
In Eqs.~(\ref{sffnucl2}) and (\ref{sffnucl3}) the upper (lower) signs 
refer to a positive (negative) value of the quantity 
$4\sin^2\theta_W F+F^s$.

We remind the reader 
 that the P--odd asymmetry in the scattering of polarized
electrons on nuclei with $S=0$ and $T=0$
is determined by the ratio $F^{NC}/F$ [see Eq.~(\ref{nuclasym})].
On the other hand the ratio of the cross sections (\ref{nucross00}) 
and (\ref{eAelas}) is determined by the ratio $(F^{NC}/F)^2$. Hence,
by comparing (\ref{sffnucl1}) with (\ref{sffnucl2}), we find the
following general relation between quantities, which are measurable 
in the scattering of neutrinos and electrons on nuclei
with $S=0$ and $T=0$:
\begin{equation}
\Acal(Q^2)= \pm \frac{\sqrt{(d\sigma_{\nu}/dQ^2)}}
{\sqrt{(d\sigma_e/dQ^2)}}
\label{relAcross}
\end{equation}
where the plus (minus) sign have the same correspondence as in 
Eqs.~(\ref{sffnucl2}) and (\ref{sffnucl3}).

Let us stress that the relation (\ref{relAcross}) takes place for any
reactions of the type (\ref{reaction1}) and (\ref{nuAscat}) in which the 
initial and final nuclei have $S=0$, $T=0$. It can be violated only 
if in the neutral current there are scalar and/or tensor terms.

The observation of the process of the scattering of neutrino on nuclei
requires the measurement of the small recoil energy of the final nucleus.
It could be easier to detect the process of scattering of neutrinos 
and electrons on nuclei if the nucleus undergoes a transition to
excited states. 

Let us consider, for example, the processes~\cite{Bernabeu92}
\bea
\nu + {^4}\mathrm{He}&& \longrightarrow \nu + {^4}\mathrm{He}^*
\nonumber\\
e + {^4}\mathrm{He}&& \longrightarrow e + {^4}\mathrm{He}^*
\nonumber
\eea
where ${^4}\mathrm{He}^*$ is the excited state of ${^4}\mathrm{He}$
with with $S=0$ and $T=0$ and excitation energy of 
$20.1$~MeV. This state can decay into $p$ and radioactive 
$^3$H~\cite{Fiarman73}.
The matrix element of the neutral current for the above processes is 
given by:
\bea
\la p'|J_{\al}^{NC}|p\ra &&= \la p'|V_{\al}^{NC}|p\ra =
\label{he4}\\
&&= \left[ 2\left( p_{\al} -\frac{p\cdot q}{q^2} q_{\al}\right) 
F_{in}(Q^2) +q_{\al} G_{in}(Q^2)\right]\,.
\nonumber
\eea

It is obvious that the contribution of the form factor $G_{in}$ can
be neglected. For the form factor $F_{in}$ we have
\begin{equation}
F_{in}^{NC}(Q^2) = -2\sin^2\theta_W F_{in}(Q^2) -\half F_{in}^s(Q^2)\,,
\end{equation}
where $F_{in}(Q^2)$  and $F_{in}^s(Q^2)$  are the {\em inelastic} 
electromagnetic and strange form factors. 
All the relations that were obtained for the
elastic case are valid also for the inelastic processes providing we 
change everywhere the elastic form factors
by the  inelastic ones and we use for $p\cdot q$ 
[e.g. in Eqs.~(\ref{nucross00}) and (\ref{eAelas})] the relation:
\begin{equation}
p\cdot q=\half Q^2+\half\left({M_A^*}^2-M_A^2\right)\,
\end{equation}
 $M_A^*$ being the mass of the excited state.
 Other $S=0,\, T=0$ nuclei, like $^{12}$C and $^{16}$O, could 
offer similar (perhaps better) possibilities of detecting $\nu$--induced
transitions to $S=0,\, T=0$ excited states.
Admittedly, the detection of the decay products of these excited states 
could be fairly difficult~\cite{Ajzenberga,Ajzenbergb}. Moreover, 
according to theoretical calculations~\cite{Molinari68}, the strength
of these isoscalar transitions is generally much weaker than the 
corresponding isovector ones. 
Yet, the measurement of the cross sections (\ref{nucross00}) 
and (\ref{eAelas}) (or the analogous ones for inelastic processes)
would allow a model independent determination of the vector strange \ff
of the nucleus.

%%%%%%%%%%%%%%%%%%%%%%%%%%%%%%%%%%%%%%%%%%%%%%%%%%%%%%%%%%%%%%%%%%%%%
\section{Neutrino (antineutrino)--nucleus inelastic scattering}
\label{sec.nuQE}

As we have seen in considering the Los Alamos future experiment and 
the BNL measurement of the axial strange form factor (see Section
\ref{sec.nufree}),
one can consider
neutrino scattering on both free nucleons and nucleons bound
inside complex nuclei. The latter are convenient targets, as many
neutrino detectors often contain nuclei as well as free protons. In fact
even the Brookhaven ``free'' scattering data were mostly obtained from
the scattering of $\nu,\nubar$ on Carbon, corrected for the Fermi
motion and other nuclear effects.

In considering bound nucleons, especially at low energies such as the
ones available at Los Alamos, it is important to be able to describe
the effects of the nuclear many body structure and to evaluate the
impact it can have on the measured quantities, in order to correctly 
interpret the experimental results in terms of single nucleon properties.
Results from quasi--elastic (QE) electron scattering, which has been
widely studied, can provide a guidance for selecting reliable
nuclear models. However NC neutrino processes involve additional
complications, with respect to the inclusive electron scattering.
In fact, as the outgoing neutrino cannot be measured experimentally,
an hadronic signature that the reaction has taken place 
has to be detected.
The corresponding cross sections are therefore exclusive
(or semi--inclusive) in the hadronic sector, si\-mi\-lar to a coincidence
electron scattering process, but inclusive in the leptonic sector,
so that a detailed balance of the energy and momentum transfer is
not possible any more.

In this Section we will summarize the studies which have been proposed
in order to evaluate the impact of nuclear uncertainties on the
extraction of the strange axial form factor of the nucleon from neutrino
nucleus scattering, in particular from the measurement of the ratio
which is under study at Los Alamos.
A general review about neutrino--nucleus scattering processes
can be found in~\cite{DonnPeccei}.

We will consider the following processes, 
\begin{equation}
\nu_\mu({\overline\nu_\mu}) + A \longrightarrow
\nu_\mu({\overline\nu_\mu}) +N + (A-1)
\label{eq.nuANCpro}
\end{equation}
in which a neutrino (antineutrino) of four--momentum $k$ interacts with a
nucleus $A$ in its ground state and a nucleon is emitted, and detected,
in the final state, with four--momentum $p_N=(E_N,p_N)$,
while both the states of the daughter nucleus and the outgoing neutrino
remain undetected.

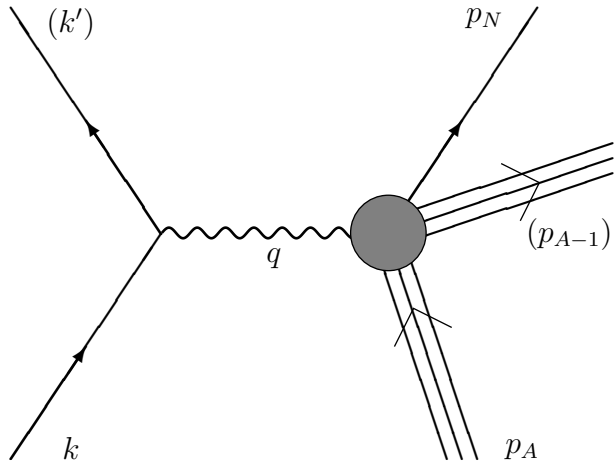
\begin{figure}[h]
\vspace{1.5cm}
\begin{minipage}[p]{\textwidth}
\begin{center}
\setlength{\unitlength}{1.0cm}
\begin{picture}(8,6)
\thicklines
\put(0,0){\line(2,3){2}}
\put(0,0){\vector(2,3){1}}
\put(2,3){\line(-2,3){2}}
\put(2,3){\vector(-2,3){1}}
\put(5,3){\line(2,3){2}}
\put(5,3){\vector(2,3){1}}
\multiput(5.8,0)(0.2,0){3}{\line(-1,3){1}}
\multiput(8,3.8)(0,0.2){3}{\line(-3,-1){3}}
\put(3.5,2.8){\makebox(0,0)[t]{$q$}}
\put(0.8,0){\makebox(0,0)[b]{$k$}}
\put(0.8,6){\makebox(0,0)[t]{$(k')$}}
\put(6.8,0){\makebox(0,0)[b]{$p_{A}$}}
\put(8,3){\makebox(0,0)[r]{$(p_{A-1})$}}
\put(6.3,6){\makebox(0,0)[t]{$p_{N}$}}
%
%% x-grid
%\thinlines
%\multiput(0,0.05)(0,1){9}{\line(1,0){6}}
%% y-grid
%\thinlines
%\multiput(0.05,0)(1,0){7}{\line(0,1){8}}
%%
\put(0,0){
}
\end{picture}
\end{center}
\end{minipage}
\vspace{1cm}
\caption{Schematic representation for the amplitude, in Born approximation, 
of  the  neutrino--nucleus  scattering.}
\label{fig.diag1}
\vspace{1cm}
\end{figure}

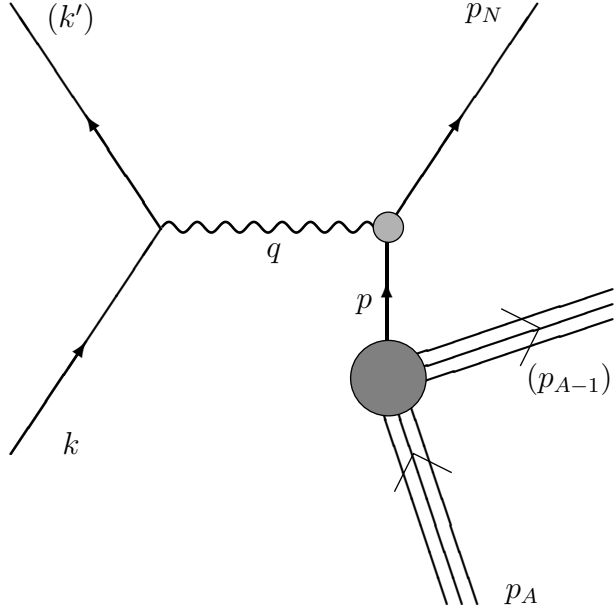
\begin{figure}[h]
\vspace{1.5cm}
\begin{minipage}[p]{\textwidth}
\begin{center}
\setlength{\unitlength}{1.0cm}
\begin{picture}(8,8)
\thicklines
\put(0,2){\line(2,3){2}}
\put(0,2){\vector(2,3){1}}
\put(2,5){\line(-2,3){2}}
\put(2,5){\vector(-2,3){1}}
\put(5,5){\line(2,3){2}}
\put(5,5){\vector(2,3){1}}
\put(5,3){\line(0,1){2}}
\put(5,3){\vector(0,1){1.3}}
\multiput(5.8,0)(0.2,0){3}{\line(-1,3){1}}
\multiput(8,3.8)(0,0.2){3}{\line(-3,-1){3}}
\put(3.5,4.8){\makebox(0,0)[t]{$q$}}
\put(0.8,2){\makebox(0,0)[b]{$k$}}
\put(0.8,8){\makebox(0,0)[t]{$(k')$}}
\put(6.8,0){\makebox(0,0)[b]{$p_{A}$}}
\put(8,3){\makebox(0,0)[r]{$(p_{A-1})$}}
\put(6.3,8){\makebox(0,0)[t]{$p_{N}$}}
\put(4.8,4){\makebox(0,0)[r]{$p$}}
%
%% x-grid
%\thinlines
%\multiput(0,0.05)(0,1){9}{\line(1,0){6}}
%% y-grid
%\thinlines
%\multiput(0.05,0)(1,0){7}{\line(0,1){8}}
%%
\put(0,0){
}
\end{picture}
\end{center}
\end{minipage}
\vspace{1cm}
\caption{
Representation of  the $\nu$--nucleus scattering in  the Plane
Wave Impulse
Approximation.}
\label{fig.diag2}
\vspace{1cm}
\end{figure}

Most of the models used in the literature describe the above quasi--free
processes within the Impulse Approximation, where the neutrino is assumed
to interact with only one nucleon, which is then emitted,
the remaining nucleons being spectators.
After the interaction with the neutrino, the struck nucleon can  be 
considered as free (Plane Wave Impulse Approximation) or the residual
interaction with the recoiling system can be taken into account, either
directly or indirectly, through distorted wave functions.
The different calculations proposed in the literature then 
differ by the models employed to describe the
bound nucleus as well as the emitted nucleons.
A schematic representation of the neutrino--nucleus scattering amplitude
and, respectively, of its description within the PWIA is given
in Figs.~\ref{fig.diag1} and \ref{fig.diag2}.

%*************************************************************************
%\newpage
%%%%%%%%%%%%%%%%%%%%%%%%%%%%%%%%%%%%%%%%%%%%%%%%%%%%%%%%%%%%%%%%%%%%%%
%Fig12
%%%%%%%%%%%%%%%%%%%%%%%%%%%%%%%%%%%%%%%%%%%%%%%%%%%%%%%%%%%%%%%%%%%%%%
%*************************************************************************
%Fig13
%*************************************************************************

A model which has been employed very often in the literature is the
Relativistic Fermi Gas (RFG), in which the bound
and the outgoing nucleon are described by plane waves.    
Although the Fermi Gas is not completely realistic, thanks to its
simplicity  it is a useful guidance to understand the behavior of the
different cross sections. Therefore, before illustrating various results,
we briefly present the relevant formalism.\footnote{
The same model was considered in Section \ref{sec.PoddAel} for the
description of the P--odd asymmetry in quasi--elastic electron--nucleus
scattering. Here, however, the kinematic situation is different, since
the final neutrino is not detected, while the nucleon ejected from the
nucleus is the only observable final particle in the process.
}

The RFG cross section is obtained by averaging over the
nucleon momentum distribution the cross section for the scattering
on a free moving nucleon of four--momentum $p$.
Its general expression can thus be written as:
\begin{eqnarray}
\left(
\frac{d^2\sigma}{d E_N d\Omega_N}
\right)_{\nu(\overline\nu)}
&&=
\frac{G_F^2}{(2\pi)^2} 
\frac{3 \Ncal}{4\pi p_F^3}
\frac{M^2 |{\vec{p}}_N|}{k_0 }
\int \frac{d^3 k'}{k_0'}\frac{d^3 p}{p_0} \times
\label{eq.sigmaA}\\
&&\times
\delta^{(3)}
\left({\vec{k}}-{\vec{k}}'+{\vec{p}}-{\vec{p}}_N\right)
\delta\left(k_0-k_0'+p_0-E_N \right)\times
\nonumber\\
&&\times \theta(p_F-|{\vec{p}}\,|)\theta(|{\vec{p}}_N|-p_F)
\left(L^{\alpha\beta} \mp L_5^{\alpha\beta}\right)
\left(W_{\alpha\beta}^{NC}\right)_{s.n.}\;,
\nonumber
\end{eqnarray}
where the integration over the momentum of the outgoing neutrino,
$\vec{k}^{\prime}$, has been included explicitly.
Here $(W_{\alpha \beta}^{NC})_{s.n.}$ is the single nucleon NC hadronic
tensor, which is given by the following expression:
\bea
\left(W_{\al\be}^{NC}\right)_{s.n.} &&=
-\left[\tau\left(\GMNC\right)^2 +(1+\tau)\left(\GANC\right)^2\right]
\left(g_{\al\be}-\frac{q_{\al}q_{\be}}{q^2}\right) +
\nonumber\\
&&+ \left[ \frac{\left(\GENC\right)^2+\tau\left(\GMNC\right)^2}{1+\tau}
+\left(\GANC\right)^2\right]\frac{X_{\al}X_{\be}}{M^2} +
\nonumber\\
&&- \left(\GANC\right)^2\frac{q_{\al}q_{\be}}{q^2} +
\frac{i}{M^2}\epsilon_{\al\be\mu\nu}p^{\mu}q^{\nu}\GANC\GMNC
\label{NCsntensor}
\eea
with\footnote{We notice that the expression (\ref{NCsntensor}) is
analogous to the sum of the ones contained in Eqs.~(\ref{nutensor3}) and
(\ref{nutensor5}), but without the energy conserving delta functions (here
explicitly included in the integral) and with $X_{\al}$ instead of
$n_{\al}=p_{\al}+{p_N}_{\al}\equiv 2p_{\al}+q_{\al}$. The two notations
are equivalent for the scattering of massless leptons, since the
contraction of $L^{\al\be}$ with terms proportional to $q_{\al}$
($q_{\be}$) vanishes.}
\[
X_{\al}=p_{\al}-\frac{\left(p\cdot q\right) q_{\al}}{q^2}\,.
\]
In Eq.~(\ref{eq.sigmaA}) 
$L^{\alpha \beta}$ and $L_5^{\alpha \beta}$ are the leptonic tensor
and pseudotensor, given in Eqs.~(\ref{lepten}) and (\ref{lepten5}),
respectively; $p_F$ is the Fermi momentum of the nucleus under
consideration, $(3 \Ncal)/(4 \pi p_F^3)\theta(p_F - |\vec{p}\,|)$
is the momentum distribution of the nucleons in the RFG, ${\Ncal=Z,N}$
being the number of protons or neutrons. We notice that the NC \ffs
which are contained in the hadronic tensor
$\left(W_{\al\be}^{NC}\right)_{s.n.}$ are different for protons and
neutrons: hence the proton form factors have to be used in Eq. 
(\ref{NCsntensor}) when the QE emission of a proton is considered and,
conversely, the neutron form factors have to be used when the emitted
particle is a neutron.
The function $\theta(|\vec{p}_N| - p_F )$ takes into account the effects
of Pauli Blocking on the ejected nucleon.
The effects of an average binding energy of the  bound nucleons can
be taken into account by subtracting
a (constant) term to the initial nucleon energy in the argument of the 
energy conserving delta function, $p_0 \rightarrow p_0 -B$.

Writing explicitly the contraction between the leptonic and the hadronic
tensors the cross section becomes:
\begin{eqnarray}
\left(
\frac{d^2\sigma}{d E_N d\Omega_N}
\right)_{\nu(\overline\nu)}
&=&
\frac{G_F^2}{(2\pi)^2} 
\frac{3 \Ncal}{4\pi p_F^3}
\frac{|{\vec{p}}_N|}{k_0 }
\int \frac{d^3 k'}{k_0'}\frac{d^3 p}{p_0}
\delta\left(k_0-k_0'+p_0-E_N\right)
\nonumber\\
&\times& \delta^{(3)}
\left({\vec{k}}-{\vec{k}}'+{\vec{p}}-{\vec{p}}_N \right)
\theta(p_F-|{\vec{p}}\,|)\theta(|{\vec{p}}_N|-p_F)
\nonumber\\
&\times& \left\{ V_M (\GMNC)^2 + V_{EM} 
\frac{(\GENC)^2 + \tau (\GMNC)^2}
{1 + \tau} +\right.
\nonumber\\
&+& \left. V_A (\GANC)^2 \pm V_{AM} \GANC\GMNC \right\}
\label{eq.sigmaA2}
\end{eqnarray}
where the form factors have been grouped as in Eq.~(\ref{nucross2}) and 
\begin{equation}
\begin{array}{l}
V_M = 2 M^2 \tau \left(k \cdot k^{\prime}\right) \\
V_{EM} = 2 \left(k \cdot p\right) \left(k^{\prime} \cdot p\right)
  - M^2 \left(k\cdot k^{\prime}\right) \\
V_A = M^2 \left(k\cdot k^{\prime}\right) +
 2 M^2 \tau \left(k \cdot k^{\prime}\right)  + 2 \left(k \cdot p \right)
 \left(k^{\prime} \cdot p\right)  \\
V_{AM} = 2 \left(k\cdot k^{\prime}\right) 
\left(k \cdot p + k^{\prime}\cdot p \right)
\end{array}
\label{eq.Vs}
\end{equation}

The interesting quantities, related to the determination of the
strange form factors of the nucleon, are then the single differential
cross section
\begin{equation}
\left(\frac{d\sigma}{d T_N}
\right)_{\nu(\overline\nu) N}\equiv
\left(\frac{d\sigma}{d E_N}
\right)_{\nu(\overline\nu) N} =
\int\, d\Omega_N 
\left(\frac{d^2\sigma}{d E_N d\Omega_N}\right)_{\nu(\overline\nu)N}\;,
\label{eq.diffT}
\end{equation}
where $T_N$ is the outgoing nucleon kinetic energy,
the total (integrated over the nucleon energy) cross section
\begin{equation}
\sigma_{\nu(\overline\nu) N}=\int\, dT_N 
\left(\frac{d\sigma}{dT_N }\right)_{\nu(\overline\nu)N}
\label{eq.total}
\end{equation}
and the corresponding ``proton over neutron'' ratios
\bea
{\mathcal R}_{\nu (\overline{\nu})} 
&=& 
\frac{\displaystyle{\left(
\frac{d\sigma}{dT_N }
\right)_{\nu(\overline\nu)p}}}
{\displaystyle{\left(
\frac{d\sigma}{dT_N }\right)_{\nu(\overline\nu)n}}},
\label{eq.ratdiff}\\
R_{\nu (\overline{\nu})}
&=& \frac{\sigma_{\nu(\overline\nu) p} }
{\sigma_{\nu(\overline\nu) n}}\,.
\label{eq.rattot}
\eea

An approximate expression for the RFG cross section can be obtained
by inserting the explicit forms [Eqs. (\ref{gaNC}), (\ref{geNC}),
(\ref{gmNC})]
of the nucleonic NC form factors into Eq. (\ref{eq.sigmaA2}) 
and neglecting both the terms proportional to 
$\left(1 - 4 \sin^2(\theta_W)\right)\simeq 0.075$ and the terms
quadratic in the strange form factors.
Under these approximations and using 
$\delta^{(3)} \left({\vec{k}}-{\vec{k}}'+{\vec{p}}-{\vec{p}}_N \right)$ 
to integrate over ${\vec{k}}'$ one obtains: 
\bea
&&\left(\frac{d^2\sigma}{d E_\N d\Omega_\N}
\right)^{NC}_{\scriptstyle{\left\{
\begin{array}{l}
\scriptstyle \nu p(n)\\
\scriptstyle {\overline\nu} p(n)
\end{array} \right. }
}
= \frac{3Z(N)}{4\pi p_F^3}\frac{G_F^2}{(2\pi)^2} 
\frac{|{\vec{p}}_\N|M^4}{ k_0 } \theta(|{\vec{p}}_\N|-p_F)\times 
\nonumber \\
&&\qquad\times
\int \frac{d^3 p}{k'_0 p_0} \delta\left(k_0-k_0'+p_0-E_\N\right)
{\mathcal I}_{p(n)}(k,p,Q^2)\, ,
\label{eq.sigmaA3}
\eea
where (the plus/minus sign refer to the $\nu$ and 
${\bar\nu}$ cases):
\bea
{\mathcal I}_{p(n)}(k,p,Q^2) &&= {\mathcal M}_{p(n)} 
+2 \tau^2 G_M^{n(p)} G_M^s +
\nonumber \\
&& +\left[(z-\tau)^2-\tau(\tau+1)\right] 
\frac{G_E^{n(p)} G_E^s +\tau G_M^{n(p)} G_M^s}{1+\tau}+ 
\nonumber \\
&&-\delta_{p(n)} \left\{\left[(z-\tau)^2+2 \tau(\tau+1)\right]
G_A G_A^s \right.+
\nonumber \\
&&\left.\pm 2 \tau (z-\tau) \left(G_M^{n(p)} G_A +G_A G_M^s-G_M^{n(p)} 
G_A^s\right) \right\},
\label{eq.Ipn}
\eea
with $\delta_p=1,\ \delta_n=-1$, $z=k\cdot p/M^2$  and, for scattering
of massless leptons,
$\tau = Q^2/4M^2 = k\cdot k'/2 M^2 =
(k\cdot p - k\cdot p_N)/2 M^2$.
The terms ${\mathcal M}_{p(n)}$ contain only the electromagnetic and axial
form factors and are given by:
\bea
{\mathcal M}_{p(n)} &&= \tau^2\left({G^{n(p)}_M}^2+G^2_A\right) +
\label{eq.Npn} \\
&& + \frac{1}{2}\left[ (z-\tau)^2-\tau(\tau+1)\right]
\left[ \frac{ {G^{n(p)}_E}^2 +\tau {G^{n(p)}_M}^2} {(1+\tau)} 
+ G_A^2 \right]\,.
\nonumber
\eea
The above equations, although approximate, can be useful
to understand the interplay between strange and non--strange 
form factors in the cross sections (\ref{eq.diffT}), (\ref{eq.total})
and thus in the ratios (\ref{eq.ratdiff}), (\ref{eq.rattot}). 
We notice, however, that all the results presented in
this Section were obtained without introducing any approximation.

The RFG was first applied to the study of neutrino--nucleon scattering
in connection with the problem of nucleon strangeness by 
Horowitz et al.~\cite{Horow93},
and later employed by other authors~\cite{Barbaro96b,ABBCGM,ABBCGM98}.
The considered nucleus was $^{12}C$, for which the values
$p_F=225$ $MeV$ and $B=25\div 27$ MeV are typically used.

A detailed analysis of the uncertainty induced by nuclear 
structure effects on the determination of axial strange form factor
in $\nu$--nucleus QE scattering 
was proposed by Barbaro et al. 
in Ref.~\cite{Barbaro96b}, both for the Los Alamos and 
the Brookhaven kinematical conditions.

In this paper the authors consider the processes 
(\ref{eq.nuANCpro}) on $^{12}C$, 
using two different nuclear models: the RFG and an Hybrid Model (HM), 
in which the bound nucleons are described by harmonic oscillator 
shell model wave functions, while the outgoing nucleon is described 
by a plane wave. 
These choices are considered as two ``extremes'' among the
available realistic nuclear models: in fact the RFG, using plane waves 
for the bound nucleons, can be seen as a ``maximally unconfined'' model,
while the HM, whose bound single nucleon wave functions decrease 
more rapidly than the expected exponential behavior, 
is somehow ``over-confined''. Moreover, while the RFG 
involves on-mass shell single nucleon amplitudes, 
the HM allows one to consider
(half)-off shell nucleonic currents, 
thus providing an estimate of off-shellness 
effects. Therefore the difference between the effects 
of the strange form factors on the considered quantities
calculated with these two models can provide an upper bound to 
the uncertainty induced by nuclear structure effects. 

These two models were used to calculate both the differential 
and total cross sections [Eqs. (\ref{eq.diffT}) and (\ref{eq.total})], 
for the neutrino--induced emission of a proton and of a neutron,
for fixed neutrino energy.
In this calculation the axial and axial strange 
form factors, $G_A$ and $G_A^s$, were assumed to have the same 
dipole $Q^2$--dependence,
while strangeness contributions to the nucleonic vector current were 
not considered.
It is important to notice that, 
contrary to what happens  at the relatively high energies of BNL, 
under the Los Alamos kinematics the correlation
between the value of the axial dipole cut-off mass 
and the strange axial constant $g_A^s$ is negligible, 
as shown in Refs.~\cite{Musolf94.rep,Horow93}, and thus  at low energies 
the dipole parameterization
can be used without introducing significant uncertainties
\footnote{Since very little is known about the $Q^2$ dependence
of strange form factors, in the literature about neutrino
scattering they are often assumed to have the same $Q^2$ dependence
as the corresponding non--strange ones. Only in Ref.~\cite{Kolbe97} 
strange form factors obtained within an SU(3) Skyrme
model were used to predict the effects of strangeness
on the Los Alamos ratio $\overline{R}_{LAMPF}$ [see Eq.~(\ref{rbar})].
}.

\begin{figure}
\begin{center}
\mbox{\epsfig{file=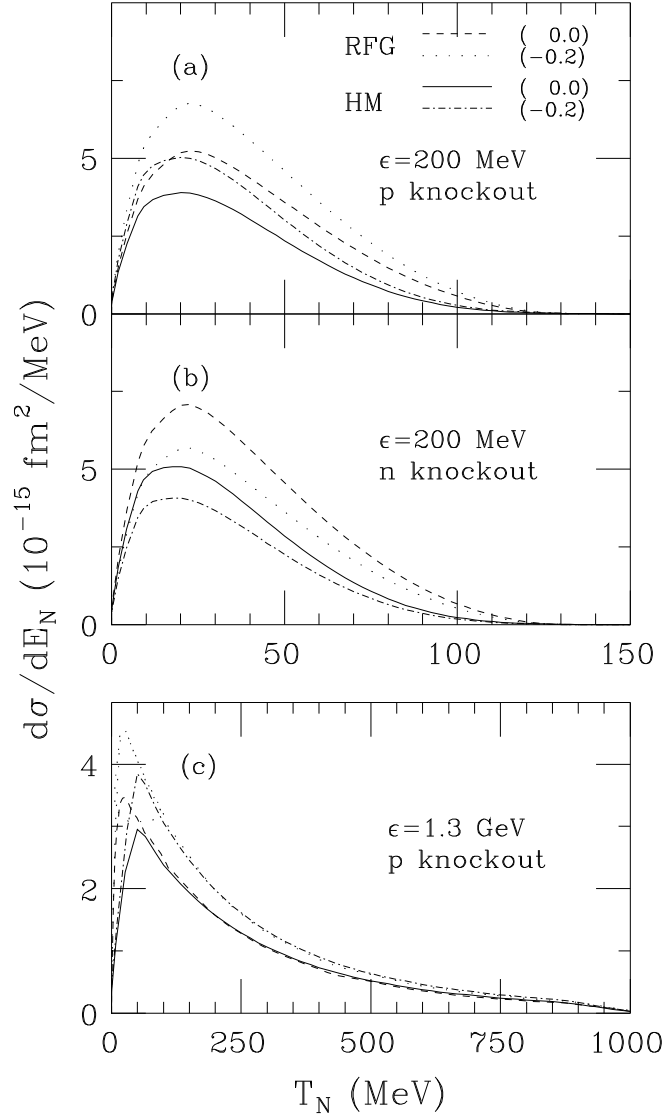,width=0.8\textwidth}}
\end{center}
\vskip -1cm
\caption{Differential cross sections 
$\left({d\sigma}/{d E_N}\right)_{\nu N}$, for the emission
of a proton and of a neutron [panels (a) and (b), respectively]
under the Los Alamos kinematic conditions, and for the emission
of a proton for the average energy $E_\nu=1.3$ GeV of Brookhaven
(panel c). The solid and dot-dashed lines correspond the the Hybrid
Model mentioned in the text, without strangeness and with the strange
axial constant set to $g_A^s=-0.2$, respectively.
The dashed and dotted lines are obtained within the RFG, again
with $g_A^s=0$ and $g_A^s=-0.2$, respectively.
(Taken from Ref.~\cite{Barbaro96b})
}
\label{fig.barbaro8}
\end{figure}

\begin{figure}
\begin{center}
\mbox{\epsfig{file=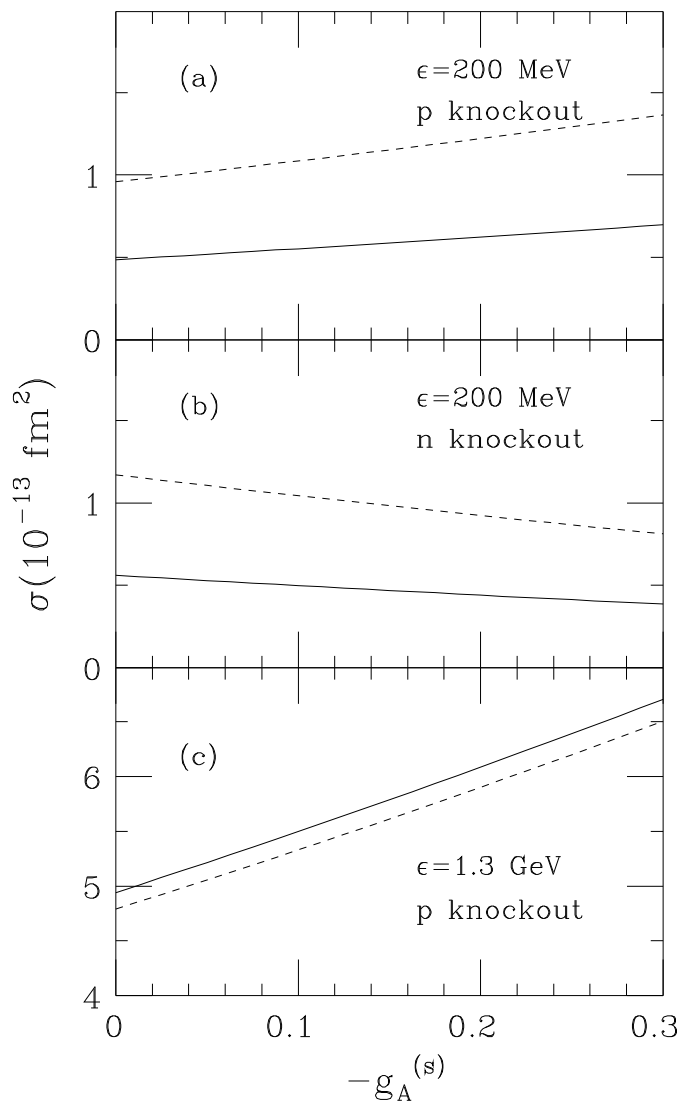,width=0.8\textwidth}}
\end{center}
\vskip -2cm
\caption{Total cross sections 
$\sigma_{\nu N}$ integrated over the nucleon kinetic energy
($T_N>60$ MeV for low energy neutrinos and $T_N>200$ MeV for higher energy 
neutrinos) as a function of the strange axial constant $-g_A^s$,
for the emission of a proton (a) and of a neutron (b) at
Los Alamos kinematics and for protons at Brookhaven kinematics (c).
The solid lines correspond to the Hybrid Model described in the text, 
while the dashed lines are the RFG results. 
(Taken from Ref.~\cite{Barbaro96b})
}
\label{fig.barbaro10}
\end{figure}

The results of Ref.~\cite{Barbaro96b} for the kinematic conditions 
of Los Alamos
($E_\nu= 200$ MeV) are shown in the upper and central panels
of Figs.~\ref{fig.barbaro8} and \ref{fig.barbaro10}.

%
%*************************************************************************
%*************************************************************************
%
%
%*************************************************************************
%*************************************************************************
%

From the comparison of their results on the 
separate cross sections at $E_\nu=200$ MeV, Barbaro and collaborators
derive an uncertainty on $g_A^s$ due to nuclear model 
dependencies given by 
$\delta_{nucl}(g_A^s) = \pm 0.25$, therefore larger than the expected 
value of $g_A^s$ itself. 
However, when the ratio of either differential or total
cross sections is considered, this uncertainty is reduced by an order 
of magnitude, down to $\delta_{nucl}(g_A^s) = \pm 0.015$. 
Including, in a worst case scenario, additional uncertainties due 
to off-shellness effects  a final estimate  $\delta_{nucl}
(g_A^s) = \pm 0.03$ is provided. 
This estimate is much smaller than the uncertainties
on measurements of $g_A^s$ in deep inelastic scattering processes, and 
therefore these authors conclude that the proposed Los Alamos measurement 
is a very good way to determine $g_A^s$. 

The same analysis was applied in~\cite{Barbaro96b} to the
single cross section for the neutrino induced QE emission of a proton
at $E_\nu=1.3$ GeV, the average energy of the BNL neutrino beam,
as shown in the lower panels of Figs.~\ref{fig.barbaro8} and
\ref{fig.barbaro10}.
In this case nuclear structure effects are much smaller 
and the estimated uncertainty on the strange axial
constant drops down to $\delta_{nucl} (g_A^s) = \pm 0.015$. 
Both low and intermediate energy results are in agreement 
with the previous analysis proposed by Horowitz at al.~\cite{Horow93} 
within the RFG, with and without binding energy effects.

The analysis of Barbaro et. al was carried on within the Plane Wave 
Impulse Approximation, thus not including the 
effects of Final State Interactions.
The authors, however, concluded that the estimated  value of $\delta_{nucl}
(g_A^s)$ for the p/n ratio can be considered as an upper bound 
also for the uncertainty associated with the use of more realistic
nuclear models, which include FSI effects. Nonetheless,
a detailed study  of these effects has to be done, in order to be able 
to extract strangeness parameters from the proposed quantities.

As already mentioned in Section \ref{sec.nufree}, FSI effects 
on the ratios (\ref{eq.ratdiff}) and (\ref{eq.rattot}) under the kinematical 
conditions typical of the Los Alamos facility were studied by
Garvey et al. in Ref.~\cite{Garvey93b}, within the random phase approximation. 
In their calculations the ground state of Carbon was described as a 
Slater determinant of Woods-Saxon wave functions, the parameters of 
the WS potential being chosen in order to reproduce the ground 
state properties of $^{12}$C. FSI were included by means of a finite 
range G-matrix interaction, derived from the Bonn $NN$ potential.
It was found that FSI can sizeably affect the separate cross sections, 
resulting in an increase of the latter of up to 40\%, but these effects 
are mainly canceled in the ratio, where they amount to less than 10\%
and do not interfere with the effects of $g_A^s$.
Similar results were obtained in a later calculation by Alberico 
et al.~\cite{ABBCGM}, who evaluated the ratio of differential 
cross sections, Eq.~(\ref{eq.ratdiff}), 
within both the RFG and a Relativistic Shell Model. 
The latter included also FSI effects through a Relativistic Optical 
Potential.
Again, FSI effects were found to be large
on the separate cross sections, resulting in a decrease of the latter
of up to 40\%, 
mainly due to the absorption into different channels described by the 
imaginary part of the optical potential. However these effects on 
the p/n ratio were reduced to less than 10\% and were found to be 
due mainly to the Coulomb repulsion, which is present for the protons 
only. In Ref.~\cite{ABBCGM} FSI effects on the p/n ratio at
Los Alamos kinematics were shown to be comparable with the effects of 
the magnetic strange form factor $G_M^s$. It is worth noticing 
that in Ref.~\cite{ABBCGM} it was found that
Final State Interactions can still 
be relevant, for the separate cross sections,  even at relatively 
high energies ($E_\nu \simeq 1$ GeV).

Here we want to make a comment about a possible source of confusion:
the authors~\cite{Garvey92,Garvey93b} who originally proposed the measurement 
of the ratio $R$ use the following definition for the nucleon kinetic energy, 
$T_N= T_p= T_n +2.77$ MeV, $2.77$ MeV being the average
Coulomb repulsion for the protons. For the total cross sections 
(\ref{eq.total}) this translates into
a different value for the lower limit of integration, 
namely $T_p^{min}=60$ MeV,  $T_n^{min}=57.23$ MeV. 
Other authors~\cite{Kolbe97,Barbaro96b,ABBCGM}, 
instead, use the same values $T_N=T_p=T_n$.
While the choice of either definition does not affect the general 
considerations on both the sensitivity to the strange form factors and 
to the nuclear model effects, it was noticed in \cite{Kolbe97}  that, 
when the experimental ratio given in Eq.~(\ref{rbar}) is considered, 
its numerical value can sensibly depend on the choice adopted. 
This has to be remembered when one is
comparing results from different authors
and especially when examining the future experimental data.  

All calculations of the Los Alamos ratio so far considered
were performed in the framework of the Impulse Approximation.
Going beyond IA, Umino {\em et al.}~\cite{Umino95PLB,Umino95PRC}
evaluated the effects of two--body relativistic meson exchange
currents (MEC) in neutrino--nucleus scattering at low and intermediate 
energies, within a soft--pion dominance model.
The exchange current effects on the single differential cross sections
(\ref{eq.diffT}) were calculated for both neutrinos and antineutrinos,
for a fixed incident energy of $200$~MeV and under
various assumptions for the strange form factors $G_A^s$
and $F_2^s$. These effects are
strongly dependent on the value of the Fermi momentum and
can be relatively large for $p_F \ge 350$~MeV. As an example,
with $p_F=300$~MeV the $(\nu,\nu',p)$ cross sections are reduced 
by MEC effects by about 20\% at the peak; these corrections however
become much less important (as already discussed for FSI) in the ratio
of cross sections, Eq.~(\ref{eq.ratdiff}). 
For Carbon, by assuming $p_F=225$~MeV, MEC contributions
were found to be small for neutrino scattering
and somewhat larger for the antineutrino cross sections, 
resulting in a reduction of the latter of about 15\%.
Moreover these effects are mainly confined to low values
of the outgoing nucleon kinetic energy, $T_N \le 50$~MeV,
which is excluded by the kinematical cuts applied in the 
proposed LAMPF experiment, in order to select quasi--elastic events.
Correspondingly, MEC corrections to the proton over neutron ratio 
were found to be limited to a few percent for the neutrino case
and about 10\% for the antineutrino one.\footnote{
We mention here that MEC effects have also been considered in connection
with the P--odd asymmetry in electron--nucleus scattering: both in the
case of $\vec{e}-d$  quasielastic processes~\cite{Schramm94} and
of $\vec{e}$--$^4$He elastic scattering~\cite{MusDonnPL93,MusDonSchia94},
the MEC contributions are small and below experimental detectability: hence
they do not affect significantly the investigation of nucleon strange 
form factors.}

The possibility to extract information on the nucleonic strange form factors 
from a measurement of the ratios in Eqs.~(\ref{eq.ratdiff}),
(\ref{eq.rattot}) at energies 
higher than the ones available at LAMPF has been studied in
Ref.~\cite{ABBCGM98}, 
for neutrino and antineutrino energies of 1 GeV. Separate neutrino and 
antineutrino ratios were considered, both for differential and integrated 
cross sections. 
It was found that while the separate cross sections can still be rather
sensitive to FSI effects, the nuclear model dependence is very weak in 
the ratios. The sensitivity of the ratios to the strange axial, magnetic 
and electric form factors as well as to the axial cut-off $M_A$ was studied 
in detail. For the axial and magnetic strange form factors the dipole
parameterizations in Eq.~(\ref{dipstrangff}) were
used, while the electric strange form factor was assumed to be:
\begin{equation}
G_E^s(Q^2) = \frac{\rho_s \tau}
{\displaystyle{\left( 1 + \frac{Q^2}{M_V^2}\right)^2}}\,.
\label{eq.gesdip}
\end{equation}
It was found that for the neutrino ratio the dominant effect 
is still due to $g_A^s$ and that
this effect can still be large enough to allow an extraction of $g_A^s$ from 
a possible experiment, although its interplay with the strange magnetic 
form factor $G_M^s$ has to be carefully considered.
On the other hand the antineutrino ratio
can be rather sensitive to the electric strange form factor, $G_E^s$.
However, although important as preliminary indications, these results
can depend on the specific kinematical cuts somehow arbitrarily used 
in the calculation, and further investigation would be needed in case
of a future experiment.

\begin{figure}[h]
\begin{center}
\mbox{\epsfig{file=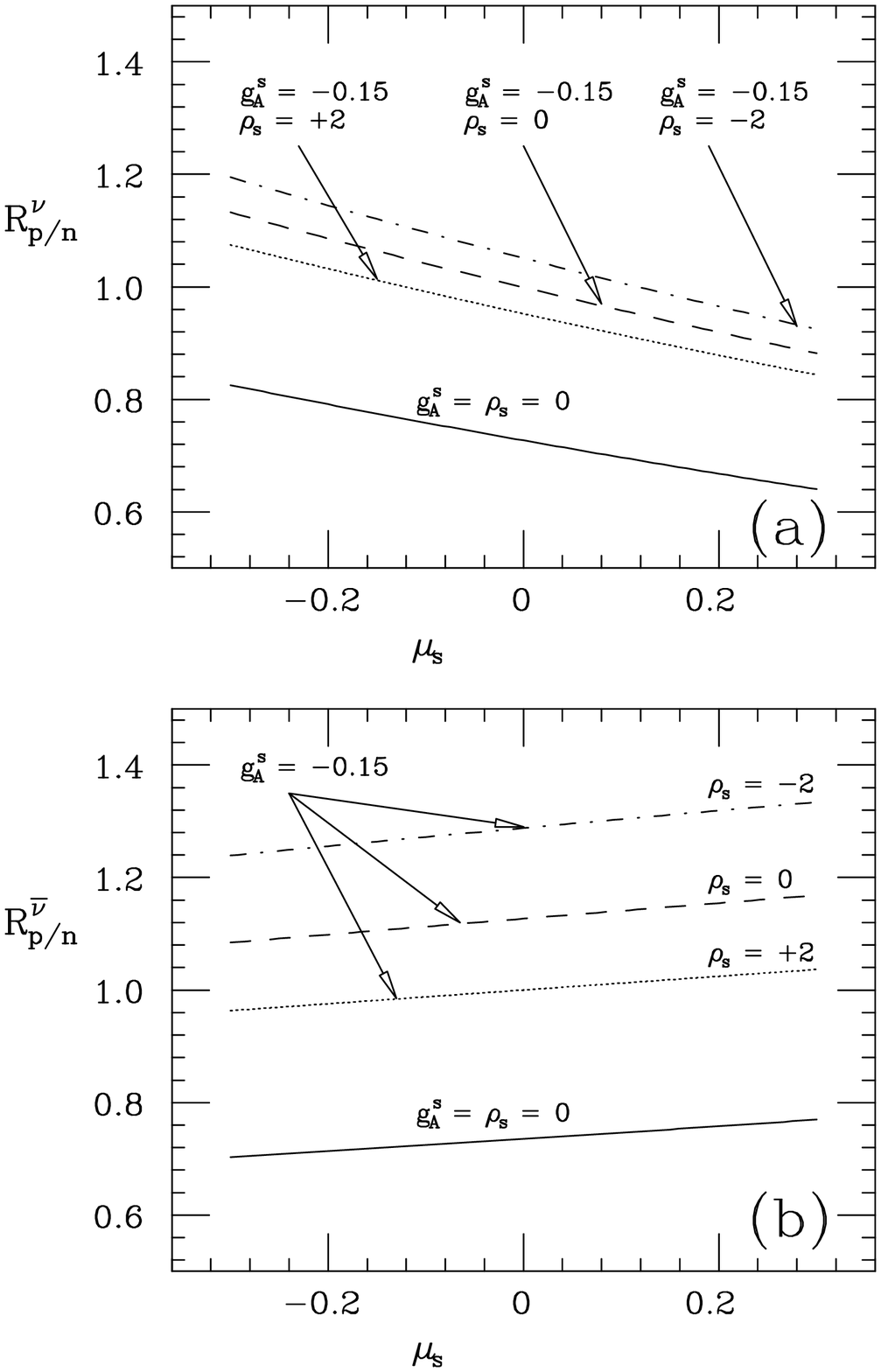,width=0.9\textwidth,
height=0.8\textheight}}
\end{center}
\vskip -1cm
\caption{
The ratios (\ref{eq.rattot}) of the
integrated NC neutrino and antineutrino--nucleus cross sections
(\ref{eq.total}), as a function of $\mu_s$, evaluated in the RFG. 
The incident energy is  $E_{\nu(\bar\nu)}=1$~GeV and the integration 
limits for the cross
sections are $100 \le T_p \equiv T_n \le 400$~MeV.  
The solid line corresponds  to $g_A^s=\rho_s=0$,
in the other three curves [both in (a) and in (b)] 
$g_A^s=-0.15$ and $\rho_s=0$ (dashed line),
$-2$ (dot--dashed line) and $+2$ (dotted line).
(Taken from Ref.~\cite{ABBCGM98})
}
\label{fig.rat1gev}
\end{figure}

An illustration of these results is given in Fig.~\ref{fig.rat1gev}, 
where the ratio (\ref{eq.rattot}) of integral cross sections,
calculated in the Relativistic Fermi Gas, 
is plotted as a function of the parameter $\mu_s$ for different values of
$g_A^s$ and $\rho_s$.
%

%*************************************************************************
%*************************************************************************

Finally let us go back to the neutrino--antineutrino asymmetry,
introduced in Section \ref{sec.freeasymm} for the case of free nucleons. 
A more realistic approach would require to use the QE NC processes
(\ref{eq.nuANCpro}) together with the CC processes 
\begin{equation}
\nu_\mu({\overline\nu_\mu}) + A \longrightarrow
\mu^-({\mu^+}) +p(n) + (A-1)\, ,
\label{eq.CCpro}
\end{equation}
for the denominator. 
Since, as already noticed, the momentum transfer 
$Q^2$ cannot be determined in QE NC processes, the 
following asymmetry should be considered
\begin{equation}
{\Acal (T_N)} = 
\frac{\displaystyle{ 
\left( \frac{d \sigma}{d T_N} \right)_{\nu}^{\mathrm{NC}} 
- 
\left( \frac{d \sigma}{d T_N}
\right)_{\overline\nu}^{\mathrm{NC}}
}} 
{\displaystyle{\left( \frac{d \sigma}{d T_N}
\right)_{\nu}^{\mathrm{CC}} - 
\left( \frac{d \sigma}{d T_N}
\right)_{\overline\nu}^{\mathrm{CC}} }
}\;,
\label{eq.inelasymm}
\end{equation}
$T_N$ being, as usual, the kinetic energy of the ejected nucleon
(proton or neutron). 
\footnote{We remind the reader that in the free nucleon case the following
relation holds: $Q^2=2 M T_N$.}

Here the CC cross sections $\left({\displaystyle\mathrm{d} \sigma/
\displaystyle{\mathrm{d}} T_N}
\right)_{\nu(\overline\nu)}^{\mathrm{CC}}$ in 
the denominator have been considered in analogy to the NC cross sections 
of Eq.~(\ref{eq.diffT}).
 
The effects of nuclear structure on this asymmetry were studied in
Ref.~\cite{ABBCGM}, by comparing the results obtained within 
two nuclear models: the RFG and a Relativistic Shell Model (RSM).
The Shell Model wave functions were obtained as 
mean field solutions of a Dirac Equation derived from
a linear Lagrangian, which includes nucleons, scalar ($\sigma$),
vector--isoscalar ($\omega$) and vector--isovector ($\rho$) mesons.
This model had been previously used in the study of $(e,e'N)$
reactions on different nuclei~\cite{Udias93a,Udias93b,Udias93c}.
The effects of Final State Interactions were described   
within the Distorted Wave Impulse Approximation,
using, for the outgoing nucleon wave functions, 
the solutions of a Dirac equation with a phenomenological 
Relativistic Optical Potential (ROP),
which includes scalar, vector and (for protons) Coulomb 
components~\cite{Hama90}.
When one is dealing with the CC cross sections entering in the denominator of 
the asymmetry (\ref{eq.inelasymm}), another type of ``final 
state interactions'' has to be considered, namely the Coulomb distortion 
of the outgoing leptons. This is expected to be small but 
it has to be taken into account since
 it is enhanced in the $\nu$--$\overline{\nu}$ difference 
in the denominator of (\ref{eq.inelasymm}) and can become relevant if other 
effects such as FSI are canceled in the ratio.
In Ref.~\cite{ABBCGM} Coulomb distortion was taken into account within 
the ``effective impulse approximation'', which prescribes 
to modify the outgoing 
lepton plane wave $e^{i{\vec{k}}'\cdot{\vec{r}}}$ according to:
\begin{equation}
e^{i{\vec{k}}'\cdot{\vec{r}}} \rightarrow 
\frac{|\vec{k}\,'_{eff}|}{|\vec{k}'|} 
e^{i{\vec{k}_{eff}}\cdot{\vec{r}}} \;\;,
\label{eq.coulomb}
\end{equation}
where 
\begin{equation}
{\vec{k}}\,'_{eff} 
= {\vec{k}}\,'\left(1 \pm\frac{3}{2}\frac{Z\alpha}{R|{\vec{k}}'|}
\right).
\label{eq.keff}
\end{equation}
Here the plus (minus) refers to the lepton
(anti-lepton), $Z$ is the number of protons
and $R\simeq 1.2 A^{1/3}$ is the effective charge radius of the
nucleus under investigation. The validity of this approximation 
has been studied in electron scattering processes~\cite{Giusti87}.

The asymmetry (\ref{eq.inelasymm}) was calculated in Ref.~\cite{ABBCGM}
for three typical 
values of the neutrino energy, 200 and 500 MeV and 1 GeV, 
using dipole--like parameterizations for both strange and non--strange 
form factors and the Galster parameterization for the electric
form factor of the neutron [see Eq.~(\ref{neutronEff})].
It was found that at low energies (200 MeV) the impact of nuclear 
uncertainties is too large to allow an unambiguous determination of 
the strange form factors from a measurement of $\Acal$. 
However at larger energies (already 500 MeV and 
especially 1 GeV) these effects are strongly reduced.

\begin{figure}[h]
\begin{center}
\mbox{\epsfig{file=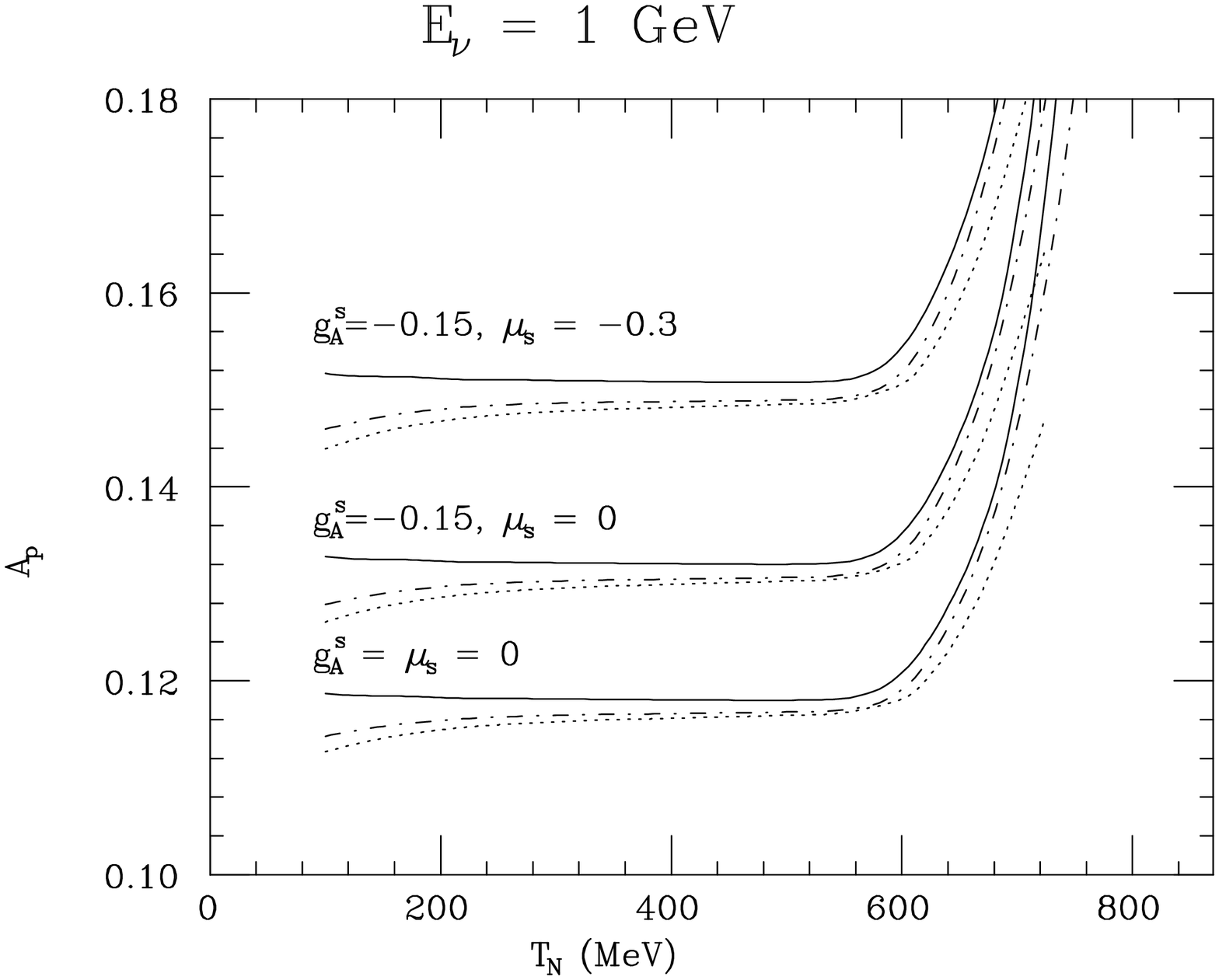,width=0.8\textwidth,angle=90}}
\end{center}
\vskip -1cm
\caption{
The  asymmetry  $\Acal$ for an ejected proton, Eq.~(\ref{eq.inelasymm}),
versus  $T_N=T_p=T_n$,  
at incident $\nu({\overline\nu})$ energy $E_\nu=1.0$~GeV. 
The solid lines correspond to the RSM calculation without
Final State Interactions, 
the dot--dashed lines include FSI effects through the ROP, 
the dotted lines  represent the RSM
corrected  by both the FSI and the Coulomb distortion of   
the outgoing muon. The three sets of curves correspond 
to different choices  of strangeness parameters: $g_A^s=\mu_s=0$   
(lower lines), $g_A^s=-0.15$, $\mu_s=0$ (intermediate  lines) and  
$g_A^s=-0.15$, $\mu_s=-0.3$ (upper lines).
(Taken from Ref.~\cite{ABBCGM})
}
\label{fig.asymQE}
\end{figure}

Results for $E_\nu=1$ GeV are shown in Fig.~\ref{fig.asymQE},
where the asymmetry ${\Acal}$, for the emission (in the NC processes) 
of a proton,  is plotted as a function 
of $T_N$ for three different choices of the strange 
form factors, as indicated in the caption. Here
the solid curves represent the Shell Model Calculations without FSI effects,
while RFG results are not displayed, since they coincide with the RSM ones.
The dot-dashed curve shows the effects of FSI as described by the 
Relativistic Optical Potential. The effect on $\Acal$ of both
FSI and Coulomb distortion is illustrated by the dotted line, 
which shows that, except for the 
region of small $T_N$, this combined effect is still small enough when 
it is compared with the effects of strangeness. 

%*************************************************************************
%*************************************************************************

A comment is required about the divergent behavior of the asymmetry
in Fig.~\ref{fig.asymQE}
for large $T_N$: this is associated with the effect of the outgoing
lepton (a muon in this case) mass, 
which brings the CC cross sections down to zero more rapidly than the NC ones.
\footnote{The outgoing lepton mass was neglected in the 
formalism presented in Section
\ref{sec.freeasymm}, its effects being irrelevant for 
the general considerations done there, 
but it has been included in the results presented in this Section.}

\begin{figure}
\begin{center}
\mbox{\epsfig{file=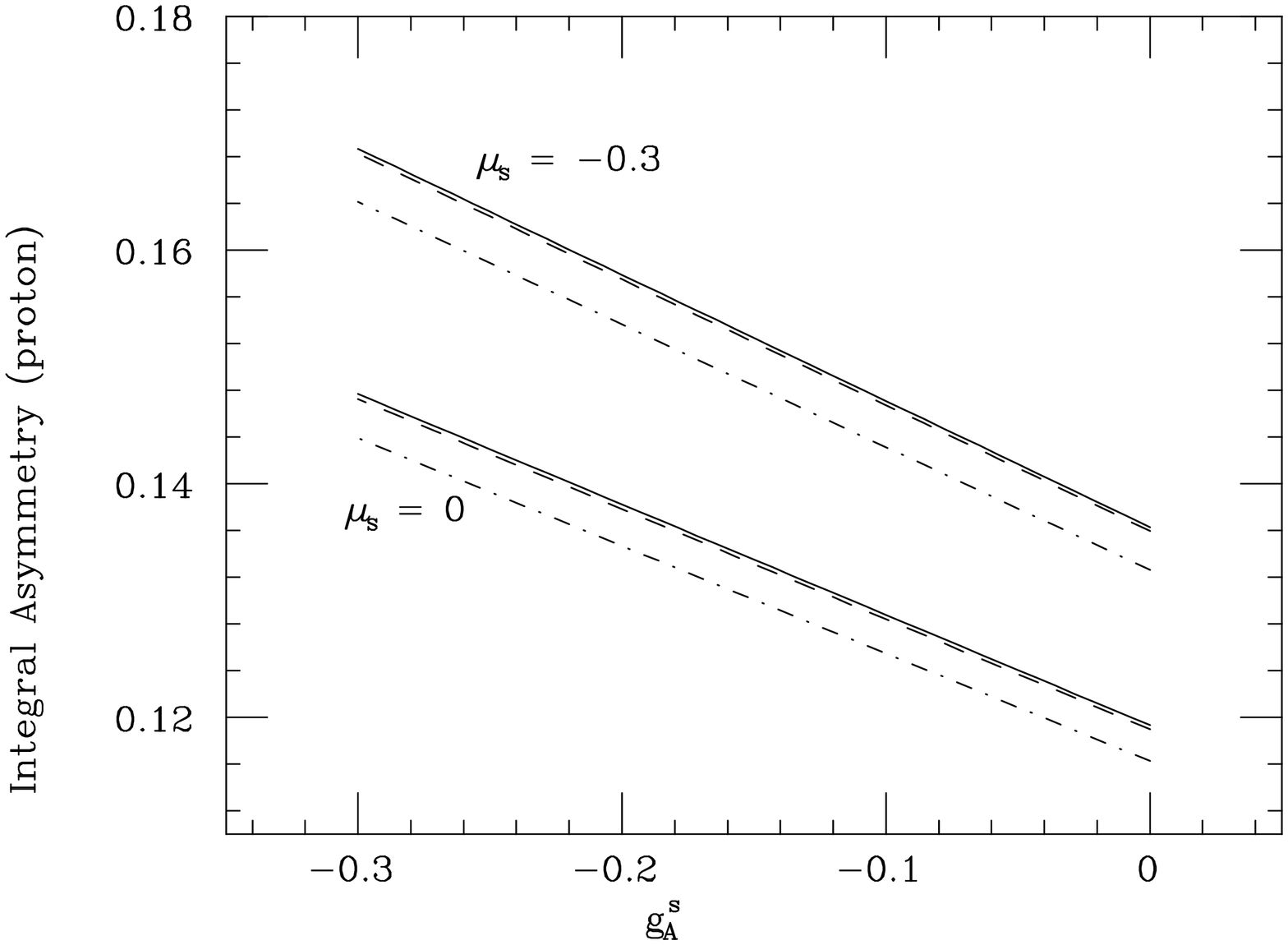,width=0.8\textwidth,angle=90}}
\end{center}
\vskip -0.9cm
\caption{
The  integral asymmetry, Eq.~(\ref{eq.intasymm}),  $\Acal^I$
for an ejected proton, versus $g_A^s$, 
at incident $\nu({\overline\nu})$ energy $E_\nu=1.0$~GeV. 
The lower limit for the integration is $T_N=100$ MeV.
The solid lines
correspond to the RSM calculation, the
dashed lines to the RFG with
average binding energy of $25$~MeV, the dot--dashed
lines include FSI effects.
The two sets of curves correspond to different choices  of the magnetic 
strangeness parameter: $\mu_s=0$ (lower lines) and  $\mu_s=-0.3$ 
(upper lines). (Taken from Ref.~\cite{ABBCGM})}
\label{fig.intasymm}
\end{figure}

The authors of Ref.~\cite{ABBCGM} considered also an ``integral asymmetry'', 
obtained as the ratio of NC and CC differences between the total cross
sections (\ref{eq.total}):
\begin{equation}
\Acal^I = 
\frac{ 
\sigma_{\nu}^{NC} - 
{\sigma}_{\overline\nu}^{NC}
}
{
\sigma_{\nu}^{CC}- \sigma_{\overline\nu}^{CC} 
}\;.
\label{eq.intasymm}
\end{equation}
The corresponding curves for the emission of a proton are shown in 
Fig.~\ref{fig.intasymm}.
%
%*************************************************************************
%*************************************************************************
%

\begin{figure}[h]
\begin{center}
%\mbox{\epsfig{file=fig1BNL.eps,width=0.8\textwidth,angle=-90}}
\mbox{\epsfig{file=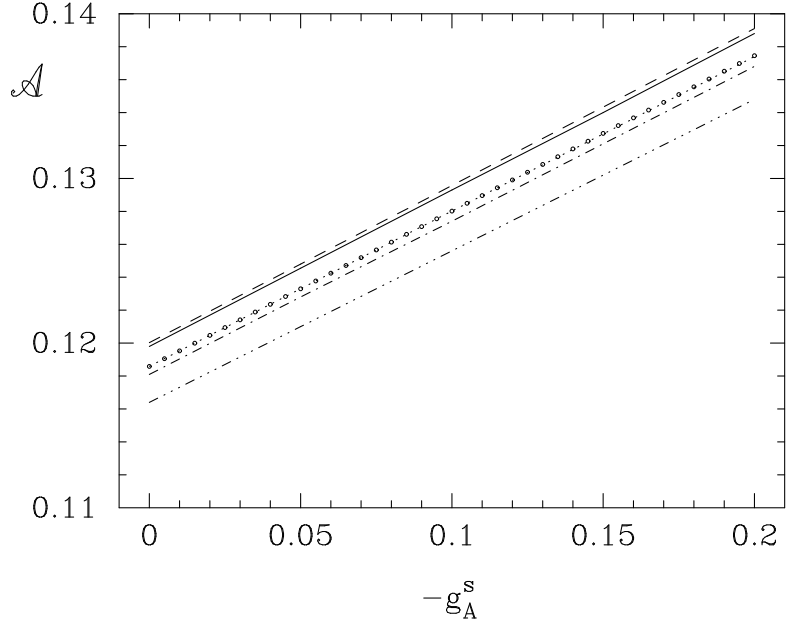,width=0.8\textwidth,angle=-90}}
\end{center}
\vskip -0.5cm
\caption{
The integral asymmetry ${\Acal}^I$ for an ejected proton
versus $-g_A^s$. The magnetic and electric strange form factors have
been set to zero, while  dipole parameterizations (with the same value
of the cut-off mass) have been assumed for the axial CC and the axial 
strange form factors. 
The solid line corresponds to the ``flux--averaged'' 
$\nu (\overline{\nu})$--proton 
elastic scattering asymmetry, the empty dots to 
elastic scattering without folding at
$E_{\nu (\overline{\nu})}=1$ GeV. 
Results for the QE asymmetry on $^{12}$C  are shown by
the following curves:
dashed line (RFG, with folding), dotted line (RFG, unfolded), 
dot--dashed line (RSM, unfolded) and 
three--dot--dashed line (RSM+ROP, unfolded); all unfolded curves
are evaluated at $E_{\nu (\overline{\nu})}=1$ GeV.
(Taken from  Ref.~\cite{ABBCGM99})}
\label{fig.BNLfolding}
\end{figure}

Since the specific sensitivity of the asymmetries in 
Eqs.~(\ref{eq.inelasymm}), (\ref{eq.intasymm})
to the parameters characterizing
the strange form factors of the nucleon 
derives from the cancellation of various contributions in the
difference between neutrino and antineutrino cross sections,
a crucial point, in considering realistic measurements, is the
role played by the different shapes of the available neutrino and
antineutrino spectra.
This issue was addressed in Ref.~\cite{ABBCGM99} for the Brookhaven
kinematical conditions: 
 the integrated cross sections 
(\ref{eq.total}) were considered both
for QE scattering on $^{12}C$ and for elastic scattering on free nucleons.
The integral asymmetry $\Acal^I$ (\ref{eq.intasymm}), calculated 
for $E_\nu=E_{\overline{\nu}}=1$~GeV, 
was compared with the following ``flux--averaged asymmetry'':
\begin{equation}
\langle {\Acal^I} \rangle = 
\frac{
\langle \sigma 
\rangle_{\nu}^{NC} - 
\langle \sigma 
\rangle_{\overline{\nu}}^{NC}
}{
\langle \sigma 
\rangle_{\nu}^{CC} - 
\langle \sigma 
\rangle_{\overline{\nu}}^{CC}
}\, .
\label{eq.foldasymm}
\end{equation}
where the quantities 
$
\langle \sigma 
\rangle_{\nu (\overline{\nu})}
$
are the total cross sections averaged over the 
BNL neutrino and antineutrino spectra 
[see Eq.~(\ref{eq.crossfold})] and
the limits for the 
integrations over the outgoing 
nucleon kinetic energy and over the neutrino
energy  correspond to the kinematics of the
BNL--734 Experiment~\cite{Ahrens87}, considered in Section \ref{sec.nufree}.
It was found that the effects of the folding amount to less than 2\%,
due to the similarity of the BNL $\nu$ and 
$\overline{\nu}$ spectra.
In Fig.~\ref{fig.BNLfolding} the integral asymmetry $\Acal^I$ 
is plotted as a function of the strange axial constant $-g_A^s$.
The solid line represents the folded asymmetry 
(\ref{eq.foldasymm}) for scattering on free
protons, which has to be compared with the unfolded elastic asymmetry
shown by the empty--dots. The dashed and dotted lines
show results for the QE emission of protons, within the RFG, with
and without folding. The curves obtained (without folding) using
the Relativistic Shell Model described previously in this Section 
are also shown, confirming
that at these energies nuclear effects on the asymmetry are small.
%

%************************************************************************
%************************************************************************

\begin{figure}
\begin{center}
\mbox{\epsfig{file=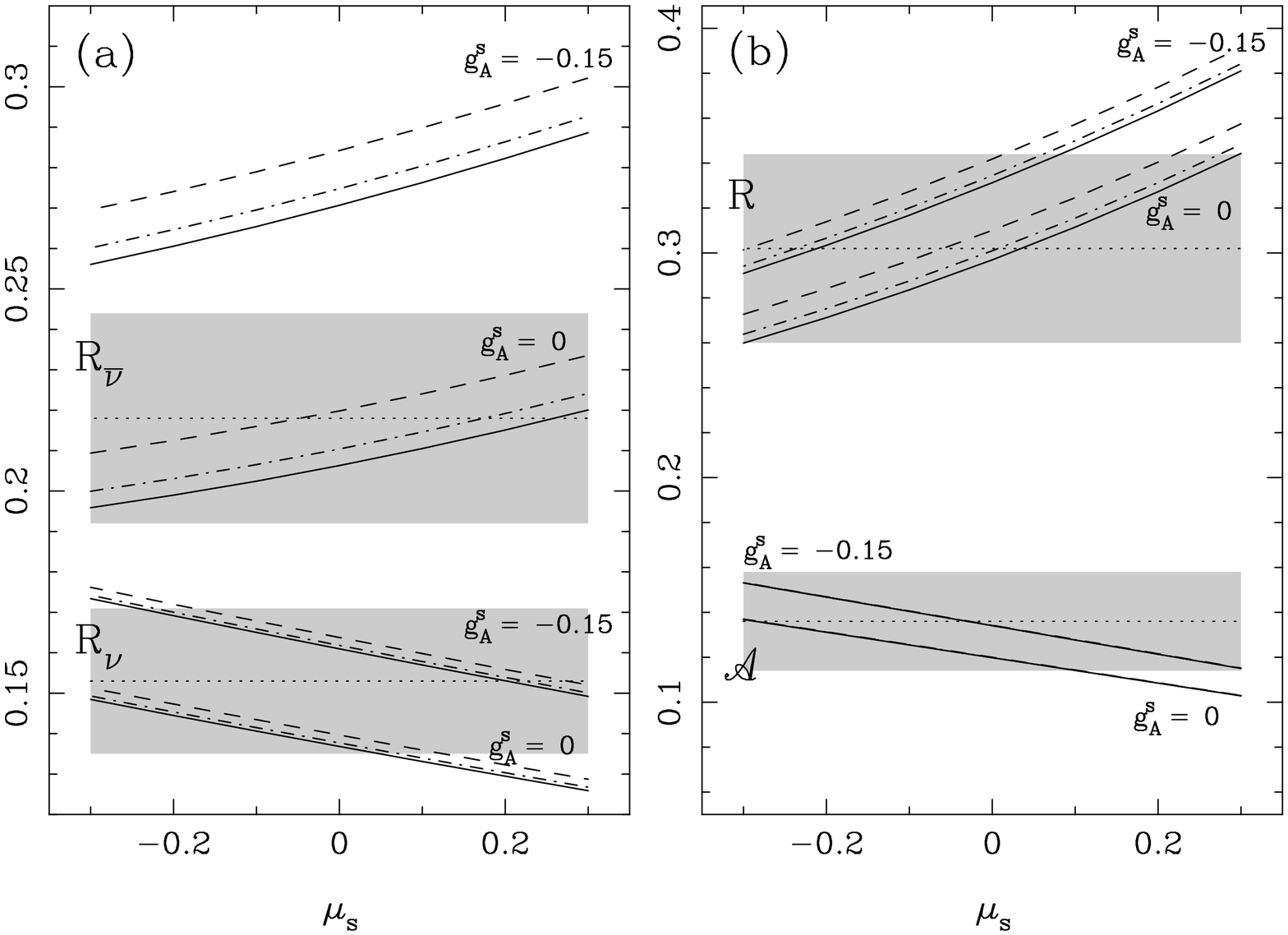,width=0.8\textwidth,angle=-90}}
%\vskip 2cm
\end{center}
\caption{The ratios $R^{BNL}_{\nu}$,  $R^{BNL}_{\overline{\nu}}$ and  
$R^{BNL}$,
here indicated without the superscript $BNL$ and the folded
 asymmetry $\la\mathcal{A}^I\ra$
as functions of $\mu_s$. All curves correspond to 
$\nu (\overline{\nu}) p$ elastic 
scattering in the kinematic conditions of Brookhaven. Results are shown for
$g_A^s=0,\,-0.15$ and for $\rho_s=0$ (solid line), 
$\rho_s=-2$ (dot--dashed line)  and $\rho_s=+2$ (dashed line). 
The shadowed regions correspond to the experimental 
data [Eqs.~(\ref{Rnu}), (\ref{Rnubar}), (\ref{RBNL}) and (\ref{eq.asymexp})] 
measured in the BNL--734 experiment.
(Taken from Ref.~\cite{ABBCGM99})
}
\label{fig.BNLexp}
\end{figure}

In Ref.~\cite{ABBCGM99} an indirect ``experimental value'' 
of the folded asymmetry
(\ref{eq.foldasymm}) for $\nu p$ elastic scattering 
was derived from the  
measured ratios of total elastic and quasi--elastic cross sections
given in Eqs.~(\ref{Rnu})--(\ref{RBNL}):
\begin{equation}
\langle \mathcal{A}^I \rangle = \frac{
\displaystyle{
R^{BNL}_{\nu} 
\left(1 - R^{BNL}\right)
 }}
{\displaystyle{
1 - R^{BNL}R^{BNL}_{\nu} /R^{BNL}_{\overline{\nu}}
}}
= 0.136 \pm 0.008 \,(\mathrm{stat})\,
\pm 0.019\,(\mathrm{syst})\,.
\label{eq.asymexp}
\end{equation}
The sensitivity of the ratios $R^{BNL}_{\nu}$,  $R^{BNL}_{\overline{\nu}}$,
$R^{BNL}$ and of the folded asymmetry $\la\mathcal{A}^I\ra$ 
to the strange \ffs of
the nucleon was then studied, by assuming the usual dipole parameterization
for the strange and non--strange form factors. The results for neutrino
and antineutrino proton scattering in the kinematical conditions of
the BNL--734 experiment are shown in Fig.~\ref{fig.BNLexp}. The 
quantities  $R^{BNL}_{\nu}$,  $R^{BNL}_{\overline{\nu}}$, $R^{BNL}$ and
$\la\Acal^I\ra$ are plotted as functions of the strange magnetic moment
$\mu_s$ for different values of the axial strange constant $g_A^s$ and 
of $\rho_s$,
while the axial cutoff mass is assumed to be $M_A=1.032$~GeV.
The horizontal dotted lines represent the experimental values and
the shaded areas are the corresponding error bands.
The latter are rather large and comparable with the
effects of strangeness and it is clear that more precise measurements 
are needed in order to extract information on the strange form factors.
Nevertheless we can observe that relatively large negative values of $g_A^s$
seem to be excluded by the antineutrino ratio $R^{BNL}_{\overline{\nu}}$.
The other ratios and the asymmetry are compatible with relatively large,
negative values of $g_A^s$: in this case a negative value of $\mu_s$
seems to be favored, in agreement with the findings of
Ref.~\cite{Garvey93a} but in contradiction with the SAMPLE results. 
In Ref.~\cite{ABBCGM99} the sensitivity of the above quantities 
to the axial cutoff mass
was also studied, in the range $M_A= 1.032 \pm 0.036$~GeV: 
in accord with previous 
analyses~\cite{Ahrens87,Garvey93a} the ratios $R^{BNL}_{\nu}$,  
$R^{BNL}_{\overline{\nu}}$, $R^{BNL}$ were found to be rather sensitive
to $M_A$, while (assuming the same
$Q^2$ dependence for $G_A$ and $G_A^s$) the asymmetry turned out 
to be practically
independent of this parameter. Noticing that, as shown in 
Fig.\ref{fig.BNLexp},
the asymmetry does not depend on the electric strange form factor $G_E^s$
we can conclude that a more precise measurement of this quantity could allow 
one
to obtain more ``clean'' information on $G_A^s$ and $G_M^s$.
\footnote{We observe that the independence of the asymmetry of 
both $M_A$ and $G_E^s$
can be understood by looking at Eq.~(\ref{nuasymm}). 
This property is maintained when
the cross sections which contribute to the asymmetry are integrated over
the momentum transfer and averaged over the BNL neutrino flux.}  

%************************************************************************
%************************************************************************

Finally we notice that most of the studies 
 of neutrino scattering QE processes
we presented here consider the  effects of negative values 
of $\mu_s$ only. This choice appeared to be favored by model calculations 
before the SAMPLE result relative to the proton data 
(see Section~\ref{sec.Poddelex}) was published. Instead, a positive
$\mu_s$ would interfere with a negative $g_A^s$, reducing 
the effect of the latter, on both the p/n neutrino ratio and the 
$\nu-\overline{\nu}$ asymmetry, the size of this reduction being dependent on 
the actual value of $\mu_s$. 
However, as we have seen considering also the deuteron data, 
the SAMPLE result is still strongly affected by 
uncertainties due to radiative corrections and the error bands of the
BNL--734 experiment do not allow to draw any definite conclusion.
More stringent results have
to be obtained before evaluating their quantitative impact on the neutrino
cross sections and ratios.

%%%%%%%%%%%%%%%%%%%%%%%%%%%%%%%%%%%%%%%%%%%%%%%%%%%%%%%%%%%%%%%%%%%%%%%
\section{Summary and Conclusions}

The strange axial and vector form factors of the nucleon
are an important issue which was under intensive
experimental and theoretical investigations during
the last decade. The first experimental evidence that
the $Q^{2}= 0$ value of the axial strange form factor of 
the nucleon, $g^{s}_{A}$, is unexpectedly large was obtained 
in 1989 in the famous EMC experiment (CERN) on the measurement
of deep inelastic scattering of polarized muons on polarized 
protons. The EMC result triggered new DIS experiments at 
CERN, SLAC and DESY and hundreds of theoretical papers
in which the connection of the constant
$g^{s}_{A}$ with the strange quarks polarization and the
polarization of gluons in the nucleon was elaborated in
detail. 

The EMC result also intensified the experimental and theoretical 
investigation of the problem of the strange form factors of the 
nucleon at relatively low values of $Q^{2}$ ($ \leq 1$~GeV). 

In order to obtain information on the strange form factors
of the nucleon it is necessary to investigate Neutral
Current induced processes. In the region of small
$Q^{2}$ only the $u$, $d$ and  $s$ parts of the NC should be
taken into account. In the Standard Model there are three 
different components of the NC:
\begin{enumerate}
\item isovector (vector and axial)  $u-d$ currents; 
\item electromagnetic current;
\item strange (vector and axial) currents.
\end{enumerate}

Thus, by using the available information on the electromagnetic
and CC axial form factors of the nucleon, from the investigation 
of the NC induced lepton (electron and/or neutrino) nucleon 
processes we can obtain information on the strange axial and 
vector form factors of the nucleon.

This strategy can be realized by the investigation of:
\begin{itemize}
\item the P--odd asymmetry in the elastic scattering of 
polarized electrons on nucleons;
\item neutrino(antineutrino) elastic scattering on nucleons;
\end{itemize}

In the case of electron--nucleon scattering
in the region of small $Q^{2}$ the diagram with the
exchange of a  $\gamma$--quantum
gives a much larger contribution to the matrix
element of the process than the diagram with $Z^{0}$
exchange. Thus, the P-odd asymmetry, which arises from the
interference between $\gamma$ and $Z^{0}$ exchanges, 
is very small ($\simeq 10^{-5}$ ). 
The measurement of such small asymmetries became
possible only when very intense beams of highly
polarized electrons and high resolution spectrometers
were developed at MIT/Bates, Jefferson Lab and MAINZ .
In Section \ref{sec.Poddelex} we have  discussed 
the interesting results that were obtained in the latest 
experiments.

The P--odd asymmetry in the polarized electron--proton 
scattering is determined by the interference of the scalar 
electromagnetic and  pseudoscalar $Z^0$--exchange parts of
the amplitude, $a V^{NC}$ and $v A^{NC}$: the latter, however,
occurs in the asymmetry multiplied by the small coefficient 
$g_{V} = -1/2 + 2 \sin^{2}\theta_{W}$ (from the electron NC).
Hence the P--odd asymmetry in $\vec{e}+p$ scattering is mainly 
sensitive to the strange {\em vector} form factors of the proton.
On the other hand, the elastic and quasi--elastic scattering of 
polarized electrons on nuclei allows one to obtain complementary
information: different choices of kinematics (e.g. backward 
versus forward scattering) and of nuclear targets can be used
to select and eventually enhance the various hadronic NC 
components.  At the moment this possibility has not been 
experimentally tested, but several experiments are under way
or foreseen in the near future, which will measure the P-odd
asymmetry in light nuclei (deuterium, $^4$He, etc.).

The NC neutrino (antineutrino)
nucleon scattering allows one to obtain information both
on the axial and vector strange form factors 
(see Section \ref{sec.nufree}).
Moreover, none of  the contributions of the strange form factors
to the cross sections of these processes is, in principle, 
suppressed. By investigating  neutrino (antineutrino) nucleon
scattering (as well as the P--odd asymmetry in electron--nucleon 
scattering) one can also hope to obtain information on the 
{\it $Q^{2}$ behavior of the strange form factors}.
The most detailed investigation of neutrino
(antineutrino) nucleon scattering was done in
1987 in the BNL--734 Brookhaven experiment, of which
we discussed here in detail the results.

In order to obtain information on the strange
form factors of the nucleon from the data of neutrino
experiments it is necessary to know with good accuracy the
$Q^{2}$ behavior of the CC axial form factor.
There is no such information at present. Yet, we have
discussed in Section \ref{sec.freeasymm} a method that allows one
to obtain model independent information on the strange form 
factors of the nucleon. For this purpose it is necessary to 
measure the neutrino--antineutrino asymmetry.
There is no doubt that with the many new neutrino
experiments and the suggested new neutrino facility, the neutrino 
factory~\cite{Albright00}, the problem of the strange
form factors of the nucleon will have new development. 

It is worth mentioning that usually in neutrino experiments nuclear 
targets are used. A large part of this review is devoted to the
detailed consideration of nuclear effects 
in the processes of neutrino (antineutrino) scattering on nuclei
(see Sections \ref{sec.nuAel} and \ref{sec.nuQE}). These effects
are usually quite relevant on the single cross sections, but they
can be drastically reduced when one considers ratios of cross
sections. This fact has been widely underlined also in connection
with the P--odd asymmetry in the (elastic or inelastic) scattering 
of polarized electrons on nuclei (see Sections \ref{sec.PoddAel}
 and \ref{sec.PoddQE}).

Finally we mention that in the introductory parts (see Sections
\ref{sec.intro}--\ref{sec.strangeFF}) 
many basic phenomenological relations are derived in sufficient
 detail, that they can  easily be followed by the reader. We
believe that this review will be useful
for many physicists who are or will be interested in the problem of
 strangeness in the nucleon.

\vspace{1cm}
\section*{Acknowledgments}
The authors are deeply grateful to A.~Molinari, T.W.~Donnelly, E.~Barone,
A.~De Pace and C.~Giunti, for helpful discussions during the preparation 
of the manuscript. S.M.B. and C.M. acknowledge support from Department of 
Theoretical Physics, University of Torino and INFN.
S.M.B. thanks the Physics Department of the Helsinki University for the
hospitality during the initial stage of this work and the Alexander von 
Humboldt Foundation for support.
This work was supported in part by Italian MURST under contract 
N.~9902198839.

%%%%%%%%%%%%%% \BIBLIOGRAPHY

\end{document}